\documentclass[aps,prd,showpacs,notitlepage,nofootinbib,superscriptaddress,floatfix,showkeys,twocolumn]{revtex4-1}

\usepackage{blindtext}
\usepackage{hyperref}
\usepackage{amsmath,amssymb}
\usepackage{float}
\usepackage{microtype}
\usepackage{ragged2e}  % Add this in your preamble
\usepackage{placeins}
\usepackage{graphicx}
\usepackage{bm}
\usepackage{latexsym}
\usepackage{epsfig}
\usepackage{psfrag}
\usepackage[dvipsnames]{xcolor}
\usepackage{subcaption}
\usepackage{graphicx}
\usepackage{multirow}
\usepackage{amssymb}
\usepackage{amsmath}
\usepackage{bm}
\usepackage{latexsym}
\usepackage{epsfig}
\usepackage{psfrag}
\usepackage[normalem]{ulem}
\usepackage{textcomp}
\usepackage{color}
\usepackage{pstricks}
\usepackage[utf8]{inputenc}
\usepackage{comment}
\usepackage{hyperref}% add hypertext capabilities
\usepackage[mathlines]{lineno}% Enable numbering of text and display math
\usepackage{hyperref}
\usepackage[all]{hypcap}
\hypersetup{  % set up for colour hyperlink
    colorlinks=true,
    linkcolor=blue,
    filecolor=red,      
    urlcolor=cyan,
    citecolor=green,
    }

\begin{document}
%opening
\title{Tracing Signatures of Modified Gravity in Redshift-Space Galaxy Bispectrum Multipoles: Prospects for \textit{Euclid}}

\author{Sourav Pal}
\email{soupal1729@gmail.com }
\affiliation{\,Physics and Applied Mathematics Unit, Indian Statistical Institute, 203 B.T. Road, Kolkata 700108, India}

\author{Debanjan Sarkar}
\email{debanjan.sarkar@mcgill.ca}
\affiliation{\,Department of Physics and Trottier Space Institute, McGill University, QC H3A 2T8, Canada}
\affiliation{\,Ciela—Montreal Institute for Astrophysical Data Analysis and Machine Learning, QC H2V0B3, Canada}

\author{Alejandro Aviles}
\email{aviles@icf.unam.mx}
\affiliation{Instituto de Ciencias Físicas, Universidad Nacional Autónoma de México, 62210, Cuernavaca, Morelos}

%\pacs{PACS}

\begin{abstract}
We study the galaxy bispectrum multipoles in the Hu–Sawicki $f(R)$ gravity model, where a scalar degree of freedom mediates a fifth force that is screened in high-density environments. The model is specified by $f_{R0}$, the present-day background value of the scalar field, which controls the strength of deviations from General Relativity (GR). Using perturbation theory, we compute the redshift-space galaxy bispectrum with the full scale- and time-dependent second-order kernels, incorporating corrections from the scale-dependent growth rate and nonlinear screening. Expanding the bispectrum in spherical harmonics, we analyze the sensitivity of the multipoles to modified gravity and forecast their detectability in a \textit{Euclid}-like survey. The monopole ($B_0^0$) and quadrupole ($B_2^0$) show the strongest signatures, with relative deviations of $2\%$--$8\%$ at $z=0.7$ and $k_1\simeq0.3\,h\,{\rm Mpc}^{-1}$ (largest side of the triangle)  for $f_{R0}=10^{-5}$. Higher multipoles provide weaker but complementary signals. For \textit{Euclid}, we forecast signal-to-noise ratios up to $\sim30$ for the monopole and $\sim15$ for the quadrupole including the Finger-of-God damping and shot noise effect. These results demonstrate that bispectrum multipoles are a powerful probe of gravity, capable of breaking degeneracies with bias and velocity effects and strengthening constraints on deviations from $\Lambda$CDM.

\end{abstract}

\maketitle
%\tableofcontents

%%%%%%%%%%%%%%%%%%%%%%%%%%%%%%%%%%%%%%%%%%%%%%%%%%%%%%%%%%%%%%%%%%%%%%%%%%%%%%%%%
%%%%%%%%%%%%%%%%%%%%%%%%%%%%%%%%%%%%%%%%%%%%%%%%%%%%%%%%%%%%%%%%%%%%%%%%%%%%%%%%%

\section{Introduction}

The discovery of the accelerated expansion of the Universe at the turn of the century has transformed cosmology into a precision science, while simultaneously posing one of its deepest puzzles~\cite{SupernovaCosmologyProject:1997zqe,SupernovaCosmologyProject:1998vns}. Within the standard cosmological model, this phenomenon is attributed to a cosmological constant, which together with cold dark matter (CDM) and General Relativity (GR) forms the concordance $\Lambda$CDM framework~\cite{Carroll:1991mt,Efstathiou:1990xe,Krauss:1995yb,Schmidt:2012xxa}. Despite its empirical success, $\Lambda$CDM suffers from severe theoretical challenges, most notably the fine-tuning and coincidence problems associated with the cosmological constant~\cite{Weinberg:1988cp,Zlatev:1998tr,Sahni:2002kh,Velten:2014nra,SolaPeracaula:2022hpd}. These shortcomings motivate the exploration of alternative explanations, including modifications to the underlying theory of gravity itself. Scalar–tensor extensions such as $f(R)$ gravity represent particularly well-studied candidates, offering self-consistent models in which cosmic acceleration arises without a finely tuned vacuum ~\cite{Koyama:2015vza,Berti:2015itd,deRham:2023byw,Bloomfield:2012ff,Gubitosi:2012hu,Horndeski:1974wa,Crisostomi:2016tcp,Langlois:2015skt,Crisostomi:2016czh,Sotiriou:2008rp,Lobo:2014ara,Linder:2010py,Turner:2007qg}. A central challenge, however, is to identify observational signatures that can distinguish these scenarios from GR while being robust to astrophysical and systematic uncertainties~\cite{Baker:2014zba,Joyce:2014kja,Berry:2011pb}.

Large-scale structure (LSS) surveys provide one of the most powerful arenas for testing gravity on cosmological scales. The growth of cosmic structures, as traced by galaxies, is directly influenced by the underlying theory of gravity and therefore offers a sensitive probe of deviations from GR~\cite{Peebles:1980yev,Huterer:2016uyq,Kazantzidis:2018rnb,Alam:2016qcl,Blake:2020mzy,Aviles:2024zlw,Li:2025mib,Viglione:2025tlh}. While two-point statistics such as the power spectrum or correlation function have long been the workhorses of cosmological analysis, they are fundamentally limited in their ability to capture the full statistical content of the galaxy distribution. The late-time Universe is highly non-Gaussian, and valuable information resides in higher-order statistics. The bispectrum—the Fourier transform of the three-point correlation function—directly probes the mode-coupling induced by nonlinear gravitational evolution and galaxy bias. As a result, it encodes complementary information that is inaccessible to the power spectrum alone. In particular, modified gravity models imprint characteristic, configuration-dependent signatures in the bispectrum through their scale- and time-dependent growth rates, providing a means to disentangle them from the predictions of GR.

Beyond its intrinsic information content, the bispectrum also plays a crucial role in mitigating degeneracies~\cite{Scoccimarro:1999ed,Verde:1998zr,Gil-Marin:2014sta,Gil-Marin:2016wya,Nandi:2024cib}. For example, constraints from the galaxy power spectrum are often limited by uncertainties in galaxy bias, redshift-space distortions, or small-scale velocity dispersions. The bispectrum, by virtue of its distinct dependence on triangle configurations and multipoles~\cite{Peacock:2001gs,Hawkins:2002sg,Guzzo:2008ac,2011ApJ...726....5O,Okumura:2023pxv,Hashimoto:2017klo,Nan:2017oaq}, provides additional handles that can help disentangle these effects. Several studies have shown that combining the power spectrum and bispectrum can significantly tighten cosmological constraints, improving sensitivity to both fundamental parameters and nuisance terms~\cite{Upadhye:2017hdl,DESI:2025ejh,Verdiani:2025znc,Yankelevich:2018uaz,Philcox:2022frc,Philcox:2022lbx,Bose:2019wuz,Gualdi:2018pyw,Gualdi:2020ymf,Slepian:2025fdx}. In the context of modified gravity, this synergy is particularly important: the bispectrum can isolate scale-dependent deviations in the growth of structure that may otherwise be masked when using only two-point statistics~\cite{Linder:2007nu,delaTorre:2016rxm,Rodriguez-Meza:2023rga,Valogiannis:2019nfz}.

The Hu–Sawicki (HS) $f(R)$~\cite{Hu:2007nk,Sotiriou:2008rp,DeFelice:2010aj,Koyama:2009me} gravity model offers a concrete and widely studied case where such effects are expected to be observable. In this model, a scalar degree of freedom mediates a fifth force that enhances structure formation on large and intermediate scales. Screening mechanisms, such as the chameleon effect~\cite{Khoury:2003aq,Li:2011qda,Li:2011vk}, ensure consistency with solar system and astrophysical tests by suppressing deviations from GR in high-density environments. This interplay gives rise to a scale- and redshift-dependent modification of structure growth, leaving distinctive signatures in higher-order clustering statistics. The bispectrum multipoles, especially in redshift space where anisotropies induced by peculiar velocities become important, therefore provide a natural laboratory to search for such effects. 

In this work, we generalize the spherical harmonic bispectrum formalism developed in Refs.~\cite{Bharadwaj:2020wkc,Mazumdar:2020bkm,Mazumdar:2022ynd} to incorporate the modifications predicted by HS-$f(R)$ gravity. We compute the redshift-space bispectrum multipoles $B_L^m(k_1,\mu,t)$, where $k_1$ denotes the longest triangle side and ($\mu,t$) encode the shape of the triangle. 
Our formalism consistently includes both the scale-dependent linear growth and the nonlinear kernel corrections induced by HS-$f(R)$ model. The resulting multipoles display distinctive redshift and shape dependence, offering robust signatures of departures from GR.  

The observational prospects for this approach are timely. Next-generation surveys like Euclid~\cite{Amendola:2016saw,Euclid:2021icp,Euclid:2024vss,Euclid:2024yrr,Euclid:2021qvm,Euclid:2021icp,Rhodes:2017nxl,Euclid:2014lvu,Vogt:2024pws,Euclid:2023tog}
% DESI~\cite{DESI:2023dwi},
and the Rubin Observatory’s LSST~\cite{LSST:2008ijt,Zhan:2017uwu,2019BAAS...51g.268J} will map the galaxy distribution over unprecedented volumes and redshift ranges, delivering high-precision measurements of clustering up to nonlinear scales. Fully capitalizing on this opportunity requires theoretical predictions that are accurate and flexible enough to account for both modified gravity dynamics and observational systematics. This work takes a step in that direction by presenting a systematic analysis of the bispectrum multipoles in the HS-$f(R)$ model. We examine how modified gravity imprints manifest across different triangle configurations, redshifts, and multipoles, and assess the signal-to-noise expected for \textit{Euclid}-like surveys. By highlighting which configurations are most sensitive to deviations from GR, we aim to clarify the role of the bispectrum as a complementary probe of gravity and to establish its potential in breaking degeneracies with bias parameters and velocity effects.

The paper is organized as follows. In Sec.~\ref{sec:formalism}, we review the $f(R)$ gravity framework and the computation of the bispectrum arising from gravitational nonlinearities in both Fourier and redshift space. In Sec.~\ref{subsec:kernel_bispectra}, we discuss the evolution of higher-order perturbative modes in HS-$f(R)$ model and present the corresponding bispectrum formalism. In Sec.~\ref{sec:multipoles_fR}, we compute in detail the redshift-space corrections to the bispectrum multipoles induced in $f(R)$ gravity, providing quantitative estimates for all multipole moments. Sec.~\ref{sec:formalism_snr} introduces the bispectrum multipole estimator used to evaluate the signal-to-noise ratios of the corresponding multipoles. In Sec.~\ref{sec:euclid}, we present predictions for the signal-to-noise of various bispectrum multipoles in the context of \textit{Euclid}, incorporating the effects of shot noise and Fingers-of-God. We identify the multipoles and triangle configurations where deviations from GR are most pronounced in the $f(R)$ model. Finally, Sec.~\ref{sec:summary} summarizes our results and discusses future prospects.
%%%%%%%%%%%%%%%%%%%%%%%%%%%%%%%%%%%%%%%%%%%%%%%%
\section{Formalism}
\label{sec:formalism}
%%%%%%%%%%%%%%%%%%%%%%%%%%%%%%%%%%%%%%%%%%%%%%%%
%%%%%%%%%%%%%%%%%%%%%%%%%%%%%%%%%%%%%%%%%%%%%%%%
\subsection{Induced Bispectrum in GR}
\label{subsec:induced_bispectrum}
%%%%%%%%%%%%%%%%%%%%%%%%%%%%%%%%%%%%%%%%%%%%%%%%
The large-scale distribution of matter in the Universe can be understood through the lens of cosmological perturbation theory, where one analyzes fluctuations around a smooth background. Specifically, the matter density field is written as a sum of a homogeneous component and a perturbation: $ \rho(\mathbf{x},\tau) = \bar{\rho}(\tau)\left[1 + \delta(\mathbf{x},\tau)\right]$,
where $\bar{\rho}(\tau)$ is the background matter density and $\delta(\mathbf{x},\tau)$ is the fractional density perturbation, or density contrast. The background quantity depends only on the time whereas, the perturbed quantity involves both spatial and temporal dependency. 
In the linear regime of $\Lambda$CDM universe, where $\delta \ll 1$, the Fourier modes of $\delta(\mathbf{x},\tau)$ evolve independently. This motivates a Fourier-space approach to structure formation, with the Fourier transform $\delta(\mathbf{k},\tau)$ capturing the mode evolution. Within the Einstein-de Sitter (EdS) approximation, valid deep in the matter-dominated era, time and scale dependence decouple, allowing the density contrast to be expressed as a perturbative series \cite{Bernardeau:2001qr,Fasiello:2022lff}:
\begin{equation}
    \delta(\mathbf{k},\tau) = \sum_{n=1}^{\infty} \delta_n(\mathbf{k},\tau),
\end{equation}
where each $\delta_n$ denotes the $n^{\rm th}$-order contribution.

Note that, each $\delta_n$ is convolution over $n$ number of linear fields ($\delta_1$), weighted by symmetrized kernels $F_n$, which encapsulate mode couplings in the mildly non-linear regime. They can be expressed as follows,
\begin{align}
    \delta_n(\mathbf{k},\tau) &= \int \prod_{i=1}^n \frac{d^3\mathbf{k}_i}{(2\pi)^3} \delta^{(3)}_{\mathrm{D}} F_n(\mathbf{k}_1,\ldots,\mathbf{k}_n) \times  \nonumber \prod_{i=1}^n \delta_1(\mathbf{k}_i).
\end{align}
Similarly, the velocity divergence field $\theta(\mathbf{k},\tau)$ admits an analogous expansion with kernels $G_n$ \cite{Scoccimarro:1995if,Scoccimarro:1999ed}:
\begin{equation}
    \theta_n(\mathbf{k},\tau) = \int \prod_{i=1}^n \frac{d^3\mathbf{k}_i}{(2\pi)^3} \delta^{(3)}_{\mathrm{D}} G_n(\mathbf{k}_1,\ldots,\mathbf{k}_n)\prod_{i=1}^n \delta_1(\mathbf{k}_i).
\end{equation}
The Dirac delta function here, enforces momentum conservation. The kernels satisfy $F_1 = G_1 = 1$ by construction in $\Lambda$CDM background.

However, the initial density perturbations are assumed to be Gaussian in standard cosmology, implying vanishing three-point correlation functions (3PCFs) by Wick's theorem~\cite{Wick:1950ee,Desjacques:2010jw}, gravitational evolution in late time can induce non-Gaussianities. As a result, higher-order correlators such as the bispectrum (the Fourier transform of the 3PCF) acquire non-zero values even at tree level.

At second order, the density contrast $\delta_2$  and the velocity perturbation $\theta_2$ involves the kernel $F_2$ and $G_2$, taking the following form in the EdS universe~\cite{Bernardeau:2001qr,Scoccimarro:1995if}:
\begin{equation}
F_2(\mathbf{k}_1,\mathbf{k}_2) = \frac{5}{7} + \frac{\mathbf{k}_1 \cdot \mathbf{k}_2}{2}\left(\frac{1}{k_1^2} + \frac{1}{k_2^2}\right) + \frac{2}{7} \frac{(\mathbf{k}_1 \cdot \mathbf{k}_2)^2}{k_1^2 k_2^2},
\end{equation}
% with the velocity counterpart given by:
\begin{equation}
G_2(\mathbf{k}_1,\mathbf{k}_2) = \frac{3}{7} + \frac{\mathbf{k}_1 \cdot \mathbf{k}_2}{2}\left(\frac{1}{k_1^2} + \frac{1}{k_2^2}\right) + \frac{4}{7} \frac{(\mathbf{k}_1 \cdot \mathbf{k}_2)^2}{k_1^2 k_2^2}.
\end{equation}

The tree-level matter bispectrum, constructed from two linear and one second-order field, is given by:
\begin{equation}
    B(\mathbf{k}_1,\mathbf{k}_2,\mathbf{k}_3) = \langle \delta_2(\mathbf{k}_1) \delta_1(\mathbf{k}_2) \delta_1(\mathbf{k}_3) \rangle + \text{cyc...},
\end{equation}
subject to the triangle closure condition $\mathbf{k}_1 + \mathbf{k}_2 + \mathbf{k}_3 = \mathbf{0}$.

To describe the triangle configurations, we define $k_1 = |\mathbf{k}_1|$, $t = k_2/k_1$, and $\mu = -\hat{\mathbf{k}}_1 \cdot \hat{\mathbf{k}}_2$. The third side is given by $k_3 = k_1 s$, with $s = \sqrt{1 - 2\mu t + t^2}$. We restrict to $k_1 \geq k_2 \geq k_3$, implying:
\begin{equation}
    0.5 \leq t \leq 1, \quad 0.5 \leq \mu \leq 1, \quad \text{and} \quad 2\mu t \geq 1.
\end{equation}
Thus, the triangle is fully specified by $(k_1, t, \mu)$ \cite{Bharadwaj:2020wkc,Pal:2025hpl}. The configuration space spanned by the parameters $(\mu, t)$ encodes distinct triangle shapes in Fourier space. In the extreme limit where $\mu \to 1$ and $t \to 1$, the triangle becomes highly squeezed, corresponding to $\mathbf{k_1} \approx -\mathbf{k_2}$ and $\mathbf{k_3} \to \mathbf{0}$. In contrast, the point $(\mu \to 1, t \to 1/2)$ marks stretched configurations, where $\mathbf{k_2} \approx \mathbf{k_3} \approx -\mathbf{k_1}/2$. Along $\mu = 1$, all three vectors, $\mathbf{k_1}$, $-\mathbf{k_2}$, and $-\mathbf{k_3}$ are aligned, forming linear triangles. Here $t = 1$ corresponds to L-isosceles configurations, where the longer sides, $\mathbf{k_1}$ and $\mathbf{k_2}$, are equal in magnitude. Meanwhile, $2\mu t = 1$ denotes S-isosceles triangles with $\mathbf{k_2}$ and $\mathbf{k_3}$ of equal length. The line $\mu = t$ identifies right-angled triangles, while the regions where $\mu > t$ and $\mu < t$ represent acute and obtuse triangles, respectively. 

In redshift-space, peculiar velocities distort galaxy positions along the line-of-sight (LoS), introducing anisotropies. Due to this, the redshift-space bispectrum $B^s$ depends on triangle shape, size, and orientation. To track this, following Ref.~\cite{Bharadwaj:2020wkc}, we define the triangle in the $x$-$z$ plane  as,
\begin{align}
    \mathbf{k}_1 &= k_1 \hat{z}, \\ \nonumber
    \mathbf{k}_2 &= k_1 t (-\mu \hat{z} + \sqrt{1-\mu^2} \hat{x}), \\ \nonumber
    \mathbf{k}_3 &= -(\mathbf{k}_1 + \mathbf{k}_2).
    \label{eq:coordinate}
\end{align}
Rotating this triangle with Euler angles $(\alpha, \beta, \gamma)$ generates all possible orientations in redshift space. The orientation with respect to the LoS, $\hat{n} = \hat{z}$, is captured by cosines $\mu_i = \hat{\mathbf{k}}_i \cdot \hat{\mathbf{n}}$, which also serves us to introduce the direction $\hat{p}$, as
\begin{align}
    \mu_1 &= \cos\beta = p_z, \\ \nonumber
    \mu_2 &= -\mu p_z - \sqrt{1-\mu^2} \sin\beta \cos\gamma = -\mu p_z - \sqrt{1-\mu^2} p_x, \\ \nonumber
    \mu_3 &= -\frac{\mu_1 k_1 + \mu_2 k_2}{k_3} = -s^{-1}[(1 - t\mu)p_z + t \sqrt{1-\mu^2}p_x].
    \label{eq:pvec}
\end{align}

The anisotropic bispectrum $B^s$ can be decomposed into spherical harmonics in the direction of $\hat{\mathbf{p}}$:
\begin{equation}
    B^m_L(k_1, t, \mu) = \sqrt{\frac{2L+1}{4\pi}} \int d\Omega_{\hat{\mathbf{p}}} \, [Y^m_L(\hat{\mathbf{p}})]^* \, B^s(k_1, t, \mu, \hat{\mathbf{p}}),
    \label{eq:Blm}
\end{equation}
with non-zero multipoles for even $L \leq 8$ and $|m| \leq 6$ \cite{Scoccimarro:1999ed,Mazumdar:2020bkm}. The components of $\hat{\mathbf{p}}$ can be computed from relations in Eq.~\ref{eq:coordinate}. These multipoles serve as the main observables in our analysis.

To compute the bispectrum of galaxies, we need the galaxy density contrast. Since galaxies trace the matter density field with a bias that encapsulates nonlinear galaxy formation physics, the galaxy density contrast $\delta_g$ is modeled as \cite{Desjacques:2016bnm,Gil-Marin:2014sta}:
\begin{equation}
    \delta_g = b_1 \delta_m + b_2\left(\delta_m^2 - \langle \delta_m^2 \rangle\right) + b_t S_m + \cdots,
\label{eq:gal_bias}
\end{equation}
where $b_1$ and $b_2$ are the linear and quadratic biases, and $S_m$ is the tidal field contribution. 
% (neglected in our analysis following Ref.~\cite{Pal:2025hpl}).

Finally, incorporating these ingredients, the redshift-space galaxy bispectrum can be written as~\cite{Mazumdar:2020bkm,Mazumdar:2022ynd}:
\begin{align}\label{eq:b_rsd}
B^s(\mathbf{k}_1,\mathbf{k}_2,\mathbf{k}_3) &=  2 b_1^{3} (1 + \beta_1 \mu_1^2)(1 + \beta_1 \mu_2^2)  \bigg[  F_2(\mathbf{k}_1,\mathbf{k}_2) +  \frac{\gamma_2}{2} \\ \nonumber  &+ \beta_1 \mu_3^2 G_2(\mathbf{k}_1,\mathbf{k}_2) - \frac{b_1 \beta_1 \mu_3 k_3}{2} (\frac{\mu_1}{k_1}(1 + \beta_1 \mu_2^2)  \\ \nonumber  &+   \frac{\mu_2}{k_2}(1 + \beta_1 \mu_1^2) ) \bigg] 
\times P(k_1)P(k_2) + \text{cyc...},
\end{align}
where $\beta_1 = f/b_1$ %\aa{Is there a reason why call it $\beta_1$ and not simply $\beta$?} 
and $\gamma_2 = b_2/b_1$. For simplicity we omit the tidal bias ($b_t$) term here, though it can be reintroduced via standard methods (see Sec.~\ref{sec:euclid} for a discussion on its impact). $f$ is the logarithmic growth rate in the corresponding background model ($\Lambda$CDM universe in this case).  
The framework described above can be straightforwardly extended beyond the $\Lambda$CDM model to extract cosmological information in alternative theories of gravity. In the following section, we illustrate this by applying it to a representative case: the Hu–Sawicki $f(R)$ gravity model.
%%%%%%%%%%%%%%%%%%%%%%%%%%%%%%%%%%%%%%%%%%%%%%%%
\subsection{$f(R)$ Gravity Model}
\label{subsec:fR_model}
%%%%%%%%%%%%%%%%%%%%%%%%%%%%%%%%%%%%%%%%%%%%%%%%
General Relativity has been the cornerstone of our understanding of gravity. Despite its success, challenges such as cosmic acceleration and the nature of dark energy have motivated exploration beyond GR \cite{Lobo:2014ara,Linder:2010py,Turner:2007qg}. One well-studied modification is $f(R)$ gravity, in which the Einstein-Hilbert action is generalized by replacing the Ricci scalar $R$ with a nonlinear function $f(R)$~\cite{Hu:2007nk,DeFelice:2010aj}. 
This leads to extended gravitational dynamics while retaining covariance and metric compatibility.
The action for $f(R)$ gravity is given by,
\begin{equation}
S = \frac{1}{2\kappa^2} \int d^4x\, \sqrt{-g}\, f(R) + S_{\rm matter}[g_{\mu\nu}, \Psi],
\label{eq:action}
\end{equation}
where $\kappa^2 = 8\pi G$ with $G$ being the gravitational constant, $g$ is the determinant of the metric $g_{\mu\nu}$, and $S_{\rm matter}$ is the matter action, which depends on the metric and matter fields $\Psi$.

Varying the action $S$ with respect to the metric yields the field equation:
\begin{equation}
f_R R_{\mu\nu} - \frac{1}{2} f g_{\mu\nu} + \left( g_{\mu\nu} \Box - \nabla_\mu \nabla_\nu \right) f_R = \kappa^2 T_{\mu\nu},
\label{eq:eom}
\end{equation}
where $f_R \equiv \dfrac{df}{dR}$ and $T_{\mu\nu}$ is the energy–momentum tensor of matter,
$\nabla_{\mu}$ the Levi-Civita covariant derivative, and $\Box\equiv g^{\alpha\beta}\nabla_{\alpha}\nabla_{\beta}$ the d’Alembertian.

Because $f_{R}$ contains second derivatives of the metric, Eq.~\ref{eq:eom} is fourth-order in $g_{\mu\nu}$ and thus propagates an extra scalar degree of freedom~\cite{Horndeski:1974wa,Sotiriou:2006hs,Zumalacarregui:2016pph,Kobayashi:2019hrl,Ji:2024gdc,Aviles:2020wme,Aviles:2018saf}.
Introducing the scalaron $\phi \equiv f_{R}$, one can recast the theory as a scalar–tensor model whose Einstein-frame potential is
\begin{equation}
V(\phi) = \frac{1}{2\kappa^2} \left( R f_R - f \right).
\end{equation}
Under a conformal transformation $g_{\mu\nu}\!\to\!\tilde g_{\mu\nu}=e^{2\omega}g_{\mu\nu}$ with $e^{2\omega}=f_{R}$, the action becomes that of GR plus a canonical scalar field $\phi$ that mediates an additional (``fifth-’’) force.  Local tests of gravity therefore impose stringent bounds on how strongly $\phi$ can couple to matter ~\cite{Chiba:2006jp,Capozziello:2007eu}. These models offer a potential explanation for cosmic acceleration without invoking a cosmological constant. Among the $f(R)$ family, a widely used form that satisfies cosmological and local constraints is
the Hu–Sawicki (HS) $f(R)$ model~\cite{DeFelice:2010aj,Hu:2007nk} with, 
\begin{equation}
f(R) = -m^2 \frac{c_1 \left( \dfrac{R}{m^2} \right)^n}{c_2 \left( \dfrac{R}{m^2} \right)^n + 1}\,.
\label{eq:func_fR}
\end{equation}
Here, $m^2 = H_0^2 \, \Omega_{m0}$, where $H_0$ is the Hubble parameter today and $\Omega_{m0}$ is the present-day matter density fraction. The parameters $c_1$, $c_2$, and $n$ control the deviation from GR. Requiring that the background expansion mimics that of $\Lambda$CDM at high curvature (i.e., $R \gg m^2$) imposes a relation among the parameters and allows the model to be re-parametrized in terms of the present-day value of the scalar field derivative:
\begin{equation}
f_{R0} \equiv \left. \frac{df}{dR} \right|_{R=R0} = -\frac{n c_1}{c_2^2} \left[ \frac{\Omega_{m0}}{3(\Omega_{m0} + \Omega_{\Lambda0})} \right]^{n+1}.
\end{equation}
This parameter $f_{R0}$ sets the strength of deviations from GR, with the GR limit recovered as $f_{R0} \to 0$ or $n \to \infty$.
Throughout our analysis, we apply $n=1$ specific case of HS-$f(R)$ model to all our results and the constants satisfy $c_1/c_2 = 6\Omega_\Lambda/\Omega_m$ to reproduce the cosmological constant at late times.

Thus, in the high-curvature limit ($R \gg M^2$), this model can be approximated by:
\begin{equation}
f(R) \approx -6 H_0^2 \Omega_\Lambda + |f_{R0}| \left(\frac{R_0^2}{R^2}\right),
\end{equation}
where $f_{R0}$ is the present-day value of the scalar field, and $R_0 = 3H_0^2(\Omega_{m,0} + 4\Omega_{\Lambda,0})$ is the current background Ricci scalar.
One of the appealing features of the HS model is its natural realization of the chameleon screening mechanism~\cite{Khoury:2003aq,Li:2011qda}. In regions of high matter density, the scalar degree of freedom acquires a large effective mass, thereby suppressing the fifth force and ensuring compatibility with Solar System tests.
In this work, we focus on the HS-$f(R)$ gravity with chameleon screening mechanism in the context of galaxy bispectrum multipoles. 
%%%%%%%%%%%%%%%%%%%%%%%%%%%%%%%%%%%%%%%%%%%%%%%%
\subsection{Kernels and Bispectrum in $f(R)$ Gravity Model }
\label{subsec:kernel_bispectra}
%%%%%%%%%%%%%%%%%%%%%%%%%%%%%%%%%%%%%%%%%%%%%%%%
%%%%%%%%%%%%%%%%%%%%%%%%%%%%%%%%%%%%%%%%%%%%%%%%
\subsubsection{Perturbative Framework}
\label{subsec:pertubative_framework}
%%%%%%%%%%%%%%%%%%%%%%%%%%%%%%%%%%%%%%%%%%%%%%%%
The perturbative dynamics of this model can be captured by a scalar–tensor formulation in the Newtonian gauge with an associated scalar perturbation $\delta f_R = f_R - f_{R0}$, sourced by matter density fluctuations $\delta\rho = \bar{\rho} \, \delta$. Here $\bar{\rho}$ and $R_0$ incorporates background density and curvature respectively. The modified Poisson equation in Fourier space can be written as,
\begin{equation}
    -\frac{k^2}{a^2} \Phi = 4\pi G\, \bar{\rho}\, \delta \, \mu(k,a)+\mathcal{S}(k)
\end{equation}
where
\begin{align}
\mu(k,a) = 1 + \frac{1}{3} \frac{k^2}{k^2 + a^2 m^2(a)}, 
\mathcal{S}(k)= -\frac{1}{6} \frac{k^2}{k^2 + a^2 m^2(a)} \delta I.
\end{align}
This enhances gravity by up to $4/3$ at small scales ($k \gg am$), but returns to GR at large scales ($k \ll am$). Here $\delta I$ encodes nonlinear self-interactions in the scalar field sector, which can be defined in a perturbative series expansion following Refs.~\cite{Aviles:2023fqx,Rodriguez-Meza:2023rga},
\begin{align}
\delta I(\delta f_R) &= \frac{1}{2} \int_{k_1 + k_2 = k} M_2(k_1, k_2)\, \delta f_R(k_1)\delta f_R(k_2) \nonumber \\
& + \frac{1}{6} \int_{\sum k_i = k} M_3(k_1, k_2, k_3) \delta f_R(k_1)\delta f_R(k_2)\delta f_R(k_3)\nonumber  \\ &
+ \cdots,
\end{align}
with $M_n = \left. \frac{d^n R}{df_R^n} \right|_{f_{R0}}$ where $R$ can be re-expressed in terms of $f_R$ following Eq.~\ref{eq:func_fR}. For the HS model with $n = 1$, these functions depend only on time through the background quantities,
\begin{align}
M_1(a) &= \frac{3}{2} H_0^2 \frac{(\Omega_{m,0} a^{-3} + 4 \Omega_{\Lambda,0})^3}{|f_{R0}|(\Omega_{m,0} + 4 \Omega_{\Lambda,0})^2}, \\
M_2(a) &= \frac{9}{4} H_0^2 \frac{(\Omega_{m,0} a^{-3} + 4 \Omega_{\Lambda,0})^5}{|f_{R0}|^2(\Omega_{m,0} + 4 \Omega_{\Lambda,0})^4},\label{eq:M2}
\end{align}
\begin{figure*}
    \centering   
    \subfloat{\includegraphics[width=0.9\columnwidth]{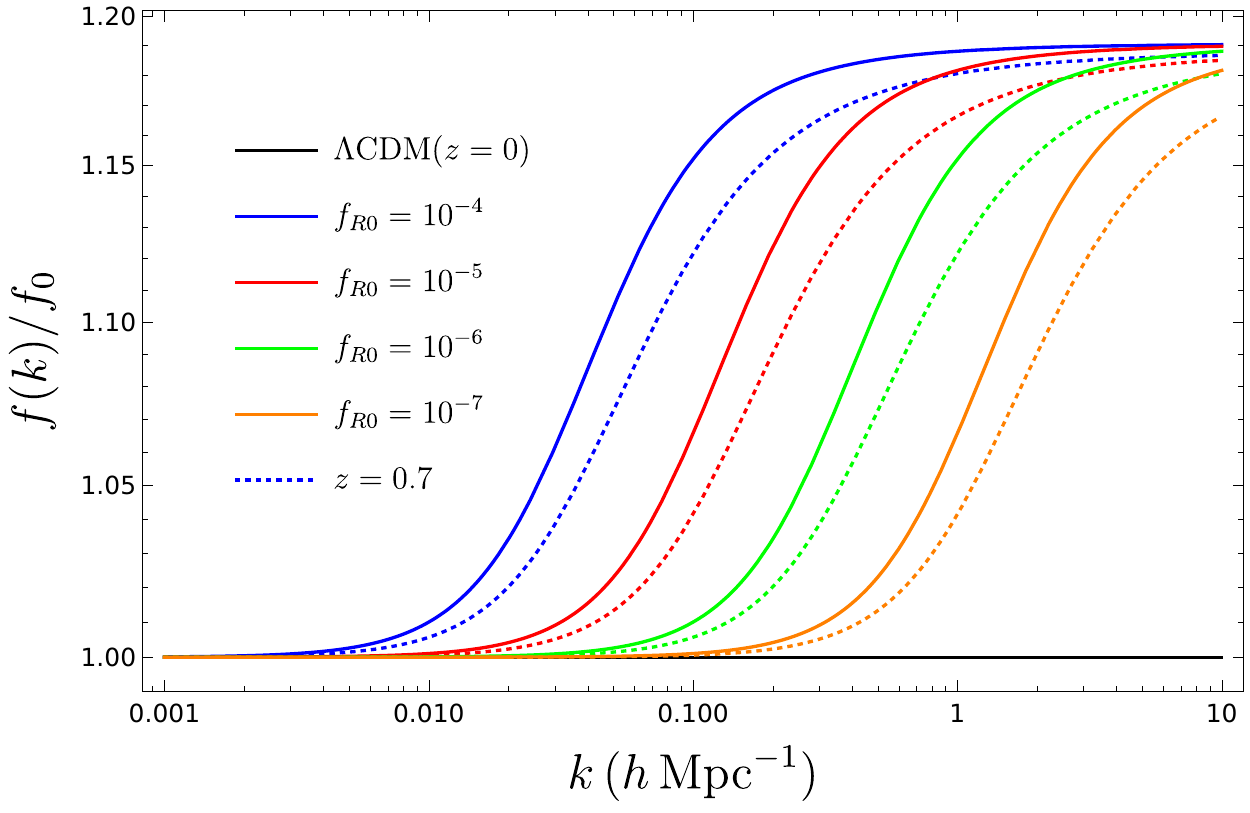}}
    \subfloat{\includegraphics[width=0.9\columnwidth]{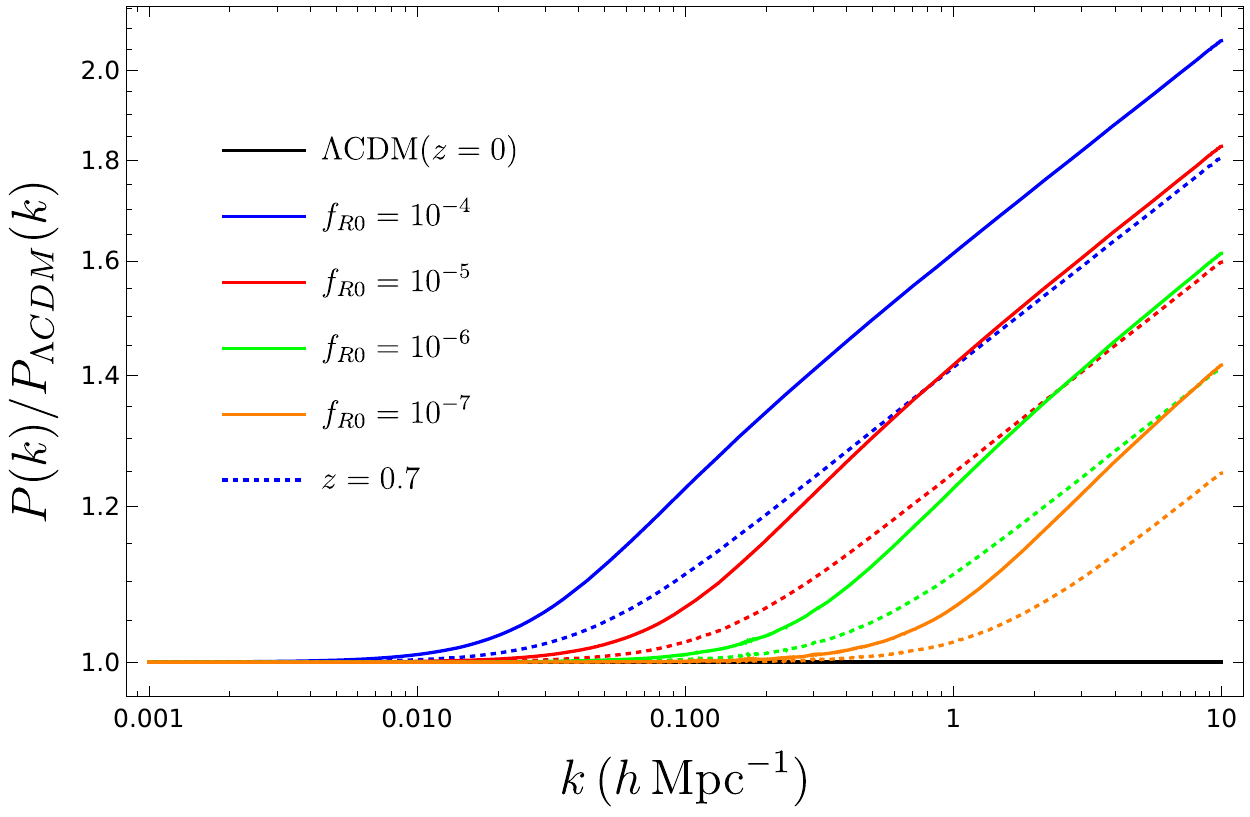}}
     \caption[]{\justifying In the \textbf{left panel}, we show the scale-dependent logarithmic growth rate $f(k)$ in the Hu–Sawicki $f(R)$ model for different values of $f_{R0}$, normalized by its large-scale value $f_0 = f(k=10^{-4}\,h\,{\rm Mpc}^{-1})$, which coincides with $\Lambda$CDM. The solid and dotted curves correspond to redshifts $z=0$ and $z=0.7$, respectively. The black line denotes the $\Lambda$CDM prediction. In the \textbf{right panel}, we show the ratio of the linear matter power spectrum in the $f(R)$ model to that in $\Lambda$CDM for the same redshifts and model parameters. Both panels illustrate that deviations grow with increasing $k$, while reaching the level of $10–20\%$ for $f(k)$ at $z=0$ for $f_{R0}\in[10^{-7},10^{-4}]$.
    }
    \label{fig:fk_pk}
\end{figure*}
These interaction terms generate a scale-dependent fifth force, which is Yukawa-like at linear order. The Compton wavelength $\lambda_C \sim m^{-1}(a) \propto |f_{R0}|^{1/2}$ determines the range of this fifth force: smaller $|f_{R0}|$ values lead to weaker deviations from GR.
These modifications to the evolution of the gravitational potential results in a different growth of perturbation modes. These modifications can be encapsulated by generalizing the linear evolution equation for the matter overdensity, $\delta^{(1)}(\mathbf{k},a)$, which now evolves under a scale-dependent gravitational strength. Specifically, the linear growth factor $D_+(k,a)$ satisfies
\begin{equation}
\left[\mathcal{T} - A(k,a)\right] D_+(k,a) = 0,
\label{eq:Dplus_evolution}
\end{equation}
where the differential operator $\mathcal{T} = \partial_t^2 + 2H\partial_t$ governs the standard background evolution, and the source term $A(k,a)$ encodes the modified gravitational response:
\begin{equation}
A(k,a) = \frac{3}{2} \Omega_m H^2 \left(1 + \frac{2\beta^2 k^2}{k^2 + m^2 a^2}\right).
\label{eq:A_def}
\end{equation}
Here, $\beta(a)$ denotes its universal coupling to matter (However, $\beta^2=1/3$ in HS-$f(R)$ gravity model).
For computational consistency and comparison across models, it is useful to normalize $D_+$ by the EdS solution at an early epoch $t_{\rm ini}$, where MG effects are negligible and $D_+(k,t_{\rm ini}) \propto a(t_{\rm ini})$. The evolved linear power spectrum in a given MG theory then reads
\begin{equation}
P_L(k,t) = \left[\frac{D_+(k,t)}{D_+^{\Lambda \mathrm{CDM}}(k,t_0)}\right]^2 P_L^{\Lambda \mathrm{CDM}}(k,t_0),
\label{eq:PL_mg}
\end{equation}
where $t_0$ is a reference time, typically today.

While $D_+$ captures linear growth, understanding non-linear structure formation requires higher-order perturbative corrections. At second order in the density contrast and velocity divergence, the non-linear evolution is governed by scale and time dependent kernels,
\begin{align}
F_2(\mathbf{k}_1,\mathbf{k}_2,t) &= \frac{1}{2} + \frac{3}{14} \mathcal{A}(\mathbf{k}_1,\mathbf{k}_2,t) + \frac{\hat{\mathbf{k}}_1 \cdot \hat{\mathbf{k}}_2}{2}\left(\frac{k_1}{k_2} + \frac{k_2}{k_1}\right) \nonumber \\
&\quad + (\hat{\mathbf{k}}_1 \cdot \hat{\mathbf{k}}_2)^2\left[\frac{1}{2} - \frac{3}{14} \mathcal{B}(\mathbf{k}_1,\mathbf{k}_2,t) \right]. \label{eq:F2_HS}\\
G_2(\mathbf{k}_1,\mathbf{k}_2,t) &= \frac{3 \left[f(k_1) + f(k_2)\right]\mathcal{A}+3 \dot{\mathcal{A}}/H}{14 f_0}  \nonumber \\
&\quad + \frac{\hat{\mathbf{k}}_1 \cdot \hat{\mathbf{k}}_2}{2} \left[ \frac{f(k_2)}{f_0} \frac{k_2}{k_1} + \frac{f(k_1)}{f_0} \frac{k_1}{k_2} \right] \nonumber \\
&\quad + \left( \hat{\mathbf{k}}_1 \cdot \hat{\mathbf{k}}_2 \right)^2 \Big[ \frac{f(k_1) + f(k_2)}{2 f_0} \nonumber \\
&\quad  - \frac{3 \left[f(k_1) + f(k_2)\right] \mathcal{B} + 3 \dot{\mathcal{B}}/H}{14 f_0}.
\label{eq:G2_HS}
\end{align}
The functions $\mathcal{A}$ and $\mathcal{B}$ are defined by
\begin{align}
\mathcal{A} &= \frac{7 D^{(2)}_{\mathcal{A}}(\mathbf{k}_1,\mathbf{k}_2)}{3 D_+(k_1) D_+(k_2)}, \quad
\mathcal{B} = \frac{7 D^{(2)}_{\mathcal{B}}(\mathbf{k}_1,\mathbf{k}_2)}{3 D_+(k_1) D_+(k_2)},
\end{align}
and they capture corrections beyond EdS evolution. In $\Lambda$CDM, $\mathcal{A}, \mathcal{B} \simeq 1$ today.

The functions $D^{(2)}_{\mathcal{A}}, D^{(2)}_{\mathcal{B}}$ solve Green's-type equations sourced by the modified gravitational dynamics:
\begin{align}
D^{(2)}_{\mathcal{A}} &= \left(\mathcal{T} - A(k)\right)^{-1} \Big[ A(k) + (A(k) - A(k_1)) \frac{\mathbf{k}_1\cdot\mathbf{k}_2}{k_2^2} \nonumber \\
&\quad + (A(k) - A(k_2)) \frac{\mathbf{k}_1\cdot\mathbf{k}_2}{k_1^2} - S_2(\mathbf{k}_1,\mathbf{k}_2) \Big] D_+(k_1) D_+(k_2),
\label{eq:DA}
\\
D^{(2)}_{\mathcal{B}} &= \left(\mathcal{T} - A(k)\right)^{-1} \left[ A(k_1) + A(k_2) - A(k) \right] D_+(k_1) D_+(k_2).
\label{eq:DB}
\end{align}
The source term $S_2$ captures non-linear modifications due to scalar interactions and plays a central role in screening mechanisms. For HS-$f(R)$ model, it takes the form
\begin{equation}
S_2 = \frac{36 \Omega_m^2 H^4 \beta^6 a^4 M_2(\mathbf{k}_1,\mathbf{k}_2)k^2}{(k^2 + m^2 a^2)(k_1^2 + m^2 a^2)(k_2^2 + m^2 a^2)},
\label{eq:S2_fR}
\end{equation}
where $M_2$ is the model-dependent non-linear coupling defined in Eq.~\ref{eq:M2}. Note that to compute the kernels up to second order, we only need source term corrected by $M_2$.
% However, to compute higher order correction to the power spectrum or bicpectrum, one need to consider higher order terms as well.
As mentioned in Sec.~\ref{subsec:induced_bispectrum}, these second-order corrections culminate in the tree-level matter bispectrum, which encapsulates non-Gaussianity induced by gravitational evolution. The same equation as Eq.~\ref{eq:b_rsd} emerges with $F_2$ and $G_2$ given by Eqs.~\ref{eq:F2_HS} and \ref{eq:G2_HS}. 
For detail calcultaions of the second and higher order kernels, one can refer to Refs.~\cite{Aviles:2017aor,Aviles:2018qot,Aviles:2018saf,Aviles:2020cax,Aviles:2020wme,Aviles:2023fqx,Rodriguez-Meza:2023rga}.
%%%%%%%%%%%%%%%%%%%%%%%%%%%%%%%%%%%%%%%%%%%%%%%%
\subsubsection{Finger of God Effect}
%%%%%%%%%%%%%%%%%%%%%%%%%%%%%%%%%%%%%%%%%%%%%%%%
\begin{figure*}
    \centering   
    \subfloat{\includegraphics[width=0.70\columnwidth]{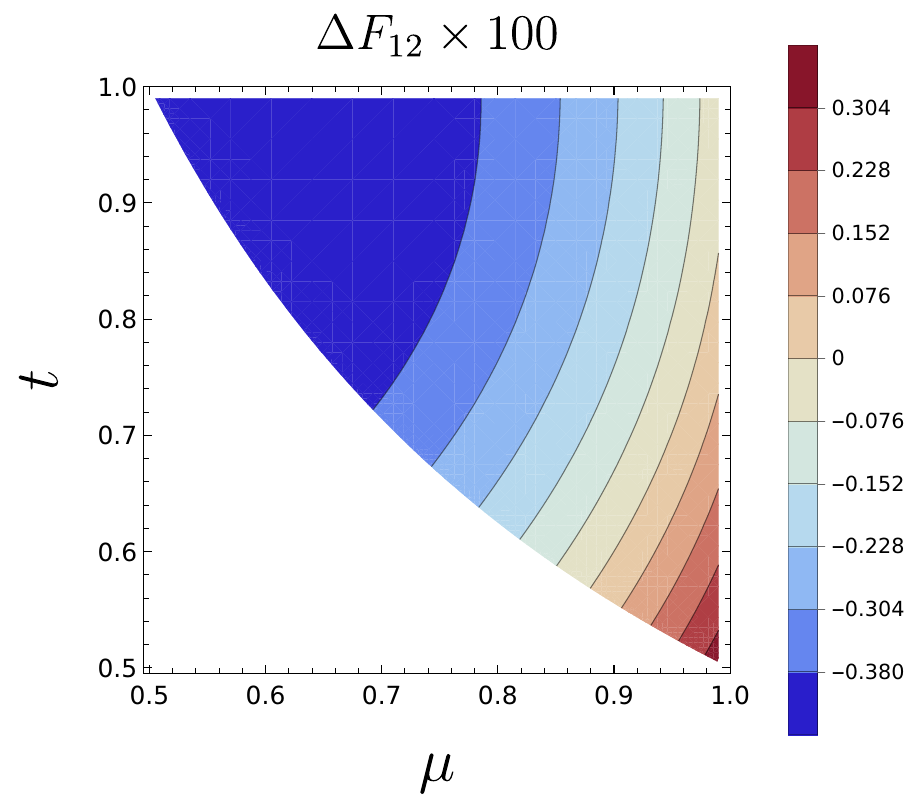}}
    \subfloat{\includegraphics[width=0.70\columnwidth]{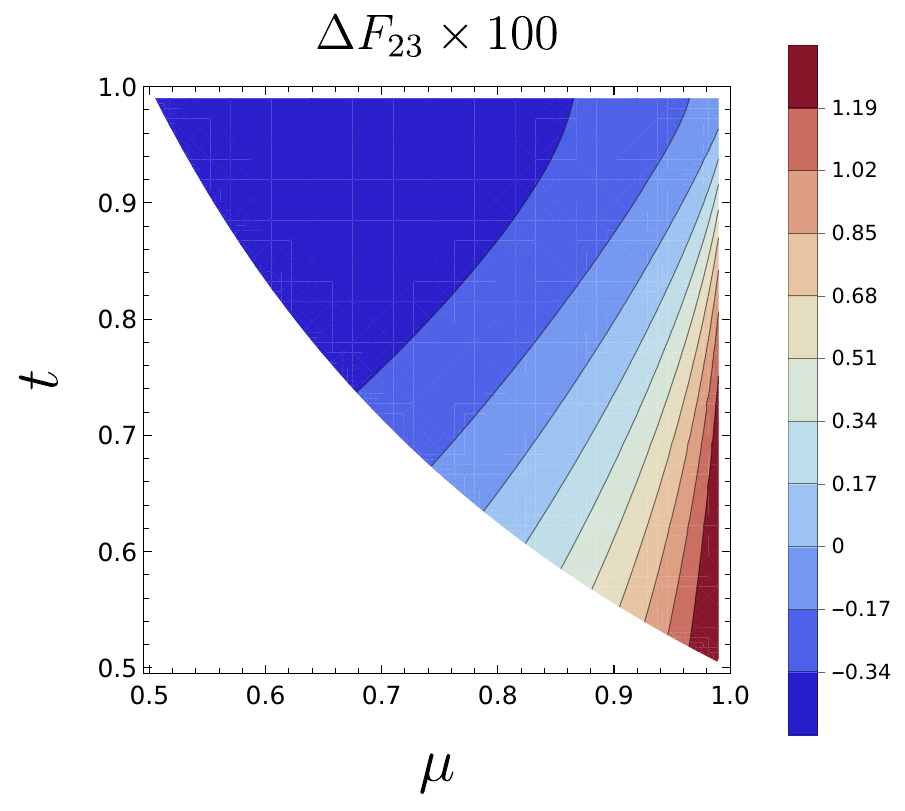}}
    \subfloat{\includegraphics[width=0.70\columnwidth]{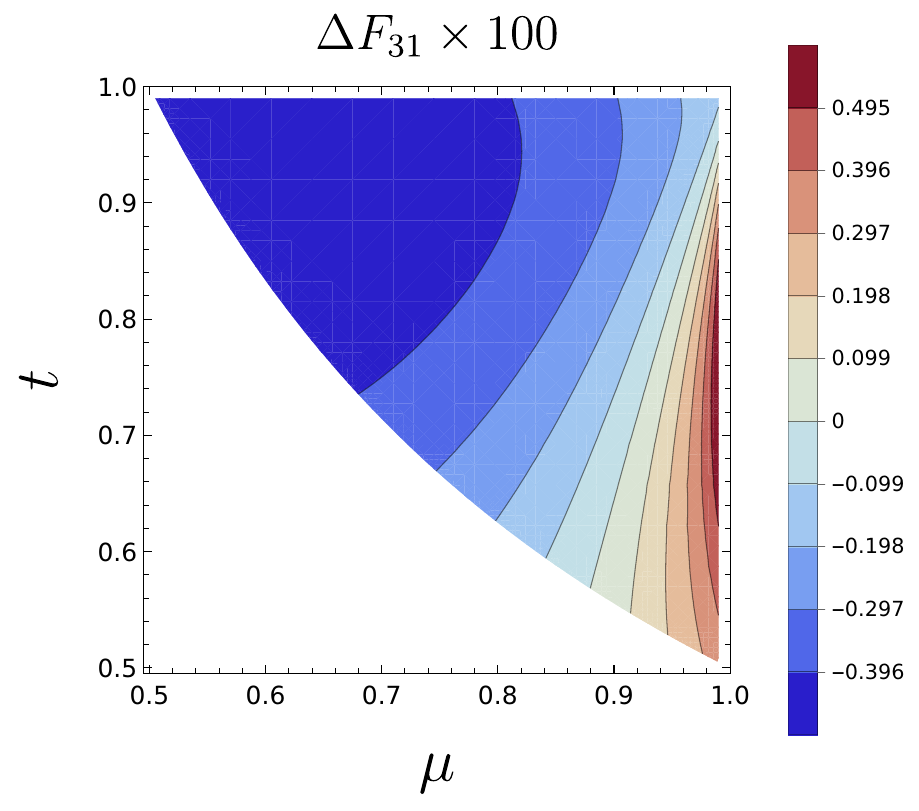}}
    \quad
    \subfloat{\includegraphics[width=0.70\columnwidth]{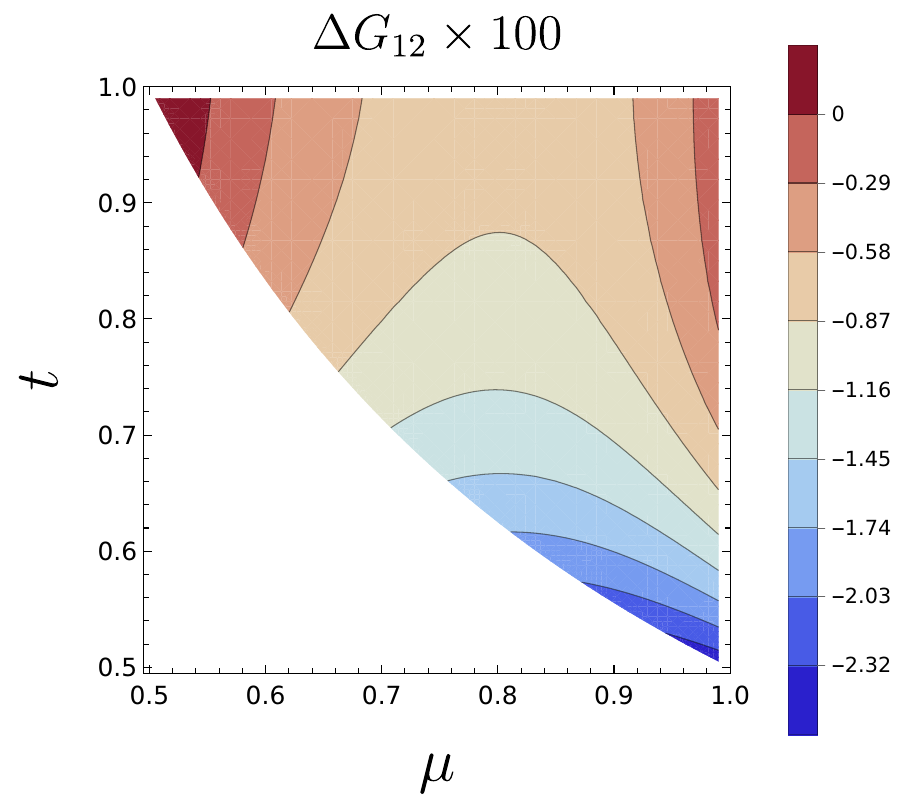}}
    \subfloat{\includegraphics[width=0.70\columnwidth]{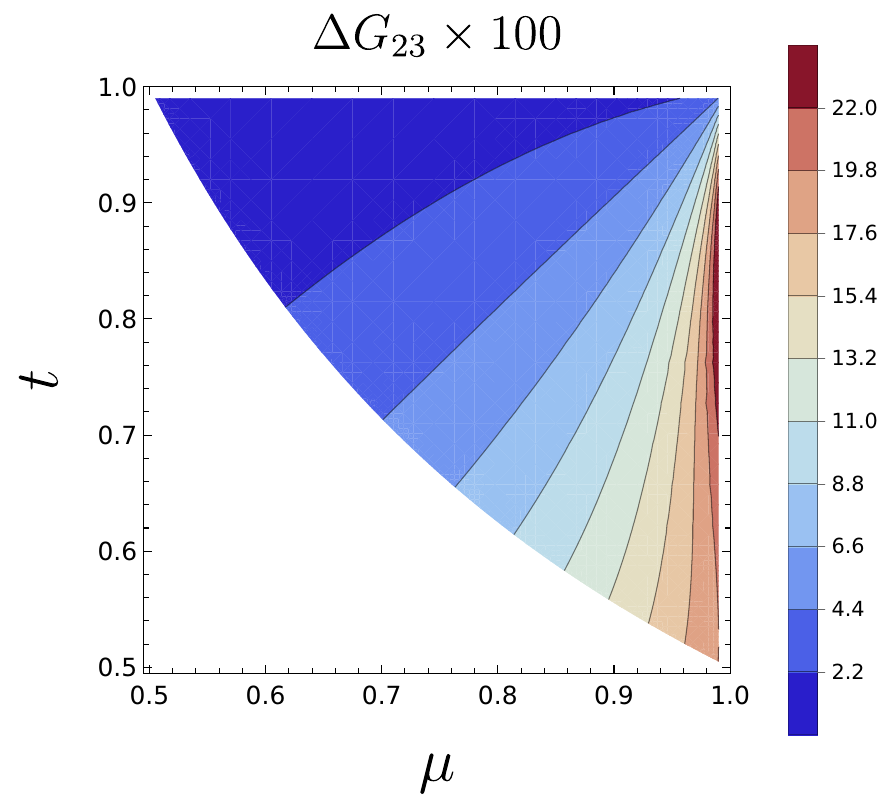}}
    \subfloat{\includegraphics[width=0.70\columnwidth]{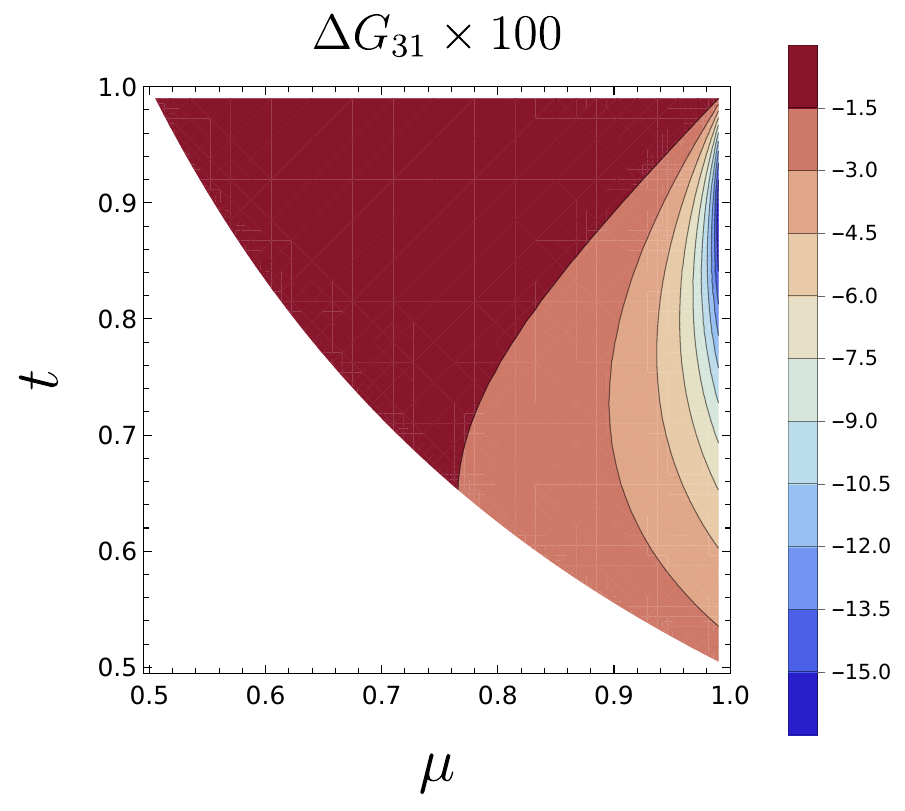}}
    \caption[]{ \justifying Differences in the second-order kernel functions, defined as $\Delta F_{ij}=F_{ij}^{\rm HS}-F_{ij}^{\Lambda{\rm CDM}}$ and 
    $\Delta G_{ij}=G_{ij}^{\rm HS}-G_{ij}^{\Lambda{\rm CDM}}$, in the 
    $\mu$--$t$ plane for the HS-$f(R)$ model with $f_{R0}=10^{-5}$. 
    The results are shown at fixed $k_1=0.3\,h\,{\rm Mpc}^{-1}$ and redshift 
    $z=0.7$, with the $\Lambda$CDM kernels evaluated using the EdS approximation 
    \cite{Bernardeau:2001qr}. The largest deviations appear in stretched 
    configurations ($\mu\simeq1, t\simeq0.5$) for $F_{12}$, $F_{23}$, and $G_{12}$, 
    while $F_{31}$, $G_{23}$, and $G_{31}$ exhibit maximal differences for linear 
    triangles ($\mu\simeq1$). These deviations are intrinsically redshift dependent, 
    decreasing at higher $z$ where the scalaron is more strongly screened. 
    Since the kernels directly feed into the tree-level bispectrum, the patterns 
    seen here translate into characteristic configuration-dependent signatures in 
    the bispectrum multipoles.
    }
    \label{fig:FG_kernel_mut}
\end{figure*}
On small scales, galaxy redshift measurements are affected by random line-of-sight motion, which produce elongated structure in redshift space --- an effect commonly referred to as the \textit{Finger-of-God}  (FoG) phenomena~\cite{Jackson:1971sky}. This leads to a suppression of both the power spectrum \cite{Sarkar:2018gcb,Sarkar:2019nak} and bispectrum, with the magnitude of the effect depending on scale and orientation.
\begin{figure*}
    \centering   
    \subfloat{\includegraphics[width=0.66\columnwidth]{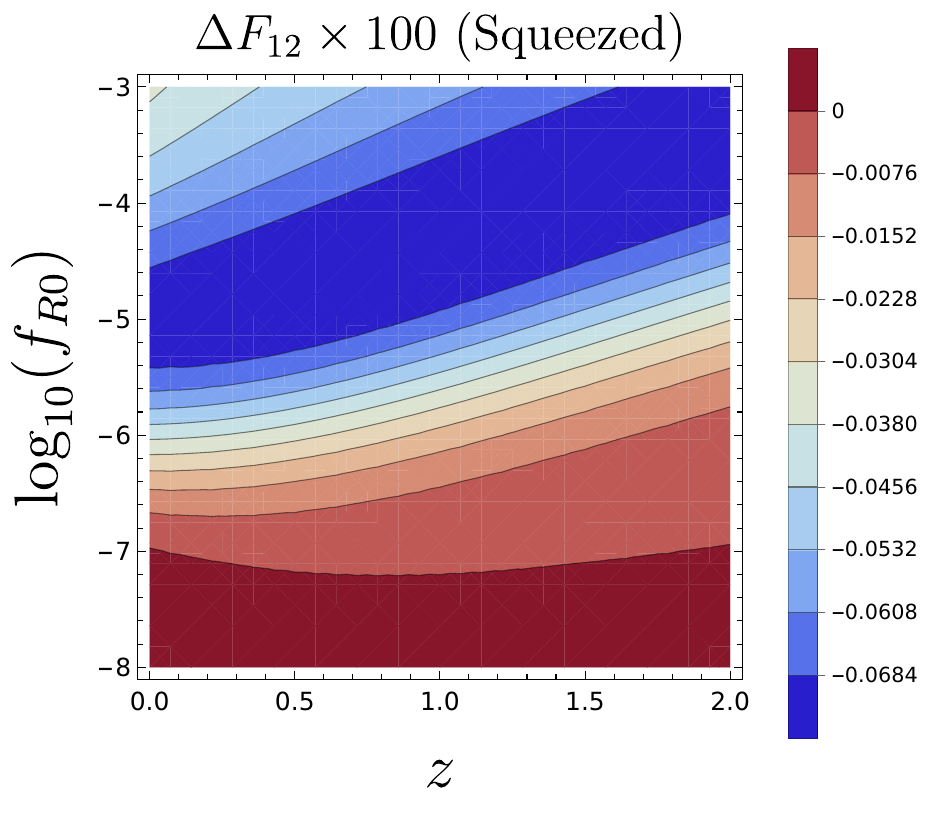}}
    \subfloat{\includegraphics[width=0.66\columnwidth]{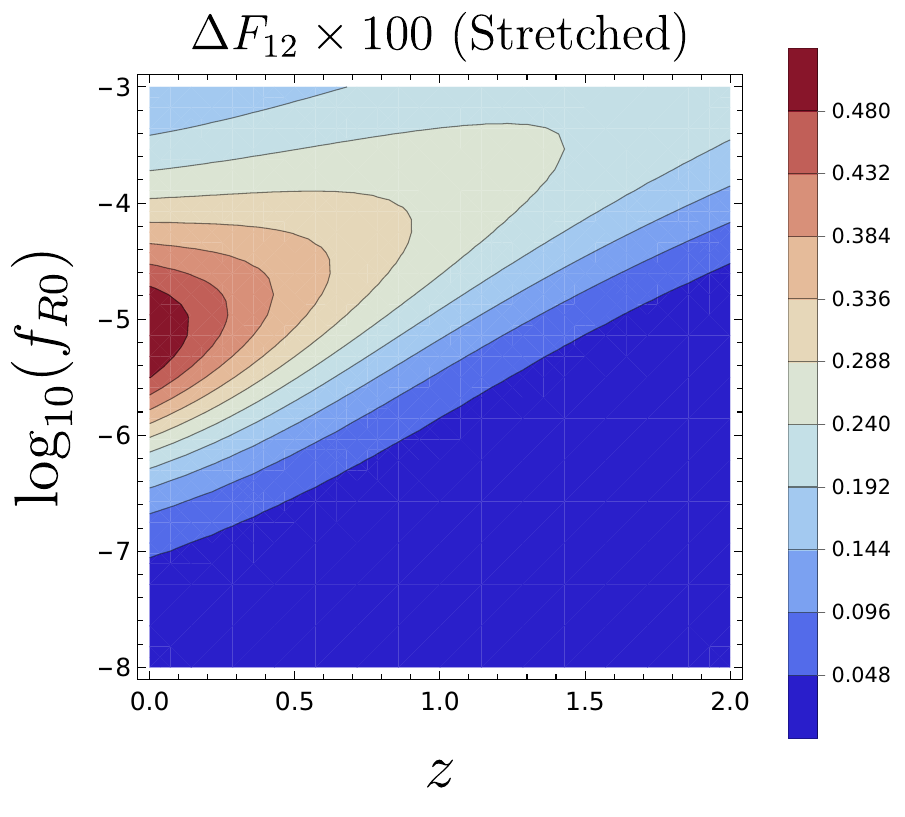}}
    \subfloat{\includegraphics[width=0.66\columnwidth]{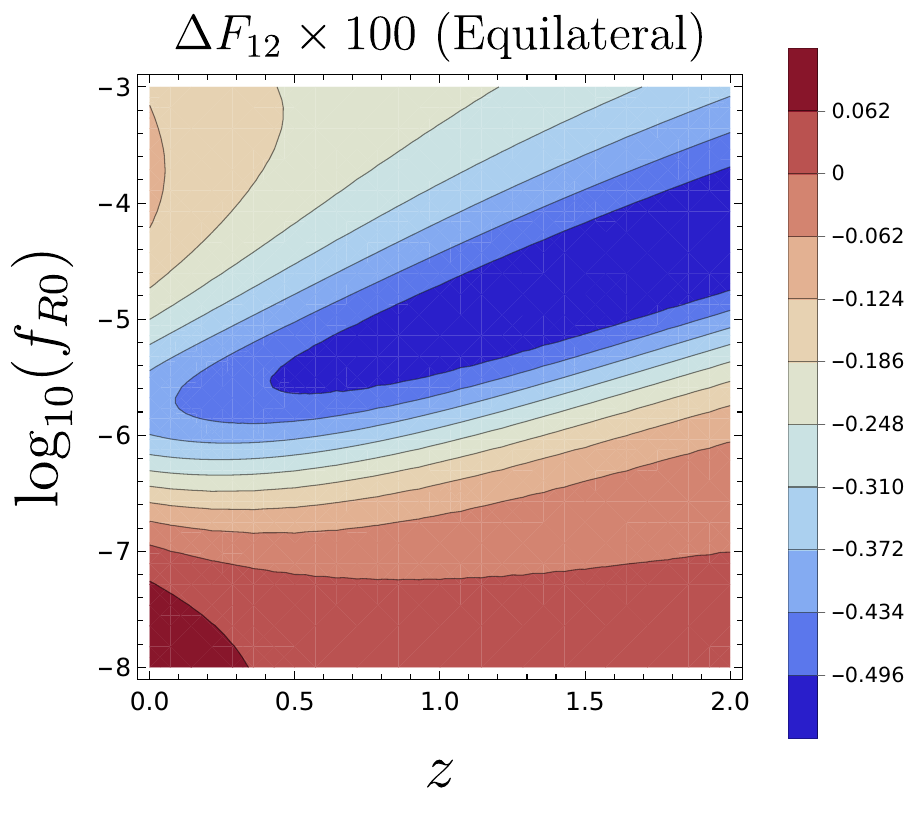}}
    \quad
    \subfloat{\includegraphics[width=0.66\columnwidth]{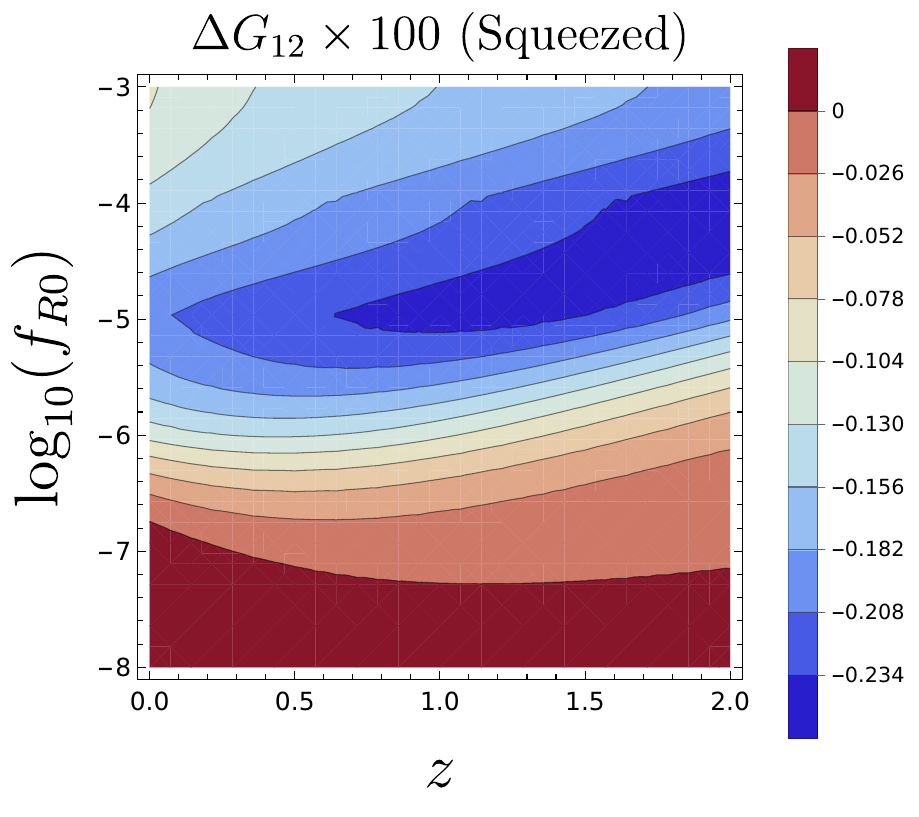}}
    \subfloat{\includegraphics[width=0.66\columnwidth]{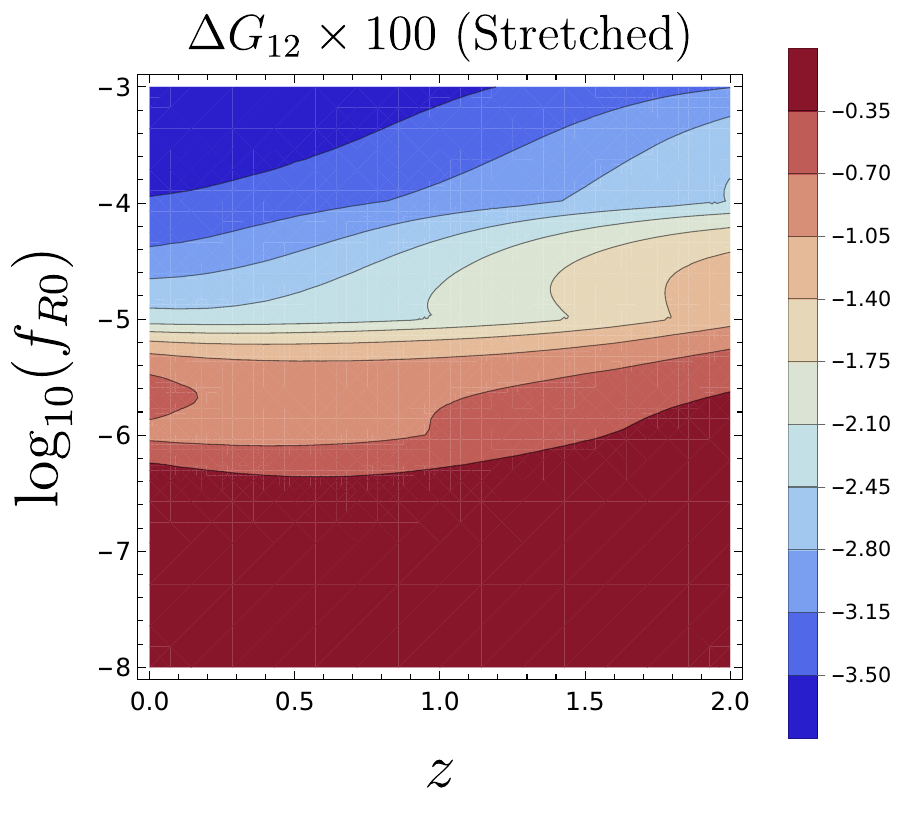}}
    \subfloat{\includegraphics[width=0.66\columnwidth]{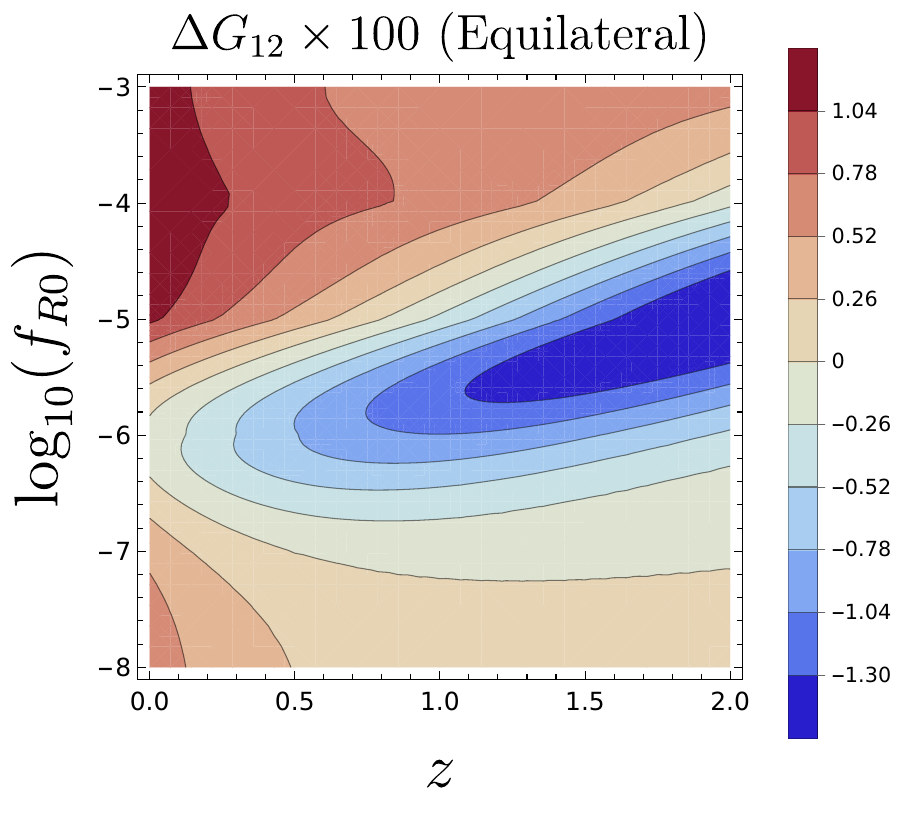}}
    \caption[]{ \justifying Contour plots of the kernel differences $\Delta F_{12}$ and $\Delta G_{12}$ 
    between the HS-$f(R)$ model and $\Lambda$CDM as functions of redshift 
    $z$ and the model parameter $f_{R0}$. Results are shown for three representative 
    triangle shapes: squeezed (left, $\mu\simeq1$, $t\simeq1$), stretched 
    (middle, $\mu\simeq1$, $t\simeq0.5$), and equilateral (right, $\mu\simeq0.5$, $t\simeq1$). 
    In the squeezed configuration, $F_{12}$ shows its largest deviations for 
    $f_{R0}\sim10^{-6}$--$10^{-3}$, with nearly linear contours in the 
    $\log_{10}(f_{R0})$--$z$ plane, while $G_{12}$ peaks for 
    $f_{R0}\sim10^{-5}$--$10^{-4}$. In the stretched case, $\Delta F_{12}$ 
    peaks at lower redshift for $f_{R0}\approx10^{-5}$, showing elliptical 
    contours with a positive tilt, while $\Delta G_{12}$ grows mainly with 
    increasing $f_{R0}$ and has weak $z$ dependence. For equilateral triangles, 
    both $\Delta F_{12}$ and $\Delta G_{12}$ reach maximum values for 
    $f_{R0}\sim10^{-6}$--$10^{-4}$, again with elliptical contours tilted 
    positively with redshift. These trends illustrate how kernel deviations 
    depend jointly on $f_{R0}$ and cosmic time, and they directly translate 
    into redshift- and parameter-dependent signatures in the bispectrum multipoles.  
    }
    \label{fig:FG_kernel_z}
\end{figure*}

A standard approach to modeling FoG is to multiply the redshift-space clustering signal by a damping function governed by the pairwise velocity dispersion, $\sigma_p$. Several functional forms have been explored in the literature, such as Lorentzian, Lorentzian-squared, and Gaussian profiles. While the Lorentzian model often provides a good match to simulations, the Gaussian form arises more naturally from physical considerations of random velocities ~\cite{Peebles:1980yev,Hikage:2015wfa,BaleatoLizancos:2025wdg}. 
In our analysis, we adopt a Gaussian damping profile, modifying the linear redshift-space power spectrum and bispectrum as: 
\begin{eqnarray}
P^s_{\rm FoG}(k,\mu) &= \exp\left(-\frac{1}{2}k^2\mu^2\sigma_p^2\right) P^s(k,\mu) \, ,\\
B^s_{\rm FoG}(\mathbf{k}_1,\mathbf{k}_2,\mathbf{k}_3) &= \exp\left(-\frac{1}{2}\sigma_p^2 \sum_{i=1}^{3} k_i^2\mu_i^2 \right) \nonumber \\
&\quad \times B^s(\mathbf{k}_1,\mathbf{k}_2,\mathbf{k}_3).
\label{eq:fog}
\end{eqnarray}
$\sigma_p$ is often considered a free parameter that tames the nonlinear relation between real and redshift spaces. In this work, and for definiteness, we adopt the values listed in Ref.~\cite{Yankelevich:2018uaz} for the pairwise velocity dispersion expected for the Euclid survey.
Note that the FoG suppression is more pronounced in $f(R)$ models due to enhanced small-scale velocities.

In what follows, we examine the observational consequences of the formulation in redshift space, discussed so far, and quantify its distinguishability from GR-based predictions.
%%%%%%%%%%%%%%%%%%%%%%%%%%%%%%%%%%%%%%%%%%%%%%%%
\subsubsection{Kernel Analysis }
\label{subsec:multipole_analysis}
%%%%%%%%%%%%%%%%%%%%%%%%%%%%%%%%%%%%%%%%%%%%%%%%
\begin{figure*}
    \centering   
    \subfloat{\includegraphics[width=0.70\columnwidth]{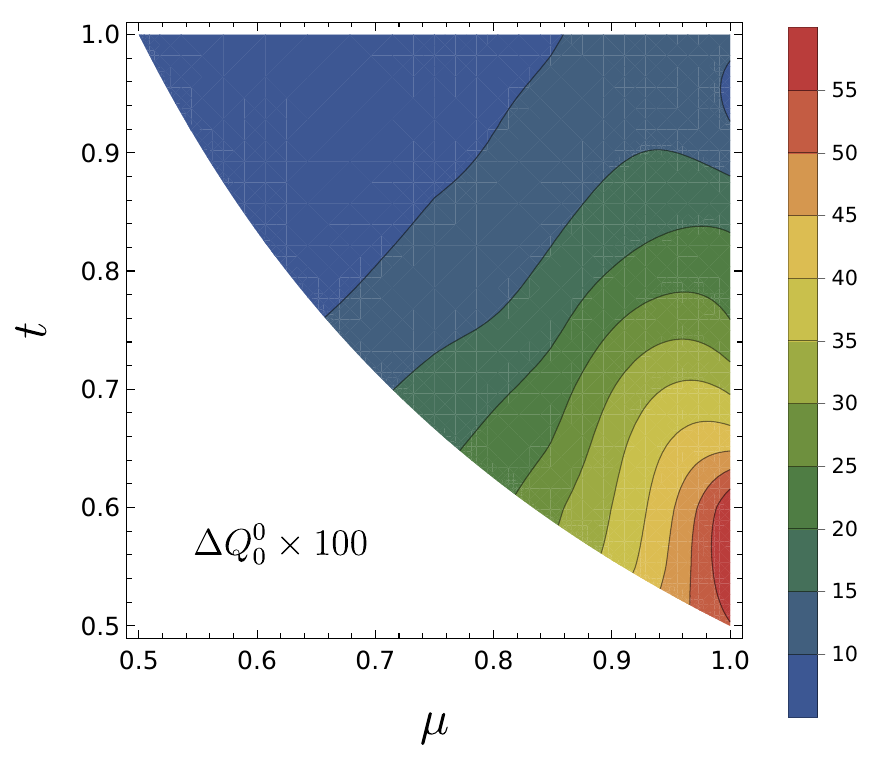}}
    \subfloat{\includegraphics[width=0.70\columnwidth]{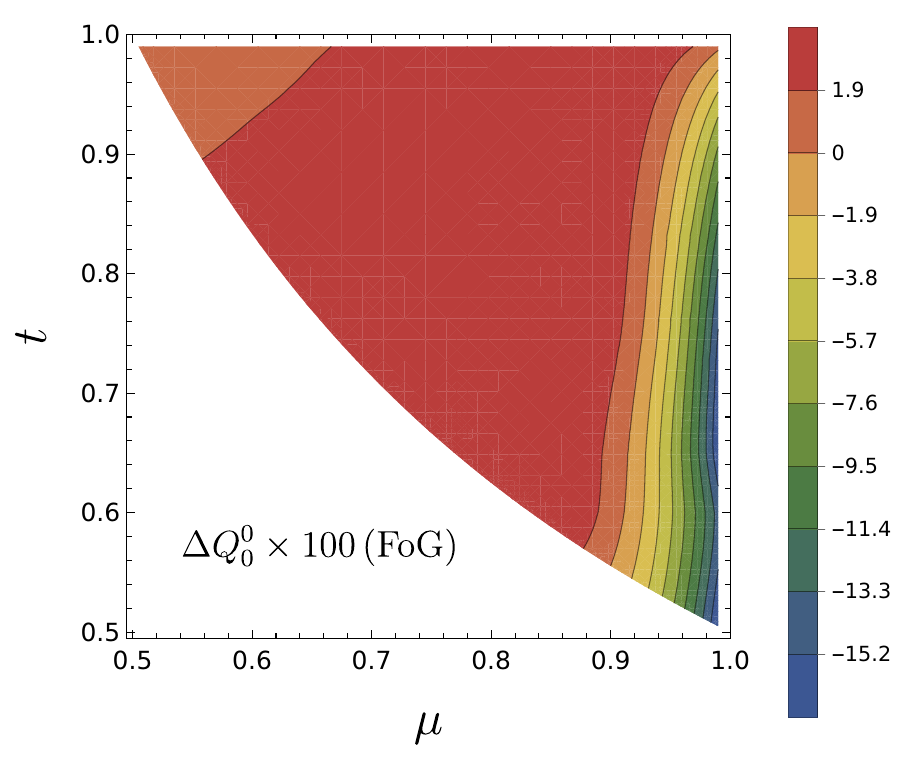}}
    \quad
    \subfloat{\includegraphics[width=0.70\columnwidth]{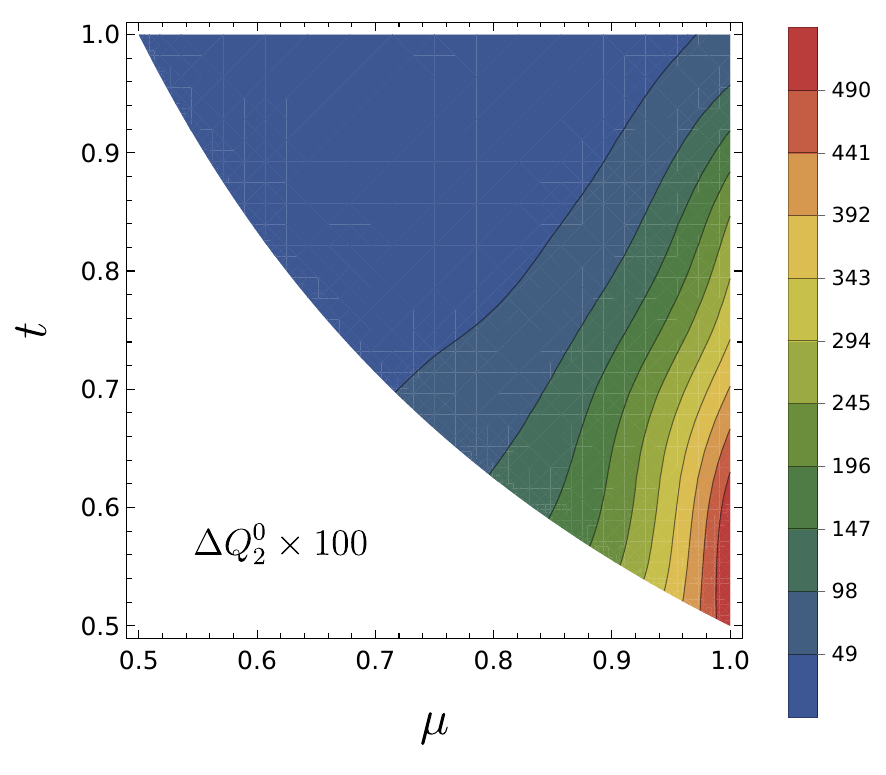}}
    \subfloat{\includegraphics[width=0.70\columnwidth]{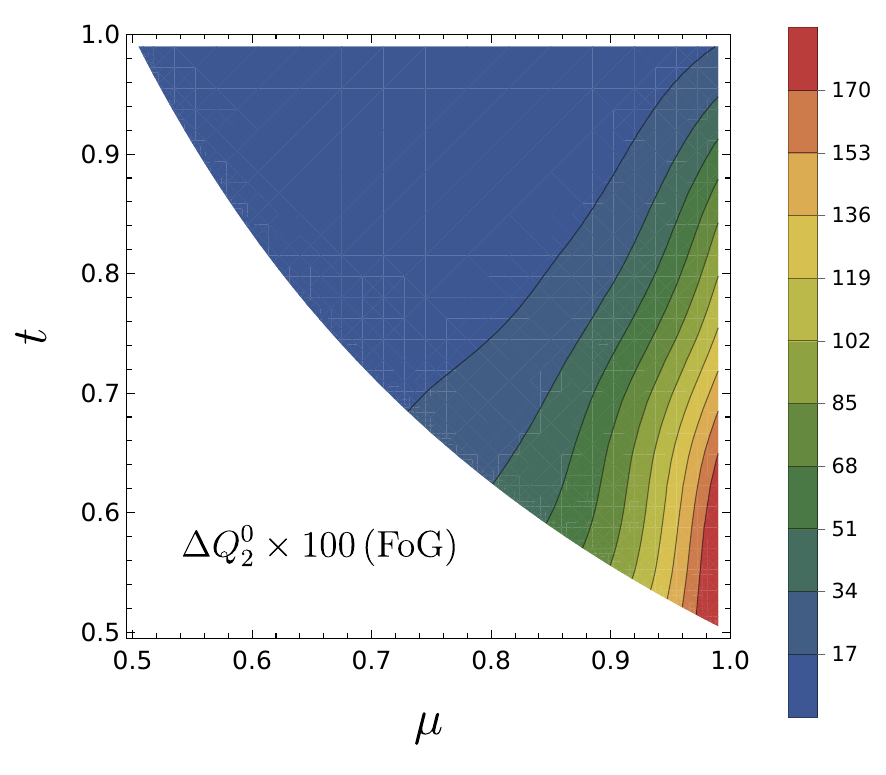}}
    \caption[]{ \justifying Heatmaps of the  signed difference $\Delta Q_L^m = (Q_L^m|_{\rm HS} - Q_L^m|_{\Lambda{\rm CDM}}) \times 100$ for the lowest bispectrum multipoles: $(L,m)=(0,0)$ (monopole, top row) and $(2,0)$ (quadrupole, bottom row). Results are shown for the HS-$f(R)$ model with 
    $f_{R0}=10^{-5}$ at redshift $z=0.7$ and $k_1=0.3\,h\,{\rm Mpc}^{-1}$. 
    The left column excludes FoG damping, while the right column includes it. 
    Without FoG, the monopole shows maximum percentage deviations of $\sim2\%$, whereas the quadrupole 
    reaches $\sim8\%$, both peaking near stretched configurations ($\mu\simeq1, t\simeq0.5$). 
    With FoG, the quadrupole retains a similar shape but is suppressed by a factor of $\sim3$, 
    while the monopole undergoes a qualitative change: the largest deviations become negative 
    along linear triangles, with a minimum at $(\mu, t)=(0.74,0.625)$, reaching $\sim2.8\%$. 
    These results demonstrate that the quadrupole $Q_2^0$ is far more sensitive to 
    $f(R)$ modifications than the monopole $Q_0^0$, and highlight how FoG effects 
    modulate the amplitude and even the sign of the MG signal in different triangle configurations.
        }
    \label{fig:QLm_00_20}
\end{figure*}
\begin{figure*}
    \centering   
    \subfloat{\includegraphics[width=0.70\columnwidth]{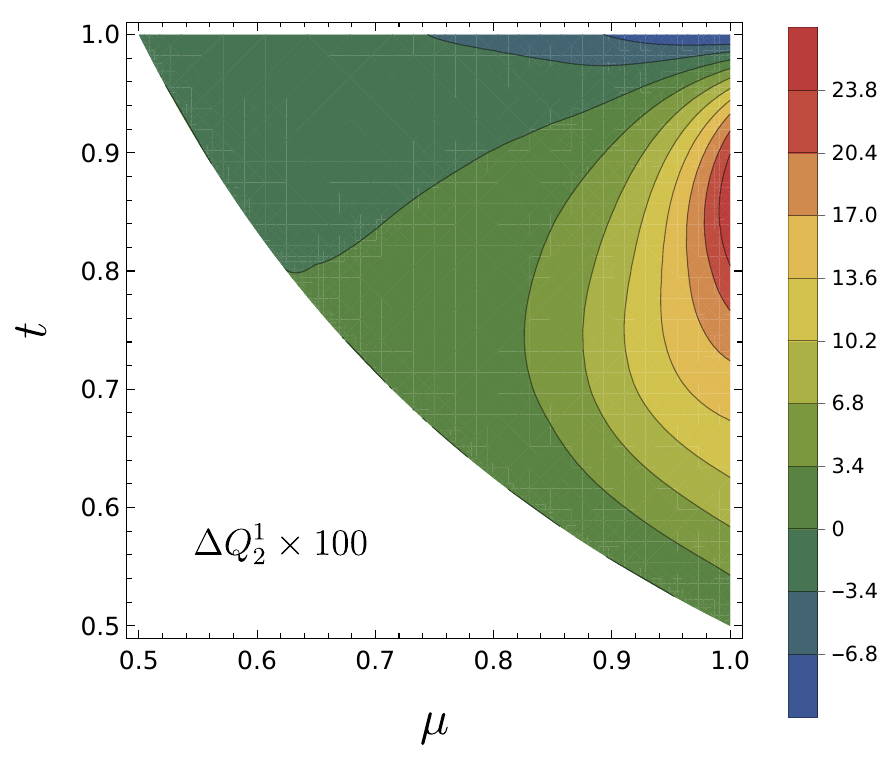}}
    \subfloat{\includegraphics[width=0.70\columnwidth]{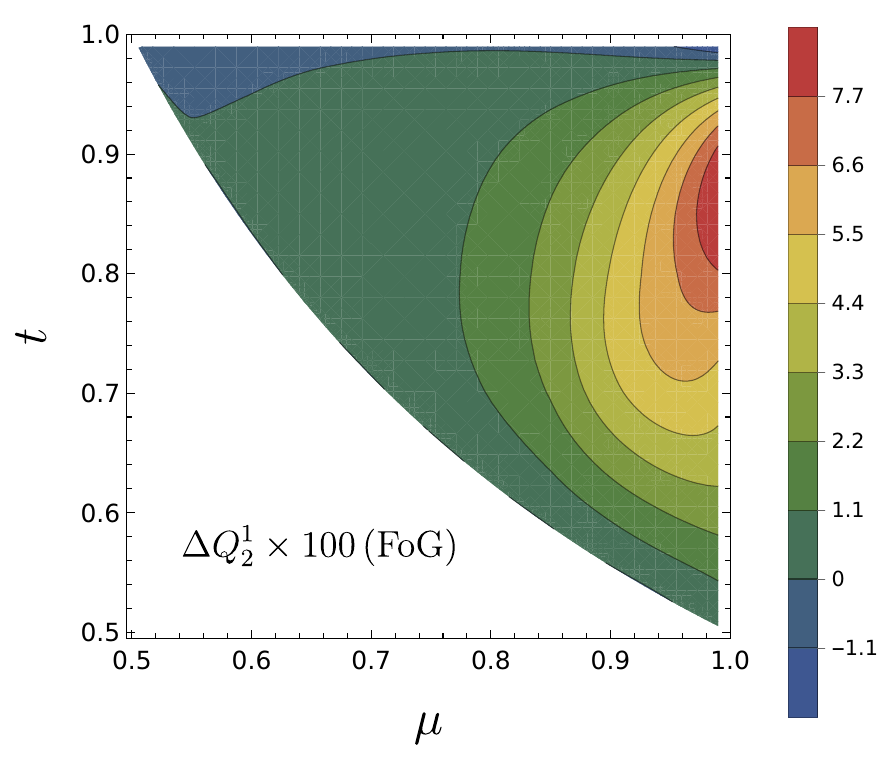}}
    \quad
    \subfloat{\includegraphics[width=0.70\columnwidth]{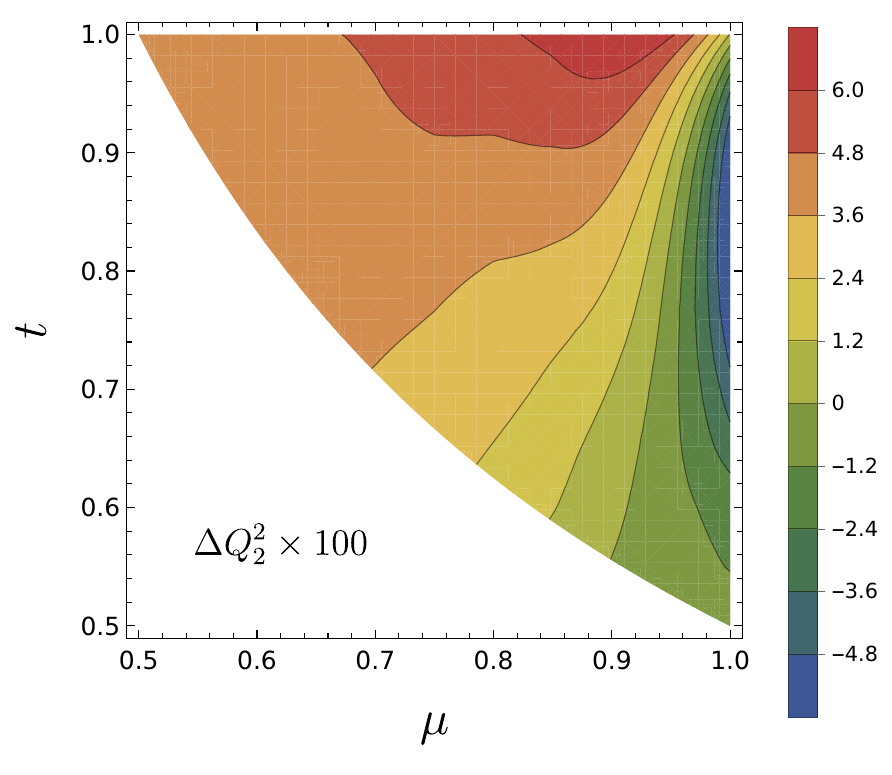}}
    \subfloat{\includegraphics[width=0.70\columnwidth]{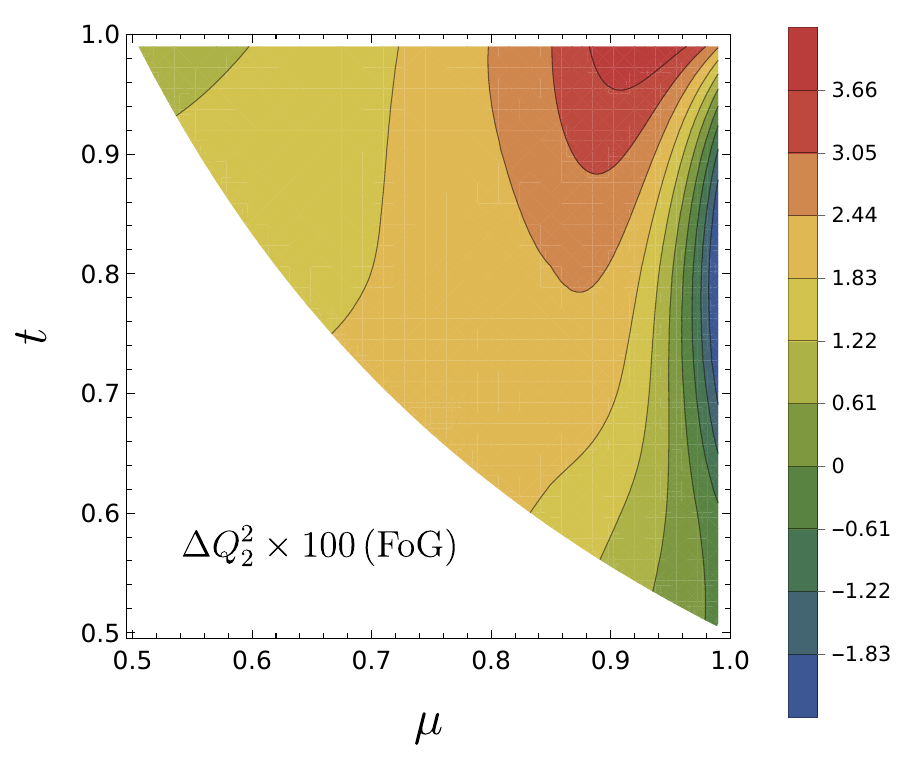}}
    \caption[]{\justifying Same as Fig.~\ref{fig:QLm_00_20}, but for the quadrupole multipoles with 
    $(L,m)=(2,1)$ (top row) and $(2,2)$ (bottom row). For $Q_2^1$, the peak deviation occurs near 
    the linear triangle ($\mu\simeq1, t\simeq0.8$), with $\sim2.7\%$ difference 
    without FoG damping and $\sim1.6\%$ with FoG. This multipole 
    is negative along L-isosceles ($t=1$) triangles but positive elsewhere. 
    For $Q_2^2$, the maximum percentage deviation ($\sim7.5\%$) arises near the L-isosceles 
    configuration ($\mu\simeq0.8, t\simeq1$), shifting slightly toward higher $\mu$ 
    when FoG is included. Compared to $Q_2^1$, the $Q_2^2$ multipole shows stronger 
    departures from $\Lambda$CDM and a distinct configuration dependence, 
    providing complementary information.
    }
    \label{fig:QLm_21_22}
\end{figure*}
\begin{figure*}
    \centering   
    \subfloat{\includegraphics[width=0.70\columnwidth]{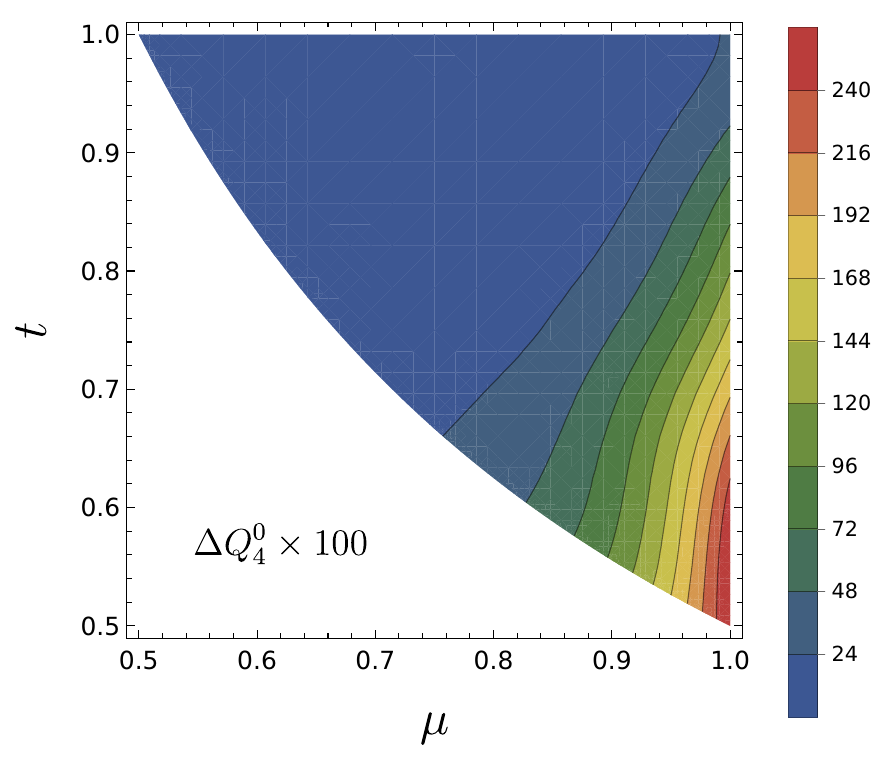}}
    \subfloat{\includegraphics[width=0.70\columnwidth]{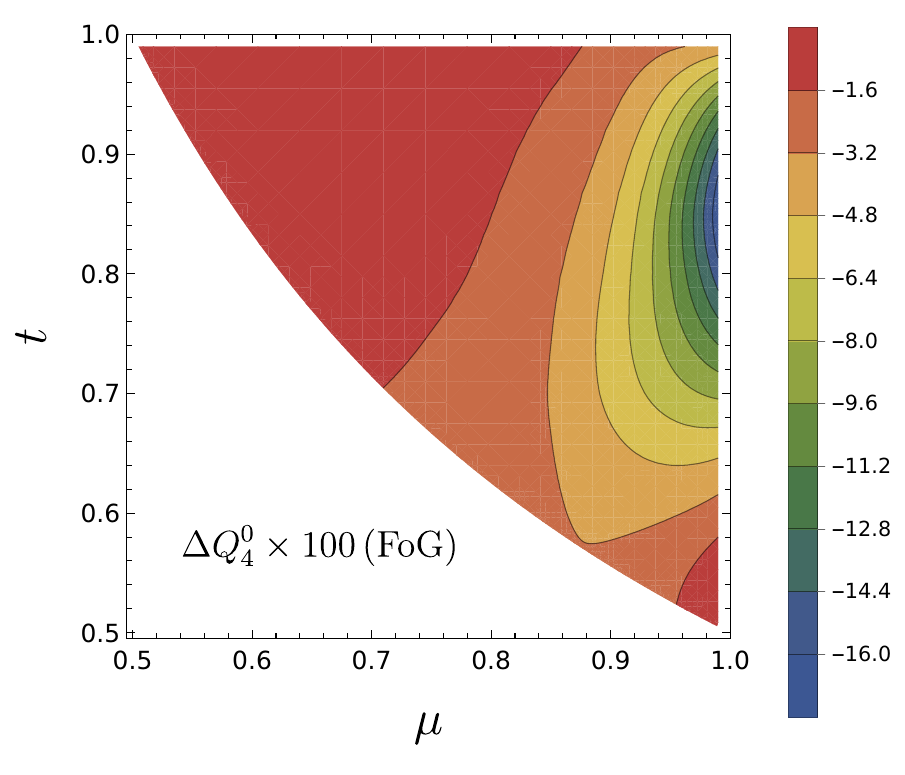}}
    \quad
    \subfloat{\includegraphics[width=0.70\columnwidth]{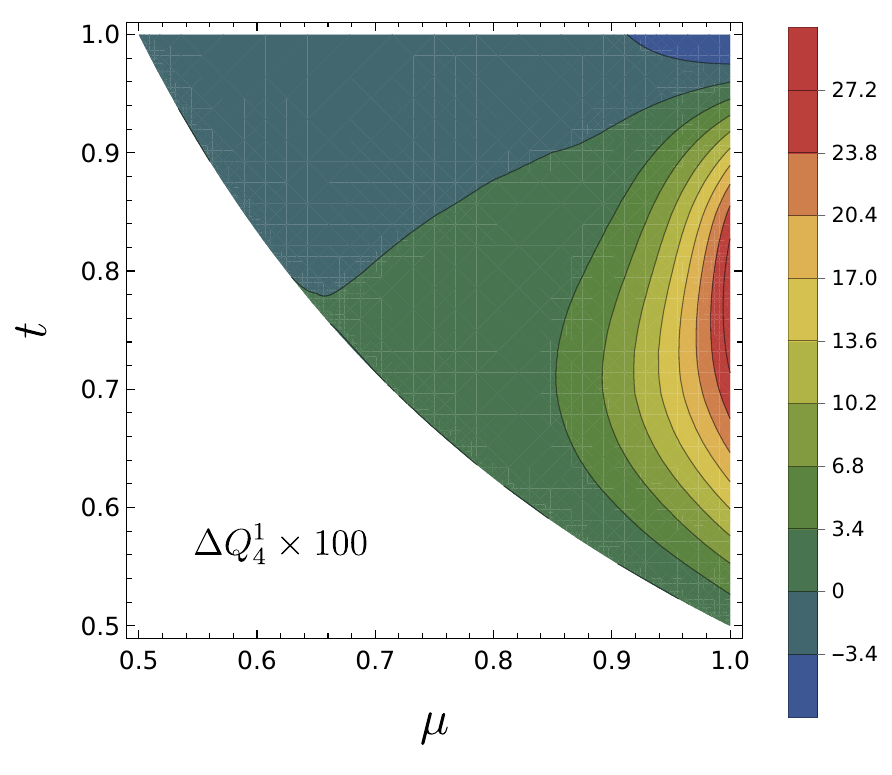}}
    \subfloat{\includegraphics[width=0.70\columnwidth]{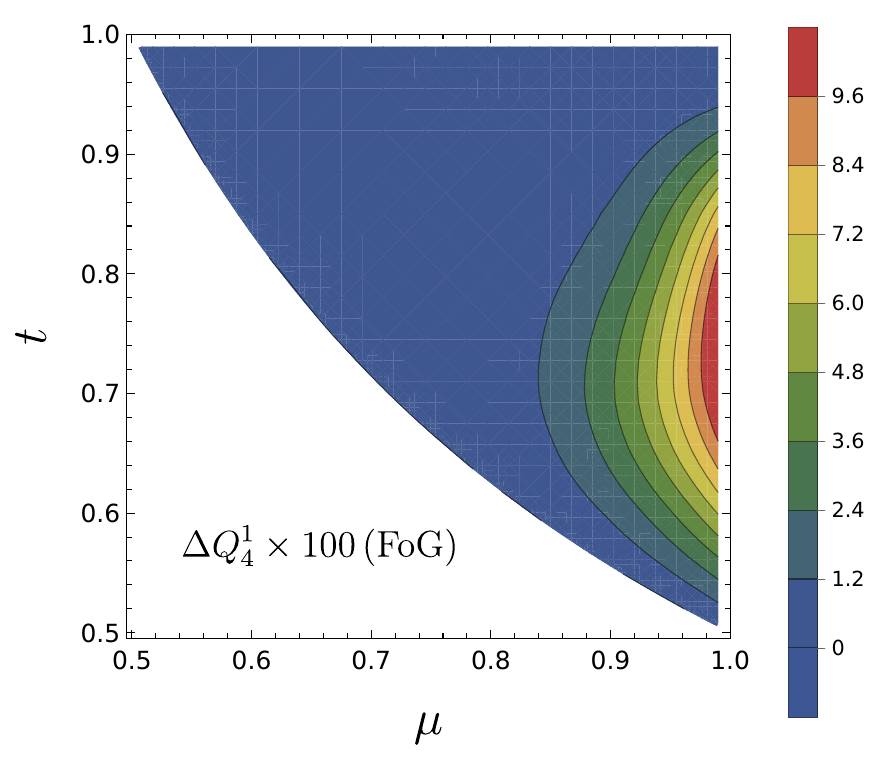}}
    \caption[]{\justifying Same as Fig.~\ref{fig:QLm_00_20}, but for the hexadecapole multipoles with 
    $(L,m)=(4,0)$ (top row) and $(4,1)$ (bottom row). For $Q_4^0$, the maximum percentage deviation without 
    FoG damping occurs near the stretched configuration 
    ($\mu\simeq1, t\simeq0.55$), reaching $\sim13\%$  compared to $\Lambda$CDM. 
    With FoG, the pattern changes qualitatively: the deviations become negative, 
    with the peak shifting to $\mu\simeq1, t\simeq0.85$, indicating that FoG 
    suppresses $Q_4^0$ more strongly in the HS model than in $\Lambda$CDM. 
    For $Q_4^1$, the contours with and without FoG look similar, with the largest 
    deviation ($\sim7\%$) near $\mu\simeq1, t\simeq0.75$ in the absence of FoG. 
    Including FoG reduces the amplitude by a factor of $\sim3$ and yields 
    negative differences for many configurations with $\mu<0.85$. Since higher 
    multipoles beyond $L=4$ have very low signal-to-noise, we do not present them here. 
    }
    \label{fig:QLm_40_41}
\end{figure*}
In an EdS universe, where $\Omega_{\rm m} = f^2$, the second-order perturbation theory kernels $F_2$ and $G_2$ have well-known analytical, time-independent solutions~\cite{Bernardeau:2001qr}. However, in a $\Lambda{\rm CDM}$ universe, the kernels become both scale- and time-dependent, necessitating a numerical approach~\cite{Fasiello:2022lff}. The situation is further complicated in the HS-$f(R)$ gravity model, where these kernels exhibit even stronger scale and time dependence due to modified gravitational dynamics, as described in the previous section. In this case, the evolution of the second-order density perturbation is governed by the kernel $F_2(\mathbf{k}_1, \mathbf{k}_2, t)$, given in Eq.~\ref{eq:F2_HS}.

The velocity divergence kernel $G_2(\mathbf{k}_1, \mathbf{k}_2, t)$, as in Eq.~\ref{eq:G2_HS}, also receives significant modifications in HS-$f(R)$ gravity model. These arise from both scale-dependent linear growth functions and additional time-dependent terms. 
As highlighted in Ref.~\cite{Aviles:2021que}, the expression of the velocity divergence kernel $G_2(\mathbf{k}_1, \mathbf{k}_2, t)$ introduces two main departures from the EdS approximation. The first arises from the scale dependence of the linear growth rate $f(k)$, which notably impacts the dipole term (proportional to $\mu$) and is associated with the non-local advection of the velocity field. The second source of deviation is the time dependence of the functions $\mathcal{A}(t)$ and $\mathcal{B}(t)$, which reflect departures from the relation $\Omega_{\rm m} = f^2$. In $f(R)$ gravity, these functions encode effects such as screening and other nonlinear features that do not arise in GR.

Interestingly, the dominant modifications to the kernel structure stem from the scale-dependent growth factor $f(k)$, while the time-dependent contributions from $\mathcal{A}$ and $\mathcal{B}$ tend to be subdominant. For instance, in the HS model, deviations due to $\mathcal{A}$ and $\mathcal{B}$ are typically at the level of $1-2\%$, as demonstrated in Ref.~\cite{Aviles:2017aor}. This is consistent with the observation that EdS-based kernels often provide sufficient accuracy even in $\Lambda$CDM analyses.

In contrast, the effect of the scale-dependent $f(k)$ can be much more significant. As shown in Fig.~\ref{fig:fk_pk}, the ratio $f(k)/f_0$ begins to depart from its $\Lambda$CDM value around $k \sim 0.01 \, h\,{\rm Mpc}^{-1}$. Here $f_0$ is the value of the growth rate at sufficiently large scale (in our case, we consider $k=10^{-4}\, h \, {\rm Mpc}^{-1}$) coinciding with the $\Lambda$CDM results.
The deviation can reach $10-20\%$ over the range $k \in [0.01, 10]\, h\,{\rm Mpc}^{-1}$ at redshift $z = 0$ for values of $f_{R0} \in [10^{-7}, 10^{-4}]$. The ratio of the power spectrum in the HS model to that of $\Lambda$CDM has been presented in the right panel of Fig.~\ref{fig:fk_pk}. Since the logarithmic growth rate $f(k)$ is the main factor, that modifies the power spectrum in HS model as evident from Eq.~\ref{eq:PL_mg}. As a result, that reflects in the power spectrum as well, as noted from the right panel of the figure. In both the plots, the solid curves are generated at redshift $z=0$, while the dotted lines are at $z=0.7$.
Some previous works~\cite{Aviles:2021que,Noriega:2022nhf,Aviles:2023fqx,Rodriguez-Meza:2023rga} have adopted an alternative approach by defining modified kernels directly in terms of the so-called \texttt{fk}-kernels. These are designed to avoid numerical instabilities and to ensure that the MG theory recovers GR behavior on large scales. However, in our analysis, we include the full time and scale dependence of both $\mathcal{A}(t)$ and $\mathcal{B}(t)$, leading to more accurate modeling of the kernel evolution.

Before turning to the spherical harmonic decomposition and its detectability in \textit{Euclid}, we first examine the structure of the second-order kernel functions in more detail.

In Fig.~\ref{fig:FG_kernel_mut}, we illustrate the changes in the kernels in the $\mu–t$ plane for the HS-$f(R)$ model with $f_{R0} = 10^{-5}$ (motivated from the analysis in Refs.~\cite{Rodriguez-Meza:2023rga,Aviles:2023fqx}) at fixed $k_1 = 0.3\, h \, \rm{Mpc}^{-1}$ and at $z=0.7$. We define $\Delta F_{ij} = F_{ij}^{\rm HS} - F_{ij}^{\Lambda\rm{CDM}}$ and $\Delta G_{ij} = G_{ij}^{\rm HS} - G_{ij}^{\Lambda\rm{CDM}}$, using the EdS approximation to compute the $\Lambda$CDM kernels as in Ref.~\cite{Bernardeau:2001qr}. The figure reveals that the largest deviations occur in the stretched triangle configuration ($\mu\approx1, t\approx0.5$) for $F_{12}$, $F_{23}$, and $G_{12}$, while $F_{31}$, $G_{23}$, and $G_{31}$ show maximal differences for linear triangle shapes ($\mu\approx1$). Note that the results are intrinsically redshift dependent, and vary depending on the redshift under consideration.
At higher $z$, $|f_{R0}|$ needs to be larger to produce the same effect as at low $z$ because the scalaron was more screened in the past. We will explore this redshift dependence next.

To further examine how the kernel deviations depend on redshift and the model parameter, Fig.~\ref{fig:FG_kernel_z} shows contour plots of $\Delta F_{12}$ and $\Delta G_{12}$ as functions of redshift $z$ and $f_{R0}$ for three representative triangle shapes: squeezed (left, $\mu\approx1$, $t\approx1$), stretched ($\mu\approx1$, $t\approx0.5$), and equilateral (right, $\mu\approx0.5$, $t\approx1$). In the squeezed configuration, the largest deviations in $F_{12}$ occur for values $f_{R0} \sim 10^{-6}$–$10^{-3}$. The constant difference contours look linear in $\log_{10}(f_{R0}) - z$ space for $f_{R0} > 10^{-7}$. A similar trend is seen for $G_{12}$ in the squeezed case, only the highest values are limited to range $f_{R0} \sim 10^{-5}$–$10^{-4}$. In the stretched configuration, $\Delta F_{12}$ peaks at lower redshift for $f_{R0} \approx 10^{-5}$. Here, the equal difference contours look elliptical in the $\log_{10}(f_{R0}) - z$ plane with a positive tilt with respect to $z$. The difference decreases along the slope. In case of $\Delta G_{12}$, the difference is larger for the higher $f_{R0}$ with a weak $z$ dependence. In the equilateral case, the highest difference for both the kernels peak for values $f_{R0} \sim 10^{-6}$–$10^{-4}$, their equal difference contours also look similar with an elliptical shape in the $\log_{10}(f_{R0}) - z$ plane and a positive slope, the difference increases along the slope with $z$. We do not show other kernel components for brevity, as they exhibit qualitatively similar behaviors as functions of $z$ and $f_{R0}$.
Having examined the underlying perturbation theory ingredients, we now proceed to analyze the bispectrum itself in redshift space.
%%%%%%%%%%%%%%%%%%%%%%%%%%%%%%%%%%%%%%%%%%%%%%%%
\section{Multipole Analysis}
\label{sec:multipoles_fR}
%%%%%%%%%%%%%%%%%%%%%%%%%%%%%%%%%%%%%%%%%%%%%%%%+
Following the framework developed in Ref.~\cite{Mazumdar:2022ynd}, we define the dimensionless reduced bispectrum in redshift space as,
\begin{eqnarray}
    Q_L^m (k_1,\mu,t)|_{\rm HS}= {b_1  B_L^m (k_1,\mu,t)|_{\rm HS}\over 3 [P(k_1)|_{\rm HS}]^2}\, . 
\label{eq:rsd_Q}
\end{eqnarray}
The power spectrum for the HS-$f(R)$ model (\textit{i.e.} $P(k_1)|_{\rm HS}$ ) has been computed following the prescription mentioned in Sec.~\ref{subsec:kernel_bispectra} and also following Ref.~\cite{Rodriguez-Meza:2023rga}\footnote{We use the publicly available code \href{https://github.com/alejandroaviles/fkpt}{https://github.com/alejandroaviles/fkpt} to compute the growth rate and power spectra in the HS $f(R)$ model}. To compute the power spectra, we fix $f_{R0}=10^{-5}$ with the other cosmological parameters are fixed to $\Lambda$CDM values\footnote{ The following fixed cosmological parameters are used to compute $\Lambda$CDM power spectrum: $\omega_{\rm cdm} =0.1201,\, \omega_{\rm b}=0.02238,\, { h}=0.6781,\, A_{s}=2.1\times 10^{-9},\, n_s=0.9660, \,\tau_{reio}=0.0543$.} . We present our results at redshift $z=0.7$. 
Instead of choosing the bias parameters as in Refs.~\cite{Aviles:2023fqx} and \cite{Rodriguez-Meza:2023rga} from the \texttt{ELEPHANT} simulation \cite{Cautun:2017tkc} or \texttt{MG-GLAM} simulations~\cite{Hernandez-Aguayo:2021kuh,Ruan:2021wup} , we adopt the bias parameters from Ref.~\cite{Yankelevich:2018uaz} in accord to \textit{Euclid} galaxy survey. 
At redshift $z=0.7$, we use $b_1 =1.18,\, b_2 = -0.76$ in absence of FoG effect. When incorporating FoG effect, we choose $\sigma_{\rm p}=4.81\, h\, {\rm Mpc}$.

Fig~\ref{fig:QLm_00_20} shows heatmaps of the signed difference between HS-$f(R)$ 
and $\Lambda$CDM for two lowest multipoles: $(L,m)=(0,0)$ (monopole)
and $(L,m)=(2,0)$ (quadrupole). The left column is without FoG, the right column includes FoG damping. The signed difference is defined as $\Delta Q_L^m = Q_L^m|_{\mathrm{HS}} - Q_L^m|_{\Lambda \mathrm{CDM}}$. However, note that we also quote the difference in percentage and the percentage difference is defined as, $\Delta Q_L^m \times 100/Q_L^m|_{\Lambda \mathrm{CDM}}$.
Considering panels without FoG, we find that the difference is almost an order of magnitude smaller 
for the monopole $Q^0_0$, as compared to the quadrupole $Q^0_2$. For both multipoles, maximum difference occurs near the stretched triangles ($\mu\approx1, t\approx0.5$). The percentage difference compared to $\Lambda$CDM is about $\sim 2\%$ for $Q^0_0$, 
and $\sim 8\%$ for $Q^0_2$.
The contours qualitatively look similar for both. Along the linear triangles ($\mu\approx1$), starting from stretched to the squeezed limit ($\mu\approx1, t\approx1$), the difference gradually decreases, and the difference is minimum for the equilateral triangles ($\mu\approx0.5, t\approx1$). Considering FoG, we see that the $Q^0_2$ contours look similar to the one without FoG, only the difference is reduced by a factor of $\sim3$. However, the maximum percentage difference now is $\sim19\%$, relative to $\Lambda$CDM. For  $Q^0_0$, we see a visible change in the contours. The highest difference is now negative and occurs along the linear triangles ($\mu\approx1$), specifically at $\mu = 0.74, t = 0.625$. This suggests that the FoG causes so much suppression in the monopole for the HS model that it takes it below the amplitude in $\Lambda$CDM. 
The maximum percentage difference is now $\sim2.8\%$ as compared to $\Lambda$CDM.
As we move from linear to the equilateral triangles, the difference gradually becomes smaller, passes through zero, becomes positive for a lot of triangles, and ultimately becomes close to zero again for the equilateral triangles ($\mu=0.5, t=1$). The results reveal that,
among the lowest multipoles, the quadrupole $Q^0_2$ is the most sensitive indicator of $f(R)$
modifications, with or without FoG, whereas the monopole $Q^0_0$ is comparatively less affected.

Next, we consider the higher-order quadrupole modes with $m\neq0$. Fig~\ref{fig:QLm_21_22} displays the difference $\Delta Q^m_{2}$ for $m=1$ and $2$, again without and with FoG (left and right columns, respectively). For $Q^1_{2}$, we find that the maximal deviations occur near the linear triangle configuration, with a peak near ($\mu \approx 1, t \approx 0.8$), for both without and with FoG. The peak percentage difference without FoG is $\sim2.7\%$ with respect to $\Lambda$CDM, while it is suppressed to $\sim1.6\%$ with FoG. For both cases, the difference $\Delta Q^1_{2}$ is negative along the L-isosceles ($t=1$) triangles, while it is positive for all other triangles. For $Q^2_{2}$, the largest deviation without FoG occurs near the L-isosceles configuration ($\mu \approx 0.8, t \approx 1$). 
With FoG, the overall deviation is decreased marginally where the percentage peak difference with respect to $\Lambda$CDM is $\sim7.5\%$ and the peak shifts marginally towards higher $\mu$. $Q^2_{2}$ shows more relative difference as compared to $Q^1_{2}$. The patterns of $\Delta Q^m_{2}$ for these two multipoles are also non-degenerate with  the effects of massive neutrinos as found out in our previous study~\cite{Pal:2025hpl}. In summary, within the quadrupole family, $m=0$ and $m=2$ offer the strongest signatures of $f(R)$ gravity.

\begin{figure*}
    \centering   
    \subfloat{\includegraphics[width=0.70\columnwidth]{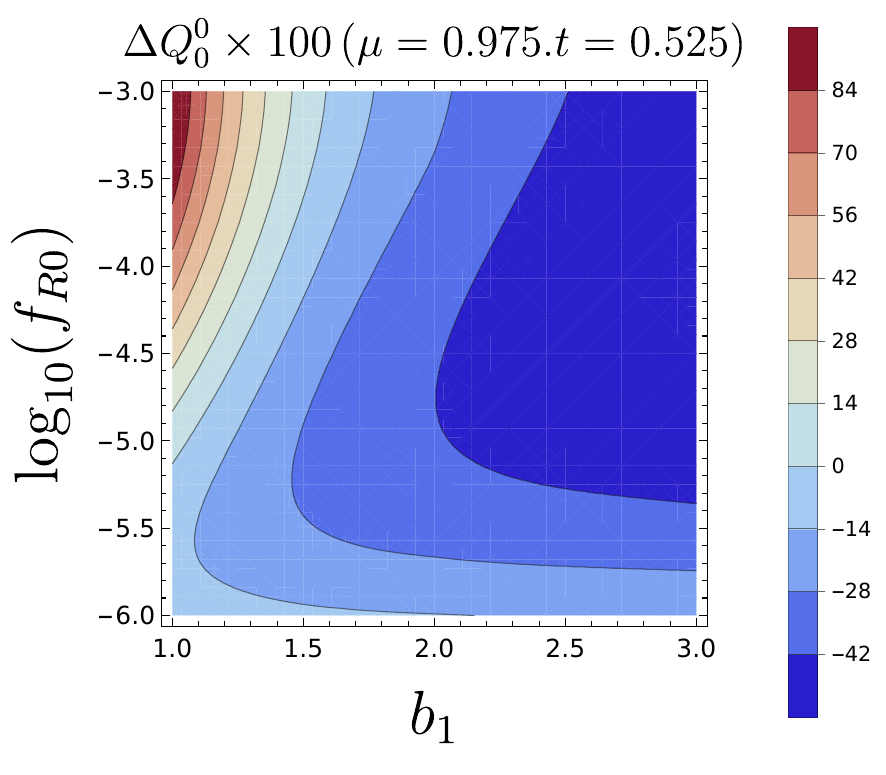}}
    \subfloat{\includegraphics[width=0.70\columnwidth]{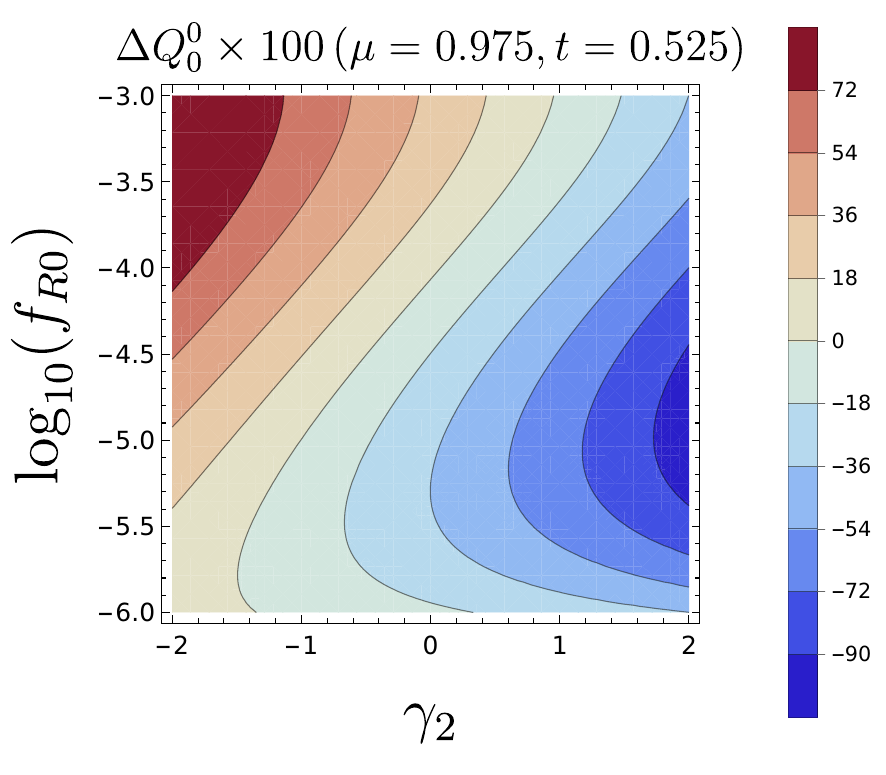}}    
    \subfloat{\includegraphics[width=0.70\columnwidth]{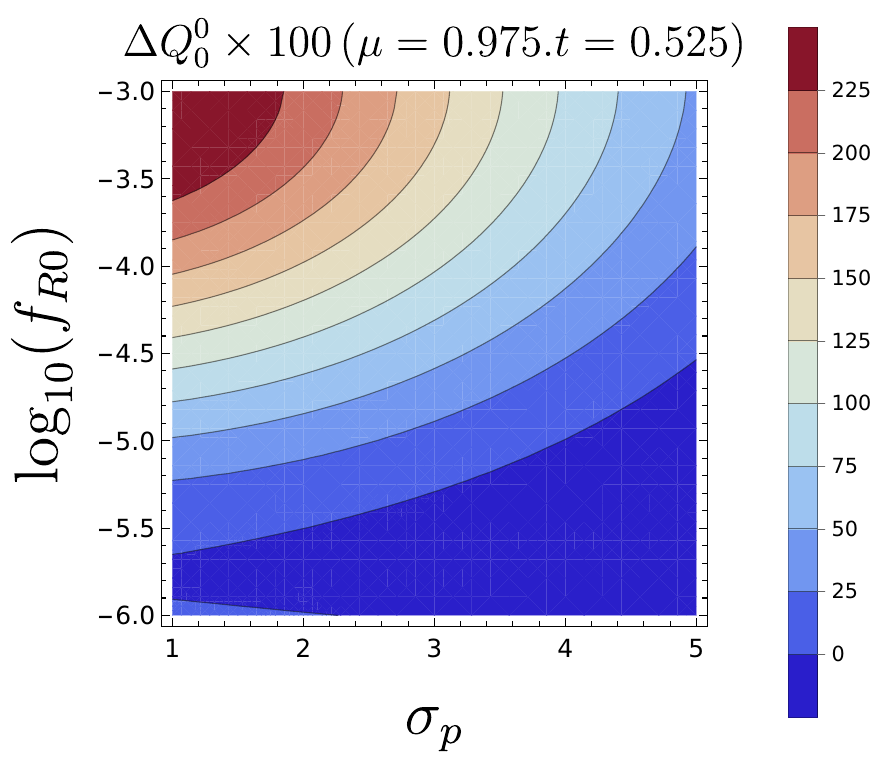}}
    \caption[]{\justifying Variation of the monopole difference $\Delta Q_0^0$ as a function of the MG parameter $f_{R0}$ and nuisance parameters, 
    for $k_1=0.3\,h\,{\rm Mpc}^{-1}$, $z=0.7$, and a stretched triangle configuration 
    with $(\mu,t)=(0.975,0.525)$. The three panels show dependence on the 
    linear bias $b_1$ (left), the nonlinear bias ratio $\gamma_2=b_2/b_1$ (middle), 
    and the velocity dispersion parameter $\sigma_p$ (right). For low $b_1$, 
    $\Delta Q_0^0$ grows with increasing $f_{R0}$, but at high $b_1$ the difference 
    eventually becomes negative. The $\gamma_2$ dependence is largely diagonal, 
    with negative $\gamma_2$ producing positive deviations and large positive 
    $\gamma_2$ driving the difference negative. In the $f_{R0}$--$\sigma_p$ plane, 
    the contours are approximately elliptical, with the largest enhancement near 
    $\sigma_p\sim1$ and $f_{R0}\sim10^{-3}$, while large $\sigma_p$ suppresses the 
    signal and makes HS-$f(R)$ nearly indistinguishable from $\Lambda$CDM. 
    These results demonstrate that the modified-gravity imprint on $Q_0^0$ is 
    correlated with galaxy bias and velocity dispersion, highlighting 
    the need for joint modeling of nuisance parameters when constraining $f(R)$ gravity.
    }
    \label{fig:bias_fR}
\end{figure*}

\begin{figure*}
    \centering   
    \subfloat{\includegraphics[width=0.70\columnwidth]{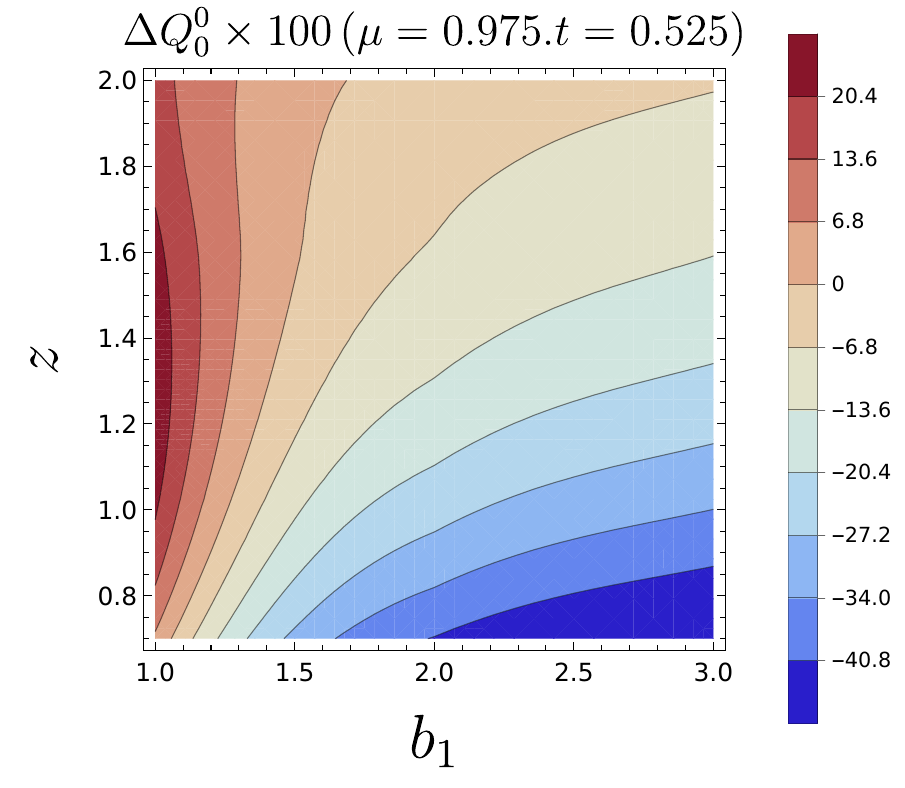}}
    \subfloat{\includegraphics[width=0.70\columnwidth]{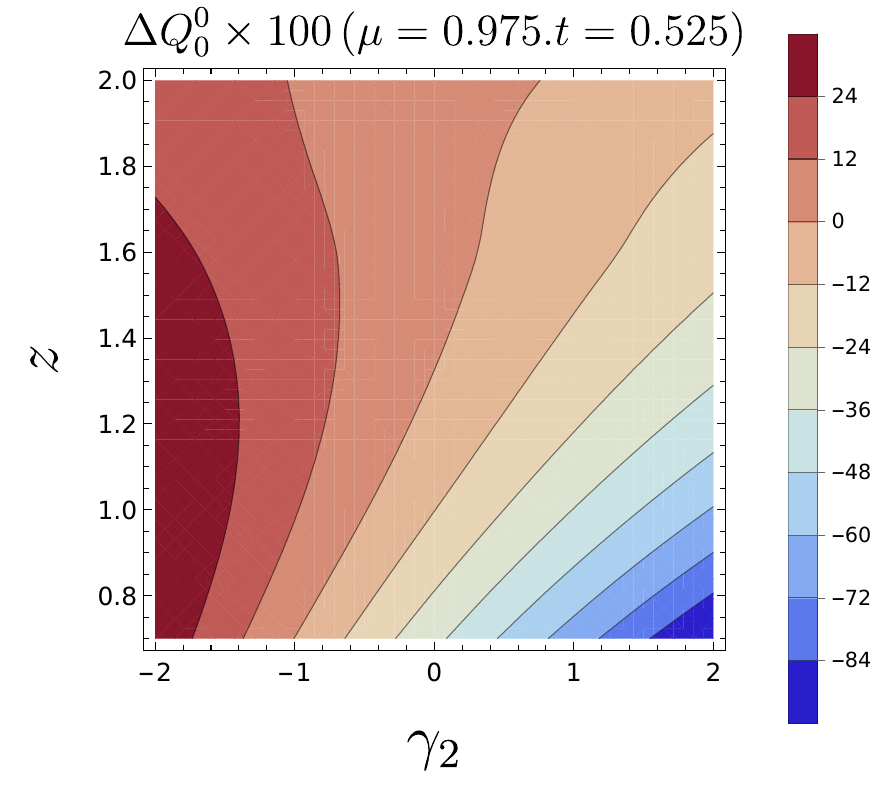}}    
    \subfloat{\includegraphics[width=0.70\columnwidth]{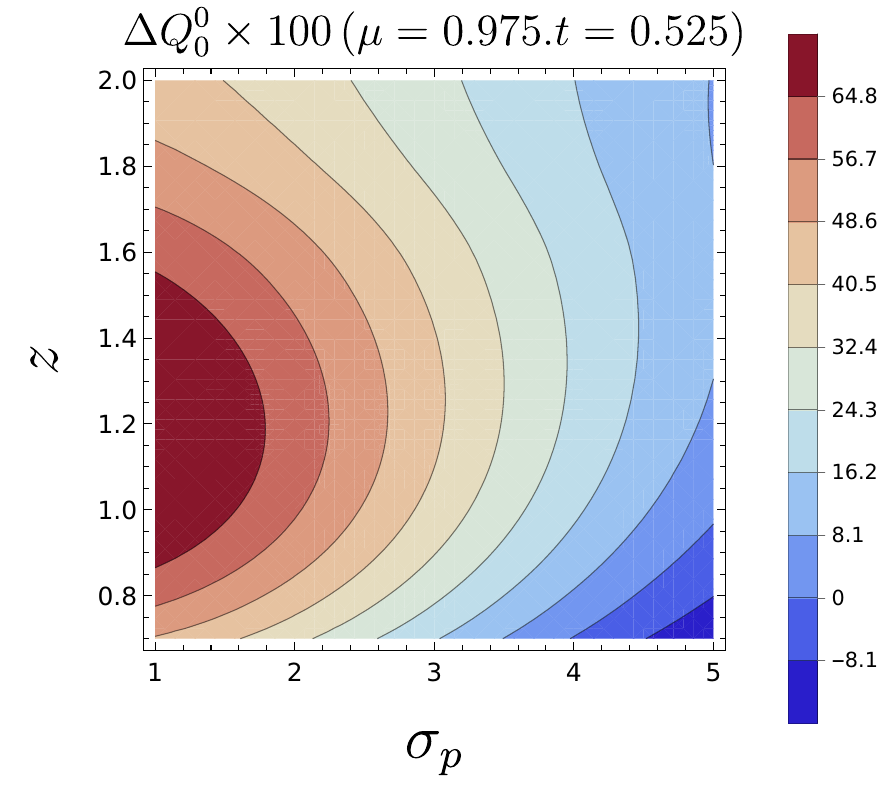}}
    \caption[]{\justifying Same as Fig.~\ref{fig:bias_fR}, but showing the redshift evolution of the monopole 
    difference $\Delta Q_0^0$. Results are plotted 
    against redshift $z$ and the nuisance parameters $b_1$ (left), 
    $\gamma_2=b_2/b_1$ (middle), and $\sigma_p$ (right), for fixed 
    $f_{R0}=10^{-5}$ and a stretched triangle configuration with 
    $(\mu,t)=(0.975,0.525)$. In the $b_1$--$z$ plane, $\Delta Q_0^0$ is positive 
    for $b_1\simeq1$ and peaks near $z\simeq1.4$, but becomes negative for 
    large $b_1$, indicating suppressed monopole amplitude in HS-$f(R)$ relative 
    to $\Lambda$CDM. In the $\gamma_2$--$z$ plane, negative $\gamma_2$ maintains 
    positive deviations out to high redshift, while large positive $\gamma_2$ 
    drives the signal negative at low $z$. In the $\sigma_p$--$z$ plane, small 
    velocity dispersions yield large positive deviations peaking around $z\simeq1.2$, 
    whereas large $\sigma_p$ suppresses the MG signal, leaving $Q_0^0$ nearly 
    indistinguishable from $\Lambda$CDM. These results emphasize the strong 
    redshift dependence of nuisance parameter degeneracies in isolating MG 
    signatures.
    }
    \label{fig:bias_z}
\end{figure*}

We have also examined the difference for the hexadecapole ($L=4$) with $m=0$ and $1$. We do not show $m\geq2$ because of the lower signal-to-noise as we shall see later. Fig.~\ref{fig:QLm_40_41} shows the results for $Q^0_{4}$ (top row) and $Q^1_{4}$ (bottom row). Considering $Q^0_{4}$, we see that the difference without FoG is maximal around the stretched limit ($\mu \approx 1, t \approx 0.55$), where the peak is $13\%$ relative to $\Lambda$CDM. The difference decreases as we go from the stretched to the equilateral limit. With FoG, we see a change in patterns. The difference becomes negative, and the maximal difference now occur around $\mu\approx1, t\approx0.85$. The FoG effect suppresses $Q^0_{4}$ stronger in $f(R)$ model than in $\Lambda$CDM, ultimately making the multipole smaller than in $\Lambda$CDM. Considering $Q^1_{4}$, we see that the contours with and without FoG overall looks similar with the maximal difference occur around $\mu\approx1, t\approx0.75$. The maximum percentage difference without FoG is $\sim 7\%$ relative to $\Lambda$CDM. With FoG, we see the difference is decreased by a factor of $\sim3$, with more triangles below $\mu<0.85$ exhibit negative values. Since higher multipoles become very subdominant beyond $L=4$, we refrain from presenting $L=6$ results here.
%%%%%%%%%%%%%%%%%%%%%%%%%%%%%%%%%%%%%%%%%%%%%%%%
\subsection{Effects of bias terms}
\label{subsec:bias_analysis}
%%%%%%%%%%%%%%%%%%%%%%%%%%%%%%%%%%%%%%%%%%%%%%%%
In the preceding section, our analysis was carried out with fixed values of the parameters $b_1$, $\gamma_2$, and $\sigma_p$. However, the bispectrum multipoles depend non-trivially on these quantities, and their signatures can be degenerate with that of the modified gravity parameter $f_{R0}$. This degeneracy is particularly relevant as the uncertainties in the nuisance parameters like the biases and $\sigma_p$ can modulate or obscure the MG signal. To robustly isolate the impact of $f_{R0}$, it is therefore important to study the response of the bispectrum multipoles to changes in each of these parameters. We address this by performing a series of controlled tests in which we vary one parameter at a time, keeping all others fixed, and compute the resulting change $\Delta Q^m_L$ between the HS model and $\Lambda$CDM. For brevity, we present results only for the monopole.

Fig.~\ref{fig:bias_fR} summarizes the variation of $\Delta Q^0_{0}$ as a function of $f_{R0}$ and three nuisance parameters: $b_1$ (left panel), $\gamma_2$ (middle), and $\sigma_p$ (right). For these tests, we fix $\mu=0.975$ and $t=0.525$ (a stretched configuration that gave a large deviation between MG and $\Lambda$CDM in $Q^0_0$), and we vary each nuisance parameter within a plausible range.

The left panel shows that $\Delta Q^0_{0}$ is positive for $b_1 \lesssim 1.5$ and $f_{R0} \gtrsim 10^{-5}$ and it peaks at the smallest value of $b_1$ and at the highest value of  $f_{R0}$. Beyond this range, $\Delta Q^0_{0}$ gradually decreases with $b_1$ for a fixed $f_{R0}$ and eventually becomes negative. For a fixed $b_1$, $\Delta Q^0_{0}$ increases with $f_{R0}$. This suggests that for high bias galaxies, the difference between GR and MG increases where the later shows lower amplitude.
In the middle panel, we see that the contours are almost diagonal, meaning increasing both $\gamma_2$ and $f_{R0}$ keeps $\Delta Q^0_{0}$ almost the same. $\Delta Q^0_{0}$ is positive for $\gamma_2<0$ and negative for $\gamma_2>0$. The maximal positive difference occurs for the lowest $\gamma_2$ and the highest $f_{R0}$, whereas the maximal negative difference occur for the highest
$\gamma_2$ and at $f_{R0}\sim10^{-5}$. 
The right panel shows that the $\Delta Q^0_{0}$ contours in the $f_{R0}-\sigma_p$ plane are almost elliptical, with the highest positive value centered at $\sigma_p\sim1$ and $f_{R0}\sim 10^{-3}$ (left-top corner of the plot).  $\Delta Q^0_{0}$ decreases as we go out diagonally from this point. For a fixed $f_{R0}$, $\Delta Q^0_{0}$ decreases with increasing $\sigma_p$, and at the maximum value of $\sigma_p$, MG effect in $Q^0_{0}$ is almost indistinguishable from GR. For $f_{R0}<10^{-5.5}$, there is little or no difference between  MG and GR for any value of $\sigma_p$ considered here. This overall suggests that the shape and scale dependence of the MG signal in higher-order statistics is intimately tied to the linear and nonlinear biases of tracers. Also, accurate modeling of FoG damping is crucial for isolating the MG signal. 

Fig.~\ref{fig:bias_z} illustrates the difference $\Delta Q_0^0$  as a function of redshift $z$ and the model parameters: $b_1$, $\gamma_2$ and $\sigma_p$. Here we use $f_{R0}=10^{-5}$ for all the panels and consider stretched triangle ($\mu=0.975, t=0.525$) to demonstrate the results. The left panel shows that the $b_1-z$ contours are almost 
linear. The values are always positive for $b_1\sim1$, and the peak positive value occurs near $z\sim1.4$. Fixing this peak redshift, if we increase $b_1$, $\Delta Q_0^0$ gradually becomes zero and then attains a negative value, suggesting that the amplitude of $Q_0^0$ is smaller in the MG than in the GR model. The peak negative value occurs for $b_1>2$ and $z \rightarrow 0$. 
The middle panel shows the effects of $\gamma_2$ and redshift. For negative $\gamma_2$, the deviation $\Delta Q_0^0$ stays positive and substantial out to high $z$, whereas for large positive
$\gamma_2$ it becomes negative at low $z$. 
The contours are almost linear with the slopes increasing with increasing $\gamma_2$. 
The right panel displays $\Delta Q_0^0$ as a function of $\sigma_p$ and $z$. We see that a small $\sigma_p$ yields a large positive deviation which peaks around $z\sim1.2$. $\Delta Q_0^0$ decreases as $\sigma_p$ increases, and gradually becomes negative. For large $\sigma_p$, we see almost no difference between MG and GR, suggesting that large velocity dispersions effectively wash out the coherent infall signals enhanced in modified gravity scenarios.

Together, these panels demonstrate that the redshift-space bispectrum monopole in $f(R)$ gravity is highly sensitive to the values of model parameters such as $b_1$, $\gamma_2$, and $\sigma_p$. While the fifth-force-induced modifications can significantly boost/suppress the bispectrum signal relative to $\Lambda$CDM under certain conditions, these effects can also be partially or fully degenerate with the bias and velocity dispersion parameters of the tracer population. Consequently, marginalizing over such parameters or jointly constraining them from data becomes essential to isolate genuine signatures of modified gravity in bispectrum analyses.

%%%%%%%%%%%%%%%%%%%%%%%%%%%%%%%%%%%%%%%%%%%%%%%%%%%%%%%%%%%
\section{SNR Calculations for Galaxy Surveys} 
\label{sec:formalism_snr}
%%%%%%%%%%%%%%%%%%%%%%%%%%%%%%%%%%%%%%%%%%%%%%%%%%%%%%%%%%
%%%%%%%%%%%%%%%%%%%%%%%%%%%%%%%%%%%%%%%%%%%%%%%%
\subsection{Bispectrum Multipole Estimator }
\label{subsec:bispectrum_estimator}
%%%%%%%%%%%%%%%%%%%%%%%%%%%%%%%%%%%%%%%%%%%%%%%%
We now assess the detectability of the $f(R)$ signatures described above by computing the signal-to-
noise ratio (SNR) for the bispectrum multipoles in a realistic survey setting. Our analysis is set in the context of the Euclid spectroscopic galaxy survey \cite{2010arXiv1001.0061R,Amendola:2016saw,Euclid:2024yvv,Euclid:2024sqd,Euclid:2024yrr,Pal:2025hpl,Mazumdar:2022ynd}, which will map approximately $10^8$ galaxies over $15{,}000$ deg$^2$ up to redshift $z \sim 2$. In any finite-volume redshift survey, the number of accessible Fourier modes is discrete and limited. The fundamental volume leads to a finite sampling of $k$-space triangles, which in turn limits how well one can measure the bispectrum at each configuration. 

Following the multipole-based bispectrum estimation methodology developed in \cite{Pal:2025hpl,Mazumdar:2022ynd}, we extend the formalism to the HS-$f(R)$ modified gravity model. For completeness, we briefly outline the estimator and its covariance properties below.

The bispectrum multipole estimator is defined as a weighted sum over all $n$ triangle triplets $(\mathbf{k}_1, \mathbf{k}_2, \mathbf{k}_3)$ that close to form a triangle within the bin $(k_1, \mu, t)$:
\begin{eqnarray}
\hat{B}_L^m (k_1,\mu, t) = \sum_{\textrm{triplets}\;n} \frac{w_L^m(\mathbf{\hat{p}}_n)}{2 V}  [\delta^s({\mathbf{k}}_n) \delta^s(\bar{\mathbf{k}}_n)
\delta^s(\tilde{\mathbf{k}}_n)\nonumber \\ + \mathrm{c.c.} ] ,
\label{eq:bispec_est}
\end{eqnarray}
where $\delta^s(\mathbf{k})$ is the redshift-space overdensity field, ``c.c.'' denotes complex conjugation, and $w_L^m(\mathbf{\hat{p}}_n)$ is a harmonic weight determined by the triangle orientation $\mathbf{\hat{p}}_n$ with respect to the line of sight{\footnote{While a direct summation is computationally expensive for the full volume of future surveys such as \textit{Euclid}, the estimator can be optimized using FFT-based expansions of the anisotropic weights $w_{L}^m(\hat{\mathbf{p}})$, analogous to the techniques commonly employed for power spectrum multipoles~\cite{Scoccimarro:2015bla}. For the purposes of the Fisher matrix forecasts presented in this work, the direct estimator provides sufficient precision to reliably assess the information content of specific triangle configurations.
}. It is defined as:
\begin{eqnarray}
w_L^m(\mathbf{\hat{p}}_n) = {\rm Re} \left[ \sqrt{\frac{2L+1}{4\pi}} \, \frac{Y_L^m(\mathbf{\hat{p}}_n)}{ \sum_{n_1} |Y_L^m(\mathbf{\hat{p}}_{n_1})|^2 } \right] \, .
\label{eq:weight}
\end{eqnarray}

The total number of triangles in a bin centered on $(k_1, \mu, t)$ with bin widths $(\Delta \ln k_1, \Delta \mu, \Delta t)$ is given by:
\begin{eqnarray}
N_{\rm tr} = \frac{(V k_1^3)^2}{8 \pi^4} \, t^2 \left[ \Delta \ln k_1 \, (t \Delta \ln k_1 + \Delta t) \, \Delta \mu \right]\, .
\label{eq:no_triangle}
\end{eqnarray}
The closure relation sets the normalization for the weights:
\begin{eqnarray}
\sum_{n_1} |Y_L^m(\mathbf{\hat{p}}_{n_1})|^2 = \frac{N_{\rm tr}}{4\pi} \, .
\end{eqnarray}

We adopt bin widths of $\Delta \ln k_1 = 0.1$, $\Delta \mu = 0.05$, and $\Delta t = 0.05$, with $k_1 = 0.3\,h \, {\rm Mpc}^{-1}$ for comparison with $\Lambda$CDM.

The error covariance for the estimator defined in Eq.~\ref{eq:bispec_est} is given by:
\begin{eqnarray}
C_{L L'}^{m m'} = \langle \Delta \hat{B}_L^m \, \Delta \hat{B}_{L'}^{m'} \rangle\, ,
\label{eq:cov}
\end{eqnarray}
where $\Delta \hat{B}_L^m = \hat{B}_L^m - \bar{B}_L^m$, and $\bar{B}_L^m$ is the ensemble average over realizations.

The analytic covariance can be explicitly written as~\cite{Gualdi:2018pyw},
\begin{eqnarray}
C_{L L'}^{m m'}(k_1,\mu,t) &=& \frac{ \sqrt{(2L+1)(2L'+1)}}{N_{\rm tr}} 
\int d \Omega_{\mathbf{\hat{p}}} \, \text{Re}[Y_{L}^m(\mathbf{\hat{p}})] \, \nonumber \\
&&  \text{Re}[Y_{L'}^{m'}(\mathbf{\hat{p}})] \times \Big[ 3 \left(B^s(k_1,\mu,t,\mathbf{\hat{p}})\right)^2  
 \nonumber \\ && + V P^s(k_1,\mu_1) P^s(k_2,\mu_2) P^s(k_3,\mu_3) \Big].
\label{eq:cov_def2}
\end{eqnarray}

\begin{figure*}
    \centering   
Features in Multipoles    \subfloat{\includegraphics[width=0.70\columnwidth]{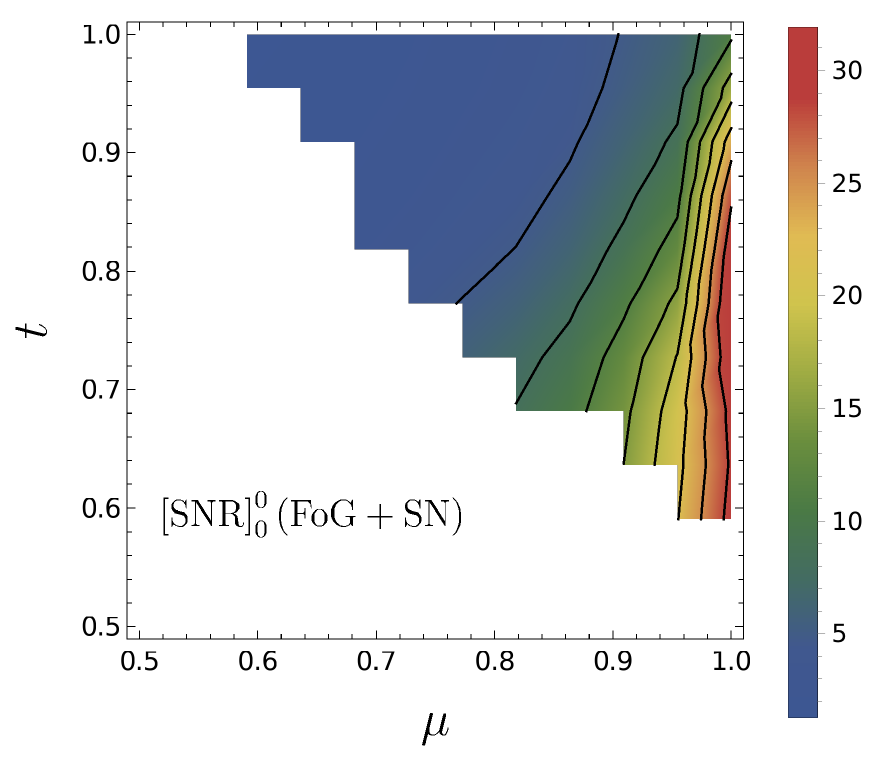}}
    \subfloat{\includegraphics[width=0.70\columnwidth]{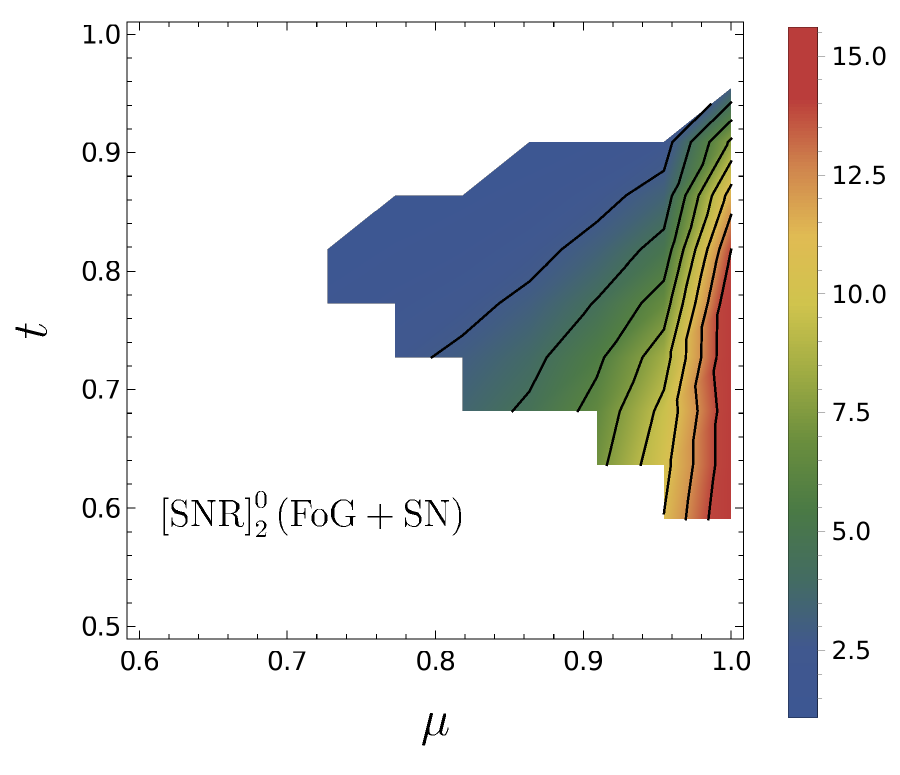}}
    \subfloat{\includegraphics[width=0.70\columnwidth]{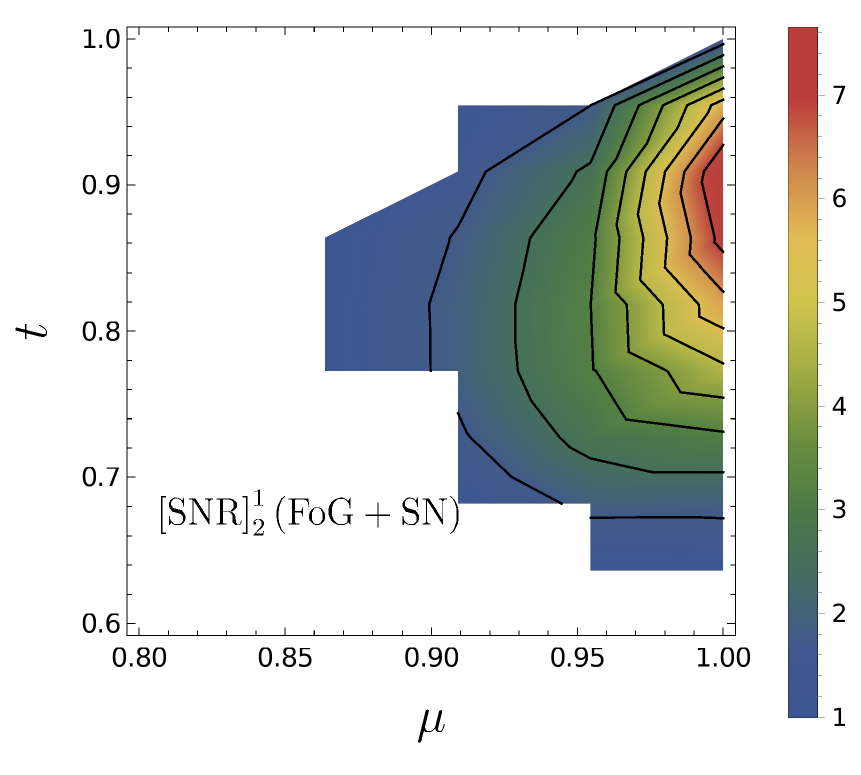}}
    \quad
    \subfloat{\includegraphics[width=0.70\columnwidth]{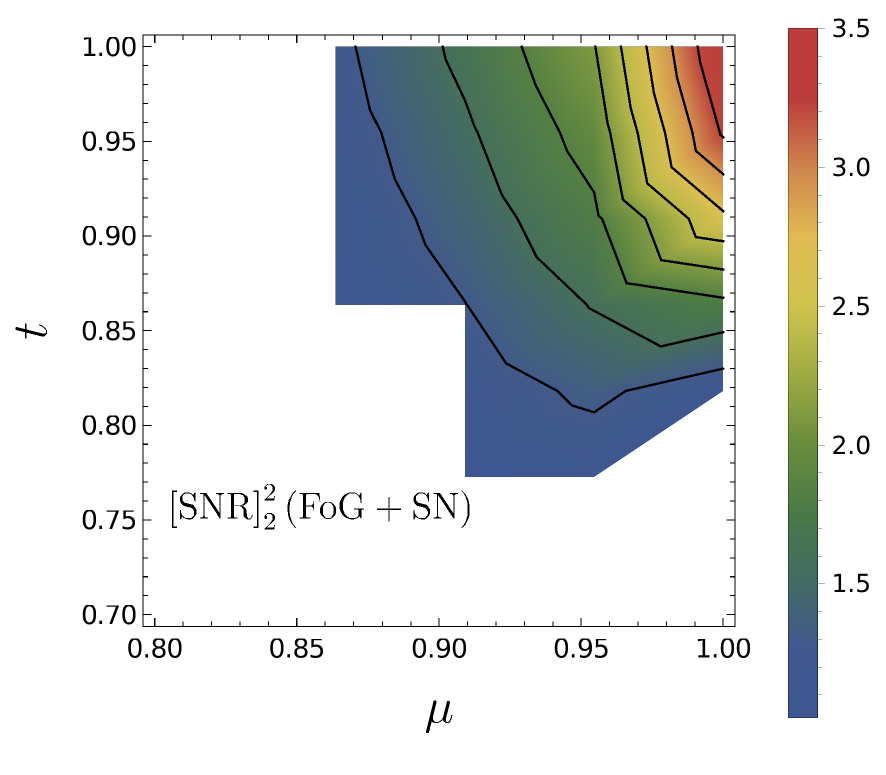}}
    \subfloat{\includegraphics[width=0.70\columnwidth]{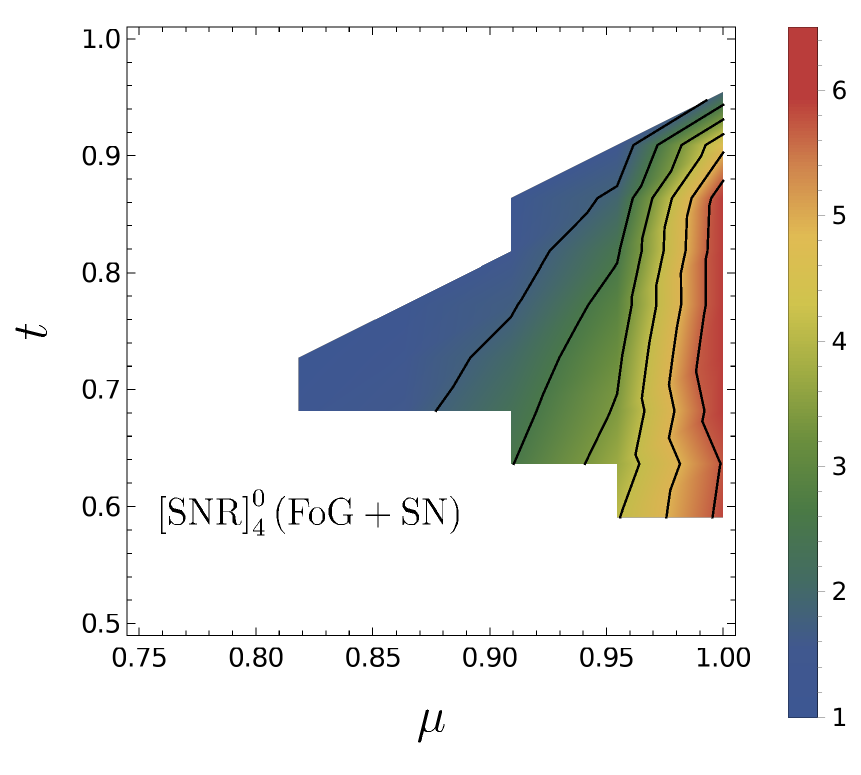}}
    \subfloat{\includegraphics[width=0.70\columnwidth]{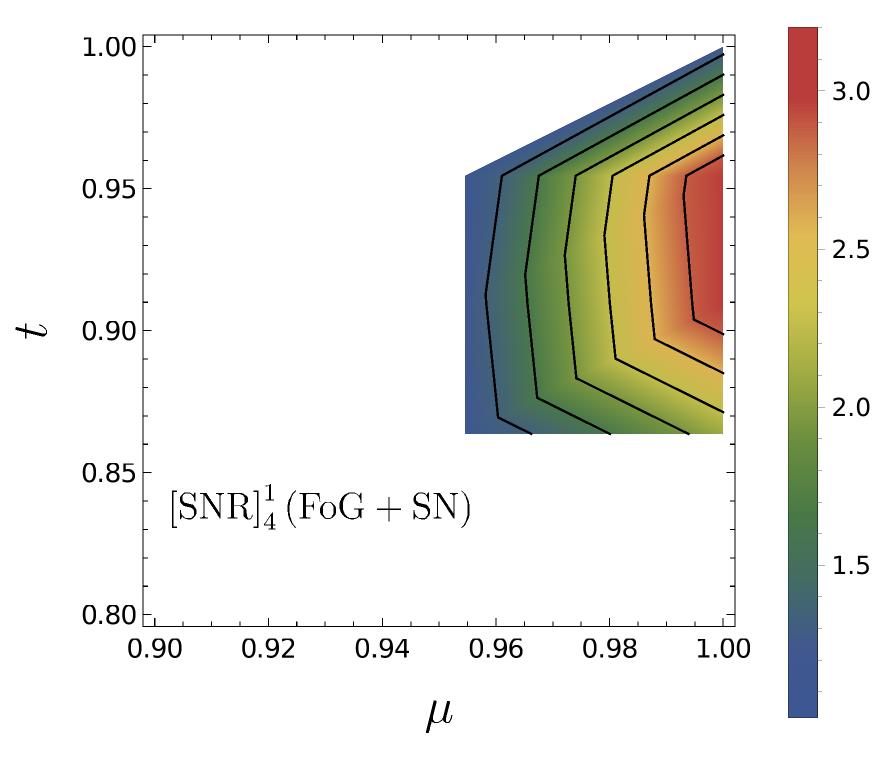}}
    \caption[]{\justifying Signal-to-noise ratio (SNR) maps for the bispectrum multipoles with 
    $(L,m)=(0,0)$ through $(4,1)$ in the HS $f(R)$ model, 
    for $f_{R0}=10^{-5}$. The SNR is shown in the $\mu$--$t$ plane at 
    redshift $z=0.7$, including the effects of both FoG 
    damping and shot noise (SN). For $L\leq2$, we consider $k_1=0.3\,h\,{\rm Mpc}^{-1}$, 
    while for the hexadecapole multipoles $(4,0)$ and $(4,1)$ we use 
    $k_1=0.4\,h\,{\rm Mpc}^{-1}$ due to their weaker overall signal. 
    Only regions with ${\rm SNR}\gtrsim1$ are displayed. The figure highlights 
    which multipoles and triangle configurations provide the strongest 
    detectability of modified gravity signatures in a \textit{Euclid}-like 
    survey. In particular, the monopole $(0,0)$ and quadrupole $(2,0)$ 
    yield the largest SNR values, while higher multipoles are suppressed 
    but still contribute complementary, configuration-dependent information.
    }
    \label{fig:snr_1}
\end{figure*}

In our forecasts, we adopt $\sigma_p \simeq 4.81\, h\, {\rm Mpc}$, consistent with \textit{Euclid} predictions \citep{Yankelevich:2018uaz}. 

Additionally, galaxies are discrete tracers, and shot noise arises due to Poisson sampling of the underlying continuous density field. The observed spectra are corrected as:
\begin{eqnarray}
P_{\rm obs}^s(k) &=& P^s(k) + P_{\rm shot}, \\
B_{\rm obs}^s(\mathbf{k}_1,\mathbf{k}_2,\mathbf{k}_3) &=& B^s + P_{\rm shot}' \sum_i P^s(k_i) + B_{\rm shot}', \\
\text{with} \quad \quad \quad P_{\rm shot}' &=& n_g^{-1}, \quad B_{\rm shot}' = n_g^{-2} \, ,
\end{eqnarray}
where $n_g$ is the galaxy number density. The corresponding correction to the covariance is given by \citep{Mazumdar:2022ynd}:
\begin{eqnarray}
[C_{L L'}^{m m'}]_{\rm shot} - C_{L L'}^{m m'} &\approx& V^{-1} \left[ n_g^{-5} + n_g^{-4} P^s + n_g^{-3} B^s \right] \nonumber\\
&& + n_g^{-2} (P^s)^2 + n_g^{-1} P^s B^s \, .
\label{eq:covg}
\end{eqnarray}
For a \textit{Euclid}-like survey at $z=0.7$, assuming $n_g \sim 10^{-3}\,h^3{\rm Mpc}^{-3}$, $P^s \sim 10^4\,h^{-3}{\rm Mpc}^3$, $V \sim 10^9\,h^{-3}{\rm Mpc}^3$, we find that the dominant term in Eq.~\ref{eq:cov_def2} scales as $C_{LL'}^{mm'} \sim 10^{21}\,h^{-12}{\rm Mpc}^{12}$, while the largest shot noise contribution scales as $n_g^{-1} P^s B^s \sim 10^{15}\,h^{-12}{\rm Mpc}^{12}$, which is negligible in comparison.
We, however, include all the terms in our signal-to-noise forecasts. The exact values used in the forecast analysis is taken from Table 1 of~\cite{Yankelevich:2018uaz} (e.g. at redshift $z=0.7$, $V=2.82\, h^{-3}\, {\rm Gpc}^3$, $n_g = 2.76 \times 10^{-3}\, h^{3}\,{\rm Mpc}^{-3} $ and $\sigma_p=4.81\, h^{-1}\, {\rm Mpc}$ ). 
%%%%%%%%%%%%%%%%%%%%%%%%%%%%%%%%%%%%%%%%%%%%%%%%
\section{Detection Prospects in Euclid}
\label{sec:euclid}
%%%%%%%%%%%%%%%%%%%%%%%%%%%%%%%%%%%%%%%%%%%%%%%%
%%%%%%%%%%%%%%%%%%%%%%%%%%%%%%%%%%%%%%%%%%%%%%%%
\subsection{Signal-to-Noise Ratio for \texorpdfstring{$f(R)$}{f(R)} Models}
\label{subsec:SNR_fR}
%%%%%%%%%%%%%%%%%%%%%%%%%%%%%%%%%%%%%%%%%%%%%%%%
Given the covariance including FoG and shot noise, the signal-to-noise ratio (SNR) for detecting deviations from $\Lambda$CDM in a given bispectrum multipole reads:
\begin{eqnarray}
[{\rm SNR}]_L^m = \frac{ \Big| B_L^m|_{\rm HS} - B_L^m|_{\Lambda{\rm CDM}} \Big| }{ \sqrt{ C_{LL}^{mm}|_{\Lambda {\rm CDM}} } }\, .
\label{eq:snr_fR}
\end{eqnarray}

This allows us to quantify the detectability of the HS-$f(R)$ signature across different triangle configurations and multipoles. $B_L^m|_{\rm HS}$ is the multipoles of the bispectrum in $f(R)$ model, calculated by using Eq.~\ref{eq:Blm}, where $B^s$ is the total tree level bispectrum in redshift space, defined through Eq.~\ref{eq:b_rsd} with modified kernels. 

\subsection{SNR prediction}
\label{subsec: SNR_results}
We present our forecasted SNR for detecting $f(R)$ gravity signals in the bispectrum multipoles for a \textit{Euclid}-like survey~\cite{Yankelevich:2018uaz}. Figure~\ref{fig:snr_1} shows the SNR as a function of triangle shape in the $\mu$–$t$ plane for each multipole, including the effects of FoG damping and shot noise, under the fiducial HS model with $f_{R0}=10^{-5}$. For multipoles with $L \leq 2$, we fix $k_1=0.3\,h\,{\rm Mpc}^{-1}$ at redshift $z=0.7$, while for the hexadecapole modes $(L,m)=(4,0)$ and $(4,1)$ we adopt $k_1=0.4\,h\,{\rm Mpc}^{-1}$ due to their lower signal strength. The galaxy bias parameters $b_1$ and $b_2$, as well as the velocity dispersion $\sigma_p$ governing FoG suppression, are taken from Ref.~\cite{Yankelevich:2018uaz} at the corresponding redshift. This figure highlights which multipoles and triangle orientations provide the largest SNR, and are therefore most sensitive to modified gravity effects.

The monopole $B_0^0$ has the highest overall signal-to-noise. The maximum SNR is found along the linear triangles ($\mu\sim1$), where the peak value is $\approx30$ for configurations with $\mu \gtrsim 0.95$ and $0.6 \lesssim t \lesssim 0.9$. We find that without FoG and shot noise, this
peak SNR exceeds $\sim100$. The small-scale velocity dispersion suppresses the signal most in the linear ($\mu\sim1$) configuration, whereas in the equilateral configuration, this effect is less pronounced. The MG effect is also maximum for the linear triangles. Therefore, a good estimate of the FoG is required in order to maximize the utility of the linear triangles.
\begin{figure*}
    \centering   
    \subfloat{\includegraphics[width=0.70\columnwidth]{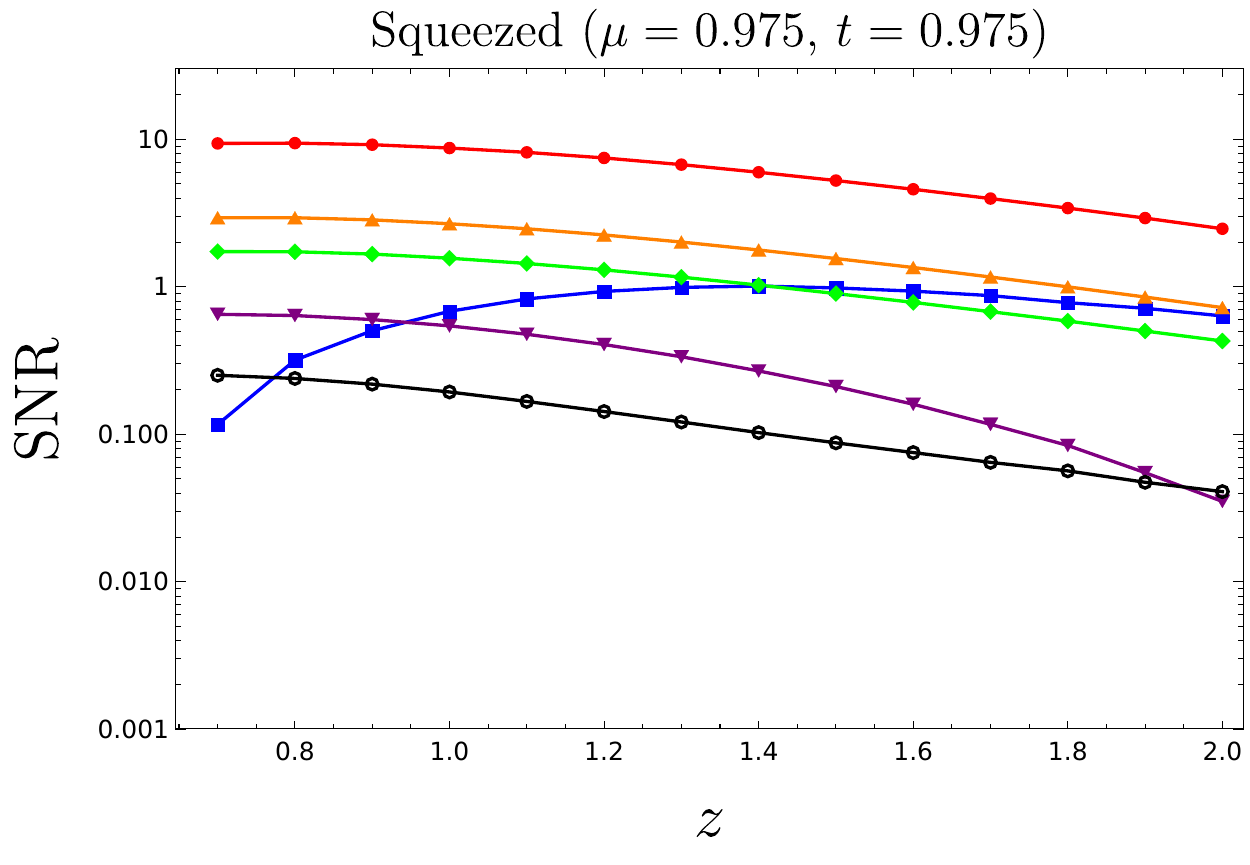}}
    \subfloat{\includegraphics[width=0.70\columnwidth]{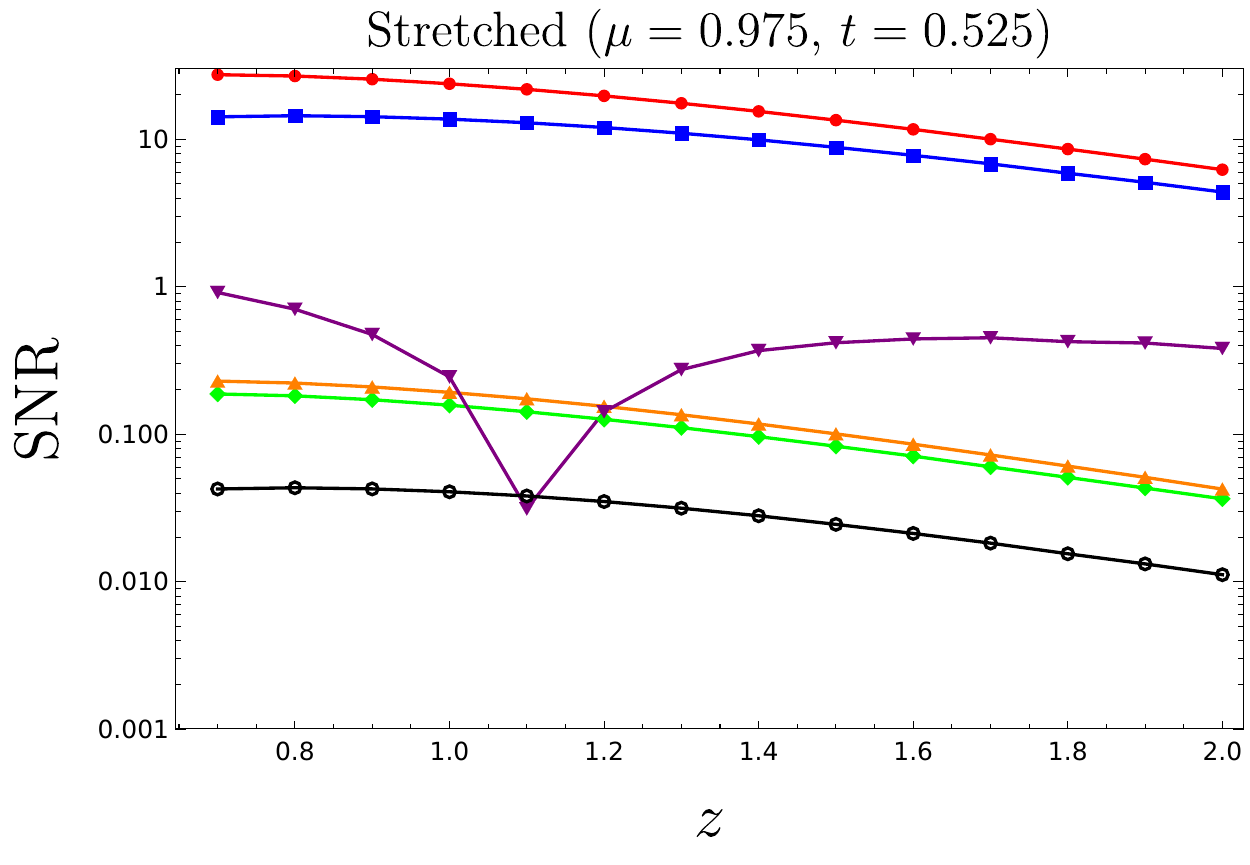}}
    \subfloat{\includegraphics[width=0.80\columnwidth]{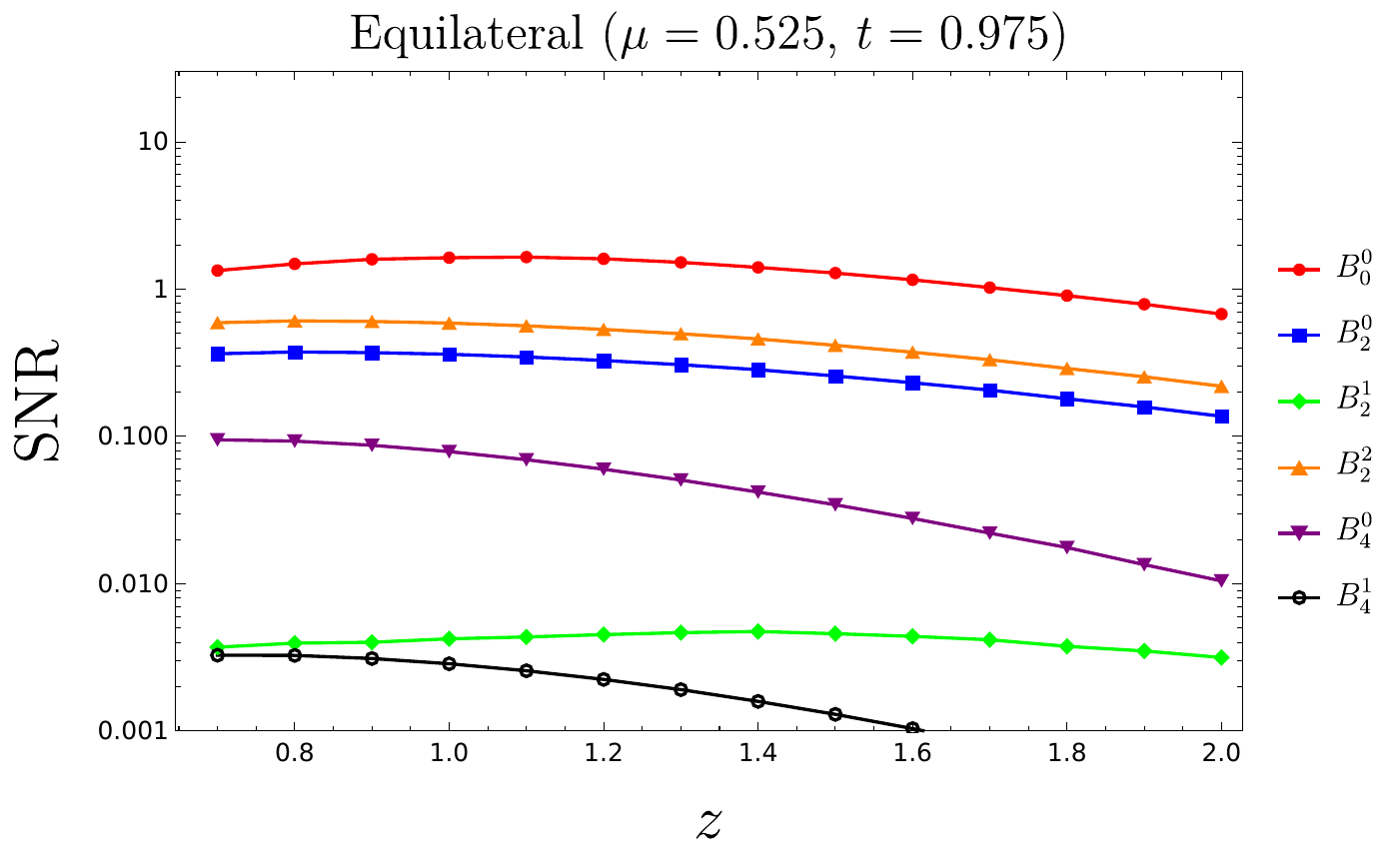}}
    \quad
    \subfloat{\includegraphics[width=0.70\columnwidth]{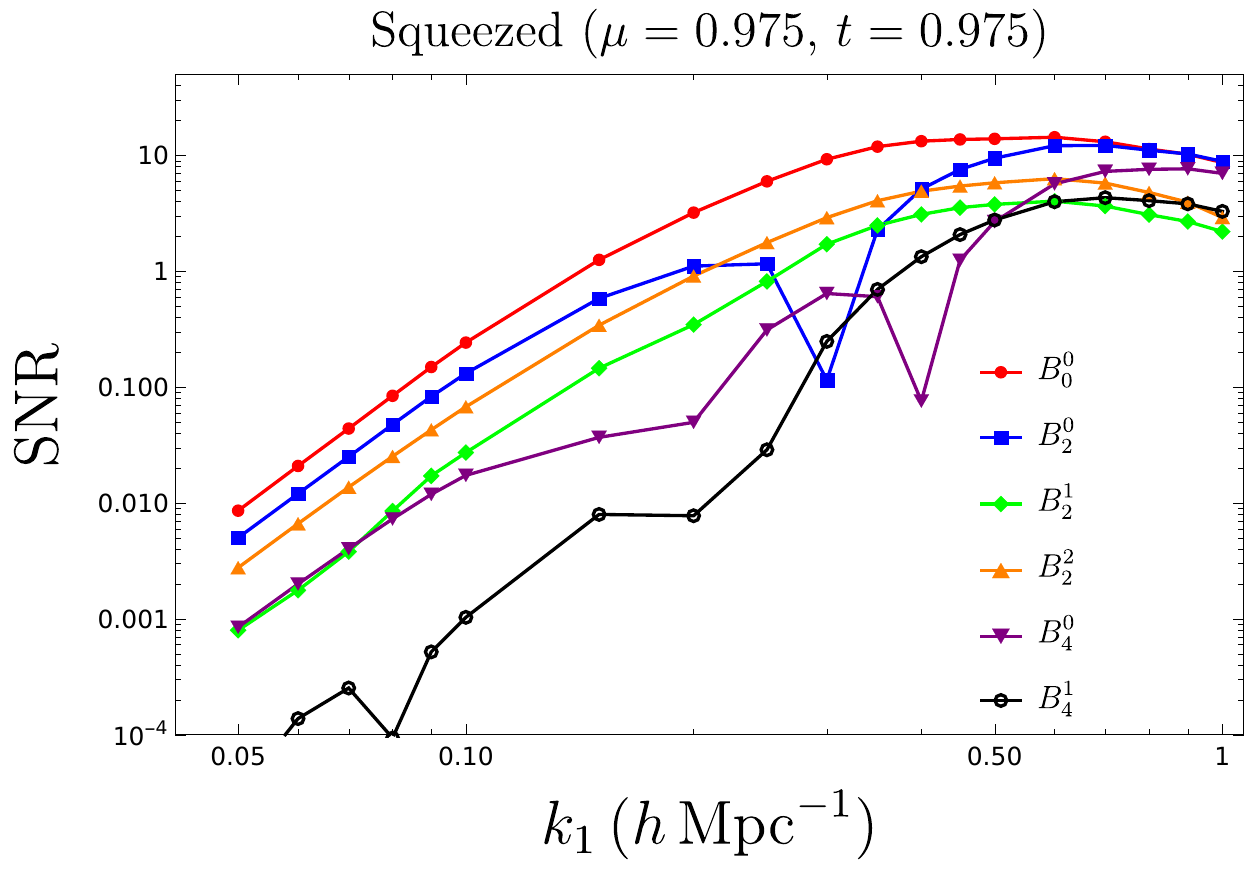}}
    \subfloat{\includegraphics[width=0.70\columnwidth]{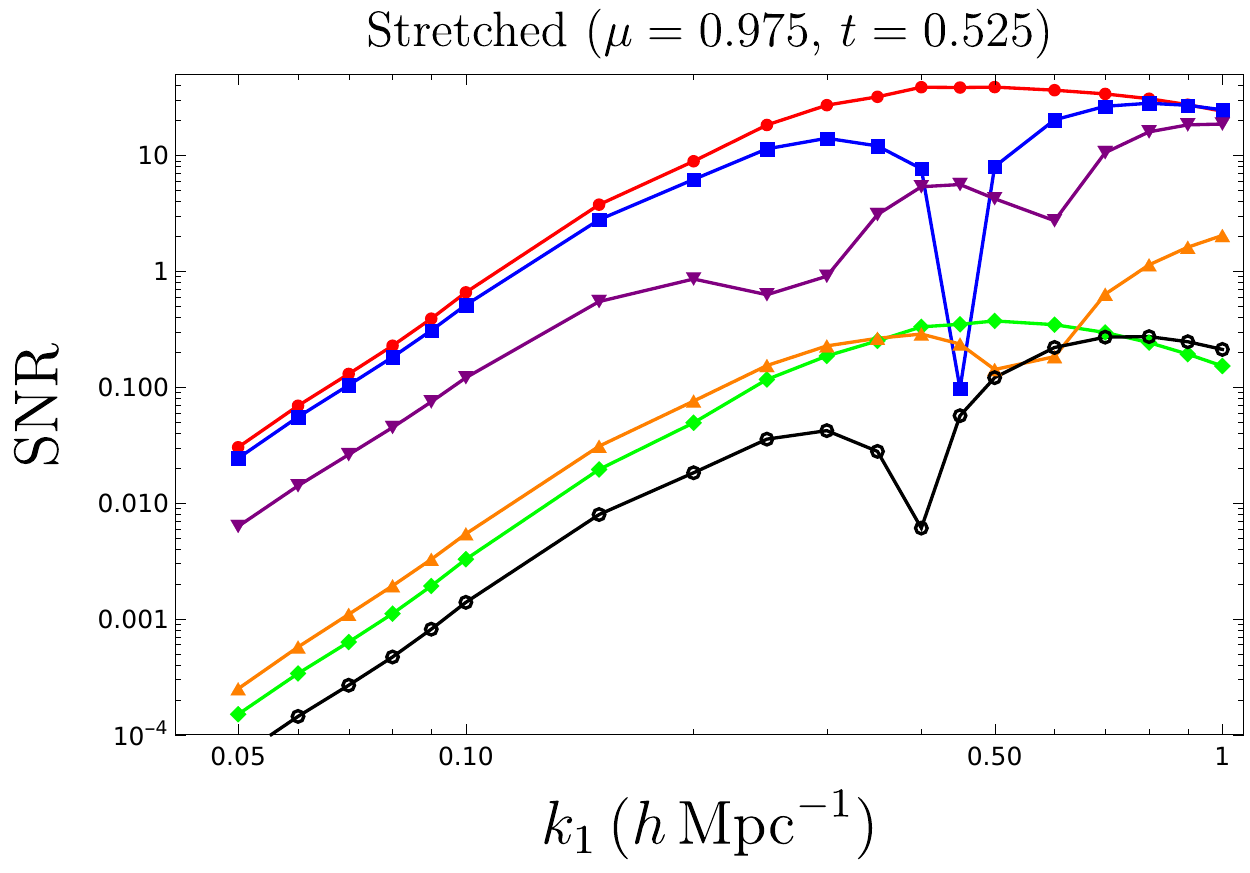}}
    \subfloat{\includegraphics[width=0.70\columnwidth]{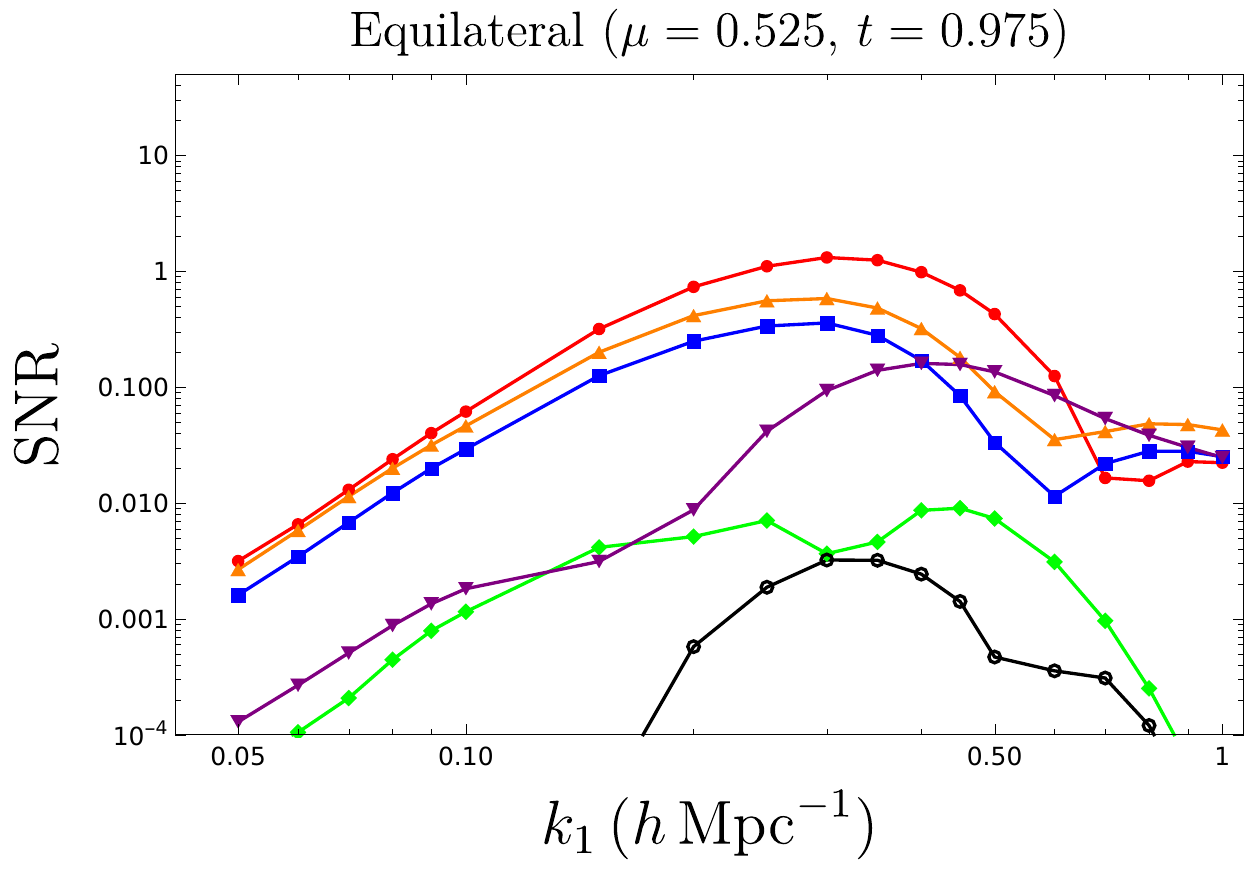}}
    \caption[]{\justifying SNR of bispectrum multipoles up to $(L,m)=(4,1)$ as a 
    function of redshift $z$ (\textbf{upper panels}) and largest triangle side $k_1$ 
    (\textbf{lower panels}), for three representative configurations: squeezed 
    (left, $\mu=0.975, t=0.975$), stretched (middle, $\mu=0.975, t=0.525$), and 
    equilateral (right, $\mu=0.525, t=0.975$). The SNR is computed for a 
    \textit{Euclid}-like survey including FoG damping and shot noise, with 
    redshift-dependent bias parameters taken from \cite{Yankelevich:2018uaz}. 
    For fixed $k_1=0.3\,h\,{\rm Mpc}^{-1}$ (upper row), the SNR of most multipoles 
    increases toward lower redshift, reflecting accumulated nonlinear growth and 
    fifth-force effects, with $B_0^0$ and $B_2^0$ dominating. In contrast, the 
    equilateral configuration shows very low SNR, rarely exceeding unity. For fixed 
    $z=0.7$ (lower row), the SNR generally increases with $k_1$ up to 
    $k_1\simeq0.5\,h\,{\rm Mpc}^{-1}$ before saturating, though some multipoles show 
    non-monotonic dips where FoG damping suppresses the MG signal, making it 
    indistinguishable from GR. These results confirm that squeezed and stretched 
    triangles provide the highest SNR, while equilateral triangles are suboptimal 
    for detecting $f(R)$ signatures.
    }
    \label{fig:snr_z}
\end{figure*}
\begin{figure*}
    \centering   
    \subfloat{\includegraphics[width=0.70\columnwidth]{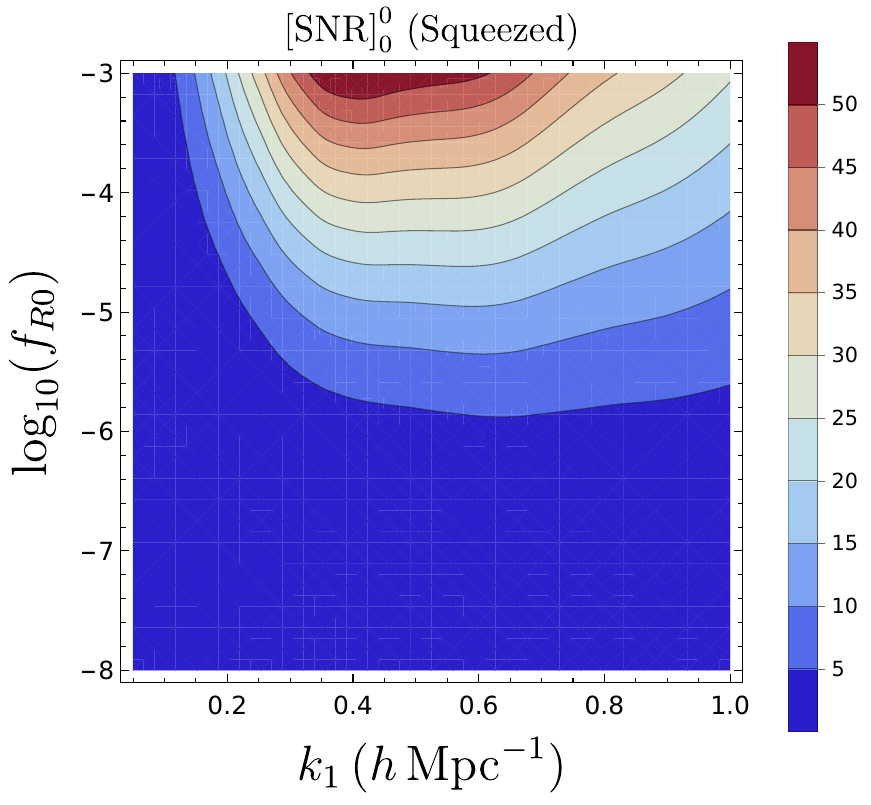}}
    \subfloat{\includegraphics[width=0.70\columnwidth]{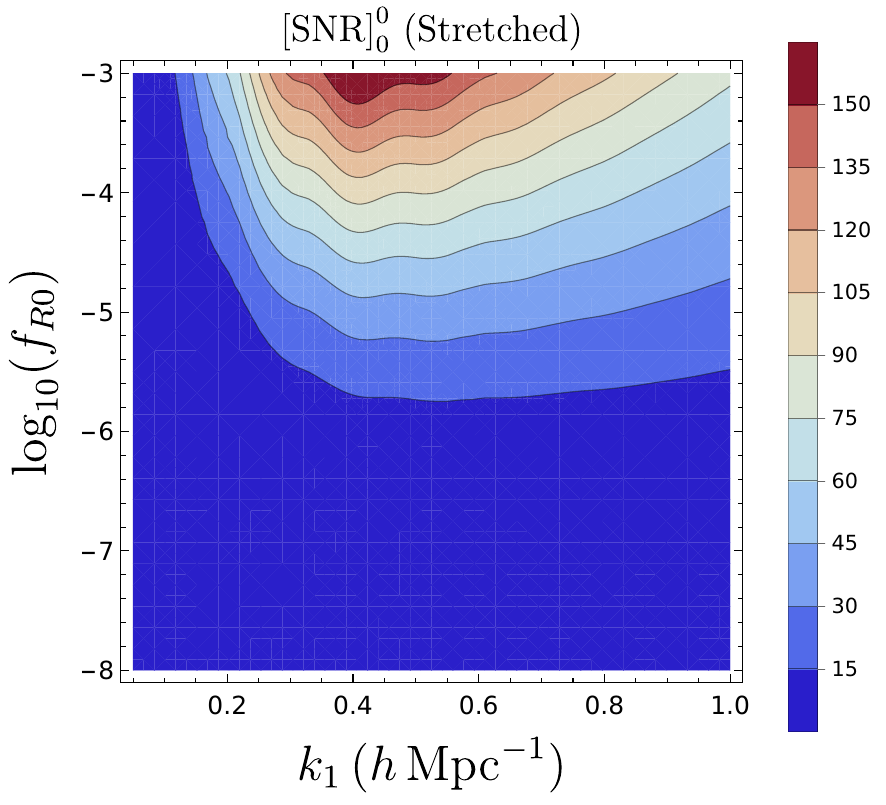}}
    \subfloat{\includegraphics[width=0.70\columnwidth]{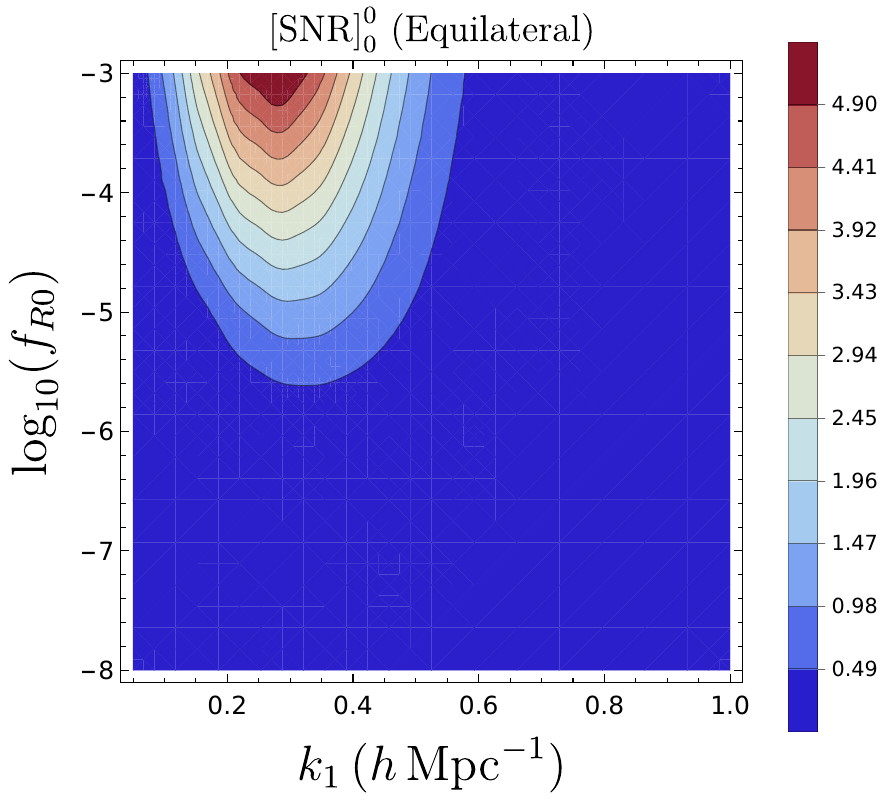}}
    \caption[]{ \justifying The SNR 
    of the monopole $B_0^0$ in the $(\log_{10}f_{R0}$--$k_1)$ plane at 
    redshift $z=0.7$. Results are presented for squeezed (left), stretched (middle), 
    and equilateral (right) triangle configurations, with all other parameters fixed. 
    The SNR increases with $f_{R0}$, remaining $\lesssim5$ for $f_{R0}<10^{-7}$ 
    where HS-$f(R)$ predictions approach GR. The maximum SNR occurs around 
    $k_1\simeq0.3$--$0.6\,h\,{\rm Mpc}^{-1}$ for squeezed and stretched triangles, 
    and $k_1\simeq0.2$--$0.4\,h\,{\rm Mpc}^{-1}$ for equilateral triangles. 
    Beyond these scales, the SNR decreases as $k_1$ grows or shrinks away from 
    the peak range. Among the three shapes, stretched triangles yield the 
    largest and broadest SNR contours, followed by squeezed, while equilateral 
    configurations remain subdominant. These trends underscore that stretched 
    and squeezed configurations provide the most favorable windows for detecting 
    $f(R)$ signatures in bispectrum monopoles.
    }
    \label{fig:snr_k1}
\end{figure*}

The quadrupoles, $B_2^0$, $B_2^1$, and $B_2^2$, exhibit progressively lower peak SNR values. Not only that, the number of triangles (area in the $\mu-t$ plane) for which ${\rm SNR}>1$ also decreases progressively with increasing $m$. Considering $B_2^0$, the peak SNR is $\sim 15$, which increases to $\sim57$ in the absence of FoG and SN. The highest values lie at $\mu\approx1$ and within $0.6\lesssim t \lesssim 0.85$. For $B_2^1$, the maximum values occur for $\mu>0.95$, and between $0.8\lesssim t \lesssim0.9$. The peak value is $\sim8$, which becomes $\sim15$ without FoG and SN. For $B_2^2$, the SNR peaks around the squeezed limit, with the peak value $\sim3.5$. These $L=2$ multipoles are sensitive to redshift-space anisotropies and scale-dependent growth, and provide complementary information for constraining modified gravity when combined with the monopole. 

Considering hexadecapole, $B_4^0$ retains modest significance, where peak SNR $\sim 6$ with FoG and SN (down from $\sim 10$ without). $B_4^1$ is similar to $B_2^1$ reaching maximum of about $\sim5$ SNR with no FoG and SN, dropping to $\sim3$ with those. Note that the number of triangles where ${\rm SNR}>1$ is even limited for the hexadecapole. We did not plot hexadecapole with $m>2$ as they do not exhibit ${\rm SNR}>1$ for any triangles.

From these results, it is clear that $B_0^0$ is by far the strongest contributor to SNR,
followed by $B_2^0$. Including $B_2^1$, $B_2^2$, $B_4^0$, $B_4^1$ adds some
complementarity, but each is at most a $\sim$few sigma effect. When combined, however, they can boost
total SNR. Across all multipoles ranges, we observe a consistent pattern. The signal-to-noise is high for the linear triangles ($\mu\approx1$), for some multipoles it peaks in the squeezed limit ($\mu\approx1$, $t \approx1$). These configurations are most sensitive to scale-dependent modifications in gravity, especially in models like HS-$f(R)$, where the fifth force is enhanced on quasi-linear scales. Hence, these are most suitable for detecting the signature of MG. These triangles, however, are mostly affected by FoG and shot noise, which reduces the detectability.

In the top row of Fig.~\ref{fig:snr_z}, we show how the SNR for different multipoles varies with redshift. The three panels display SNR for the various multipoles for three triangle configurations: squeezed ($\mu=0.975, t=0.975$), stretched ($\mu=0.975, t=0.525$) and equilateral ($\mu=0.525, t=0.975$). Note that the biases, $b_1$ and $b_2$, are in general redshift dependent and the values of these at respective redshifts are taken from~\cite{Yankelevich:2018uaz}. We consider fixed $k_1=0.3\, h\,{\rm Mpc}^{-1}$ to demonstrate the results. The effects of FoG and SN are included. For the squeezed configuration, all multipoles show a monotonically increasing trend in SNR toward lower redshift, barring $B_2^0$, which decreases with decreasing $z$ below $z\sim1.5$. The results overall reflect the accumulation of nonlinear growth and fifth-force effects in the late Universe. The monopole $B_0^0$ dominates over all redshifts, followed by the quadrupole $B_2^0$, with higher multipoles suppressed by almost an order of magnitude. For the stretched triangles, $B_0^0$ and $B_2^0$ show the highest SNR across redshifts. All other multipoles have almost an order of magnitude or more lower SNR, which mostly increases as $z$ decreases. $B_4^0$ shows a distinct dip near $z\sim1.1$.  
For the equilateral triangles, $B_0^0$ and $B_2^2$ shows SNR close to or slightly above $\sim 1$ across redshifts. For $B_2^0$, SNR stays close to $\sim0.5$, whereas for other multipoles the SNR is extremely small. This overall underscores the limited utility of equilateral configuration for detecting signatures of MG.

The bottom row of Fig.~\ref{fig:snr_z} shows how the SNR for different multipoles varies with the largest wave vector $k_1$, or in other words, varies with the size of the triangles. We keep the redshift fixed at $z=0.7$. The three panels show results for the same three triangle configurations as in the top row. Considering squeezed and stretched triangles, we overall see that the SNR for multipoles, from the smallest $k_1$, first grows with $k_1$ and beyond $k_1\sim0.5 \, h \, {\rm Mpc}^{-1}$ SNR almost saturates. SNR is maximum for $B_0^0$ and $B_2^0$ throughout $k_1$. Some multipoles, like $B_2^0$ and $B_4^0$ for squeezed, and $B_2^0$, $B_2^2$ and $B_4^1$ for stretched triangles, show a non-monotonic behavior, where the SNR values show a sudden dip at particular $k_1$ values. These dips correspond to transitions through zero. The FoG damping suppresses the effect of MG, and around these dips, the multipoles in the MG model match those predicted by GR. Considering the equilateral triangles, we see that the SNR for most of the multipoles is well below $1$ through the entire $k_1$ range, barring $B_0^0$ which exhibits SNR$\gtrsim 1$ around $k_1\sim0.4 \, h \, {\rm Mpc}^{-1}$. The SNR for all the multipoles shows a non-monotonic behavior, where it increases with $k_1$, attains a plateau, and finally dips sharply with $k_1$. These results also confirm that equilateral configurations are suboptimal for probing modified gravity signals in the bispectrum.

Finally, Fig.~\ref{fig:snr_k1} shows the SNR maps of the monopole $B_0^0$ in $(\log_{10}|\!f_{R0}|,\,k_1)$ plane for the same three three triangle types as in Fig.~\ref{fig:snr_z}. The parameter values used in this plot are also the same as in Fig.~\ref{fig:snr_z}. We overall see that the SNR increases with increasing $f_{R0}$, which is expected. SNR is $\lesssim 5$ for $f_{R0}<10^{-7}$, when the MG predictions are close to the GR. 
The SNR is maximum around $k_1\sim0.3-0.6 \, h \, {\rm Mpc}^{-1}$ for squeezed and stretched triangles, and $k_1\sim0.2-0.4{\rm Mpc}^{-1}$ for equilateral triangles. The SNR decreases as we move on either side of this $k_1$ range. The stretched triangles show maximum SNR, followed by squeezed and equilateral. The equal SNR contours are broader for stretched and squeezed triangles in comparison to the equilateral triangles.

Finally, while we have omitted the tidal bias ($b_t$) term~\cite{Desjacques:2016bnm} in our main equations for simplicity, it is important to note that the tidal bias introduces a notable shift in the bispectrum at large scales, where it plays a role comparable to the non-linear bias ($b_2$) in controlling the amplitude. While omitted in our primary framework, it can be readily reintroduced using standard methods. 

To explicitly demonstrate its impact, Fig.~\ref{fig:snr_bt} shows the difference in the signal-to-noise ratio (SNR) of the $B_0^0$ and $B_2^0$ multipoles when tidal bias is included. For demonstration purposes, we present results for these two representative multipoles. We define the difference in SNR as
\begin{equation*}
    \Delta[\mathrm{SNR}]_L^m=\left.[\mathrm{SNR}]_L^m\right|_{b_t=0}-
    \left.[\mathrm{SNR}]_L^m\right|_{b_t\neq 0}\, .
\end{equation*}
For the results shown in Fig.~\ref{fig:snr_bt}, we fix the redshift to $z=0.7$, choose $k_1 = 0.3\,h\,\mathrm{Mpc}^{-1}$, $f_{R0}=10^{-5}$, and $b_t=-0.1$ and set all other parameters ($b_1$, $b_2$, $\sigma_p$, and $n_g$) to the fiducial values listed in Table~1 of~\cite{Yankelevich:2018uaz}.

\begin{figure*}
    \centering
    \subfloat{\includegraphics[width=0.40\textwidth]{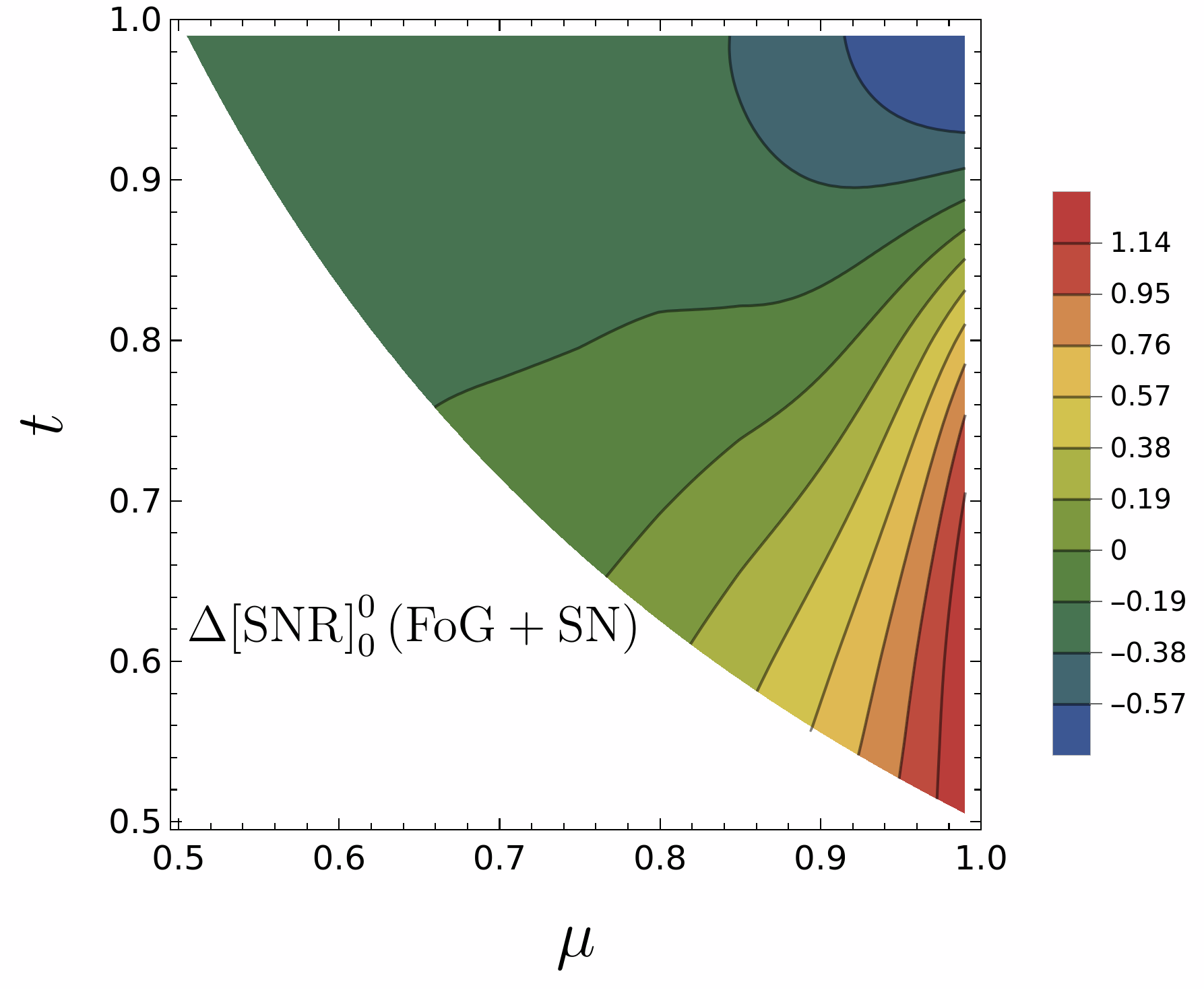}}
    \subfloat{\includegraphics[width=0.40\textwidth]{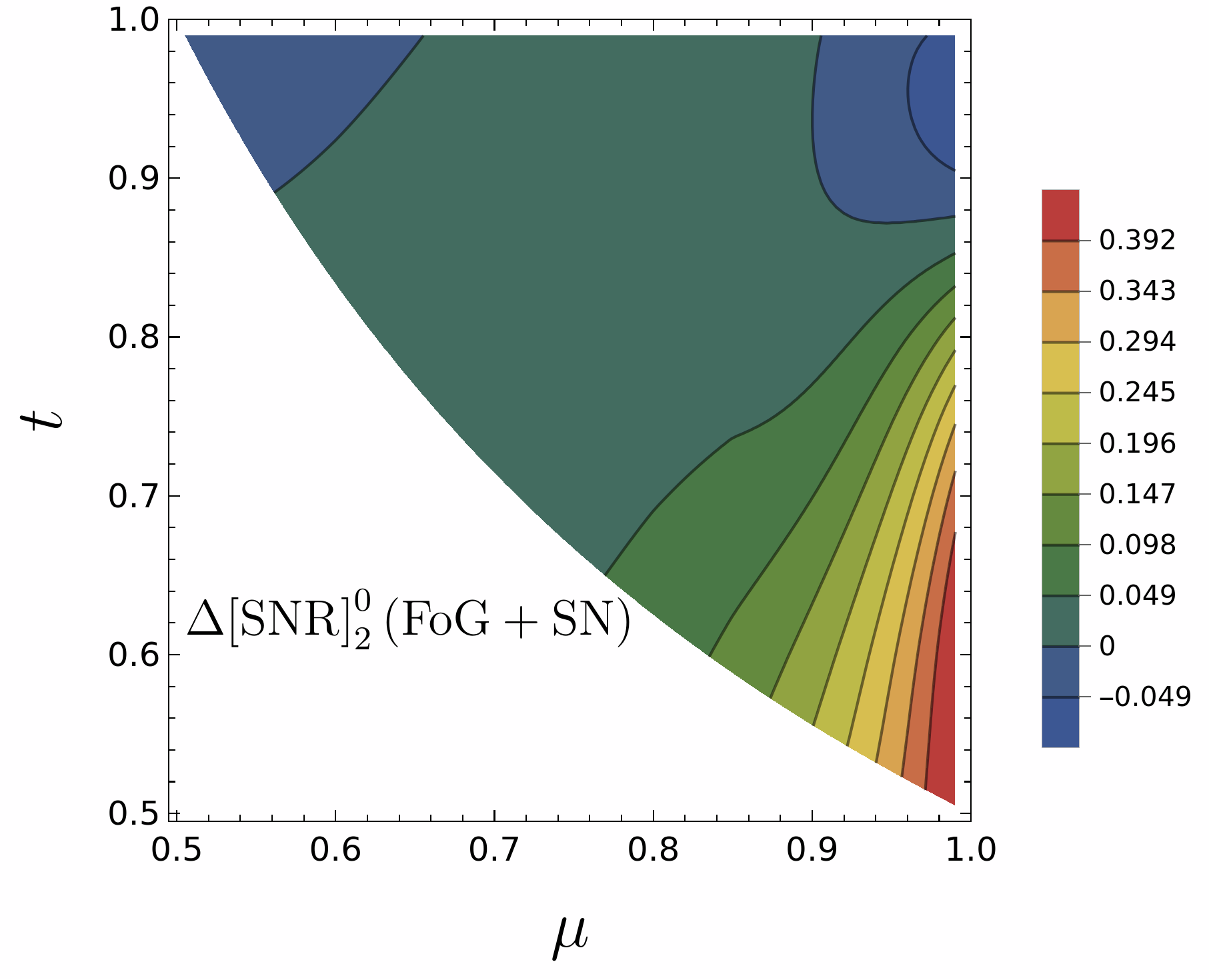}}
    \caption{\justifying
    Change in the signal-to-noise ratio with and without the tidal bias term $(b_t =-0.1)$ for the $B_0^0$ (left) and $B_2^0$ (right) multipoles at redshift $z=0.7$ and $k_1 = 0.3\,h\,\mathrm{Mpc}^{-1}$. The remaining parameters ($b_1$, $b_2$, $\sigma_p$, and $n_g$) are fixed to the values given in Table~1 of~\cite{Yankelevich:2018uaz} at $z=0.7$.}
    \label{fig:snr_bt}
\end{figure*}

%%%%%%%%%%%%%%%%%%%%%%%%%%%%%%%%%%%%%%%%%%%%%%%%
\section{summary and discussion}
\label{sec:summary}
%%%%%%%%%%%%%%%%%%%%%%%%%%%%%%%%%%%%%%%%%%%%%%%%
In this work, we present a detailed perturbative analysis of the galaxy bispectrum multipoles in the Hu–Sawicki $f(R)$ gravity model. We focus on their detectability in redshift space for a \textit{Euclid}-like galaxy survey. We adopt a spherical-harmonic decomposition~\cite{Mazumdar:2022ynd,Pal:2025hpl} technique that directly captures the anisotropies induced by redshift-space distortions, and allows us to study the impact of modified gravity across the full $(L,m)$ multipole hierarchy and for different triangle configurations.

Our formalism includes the scale- and time-dependent modifications to the second-order perturbation theory kernels $F_2$ and $G_2$ in HS-$f(R)$ gravity, as well as the nonlinear source term corrections responsible for chameleon screening. At linear order, the growth factor $f(k)$ deviates from its GR counterpart at scales $k\gtrsim 0.01 \, h \,{\rm Mpc}^{-1}$, with differences of $10$–$20\%$ for $f_{R0}\in[10^{-7},10^{-4}]$ at $z=0$. These deviations propagate into the non-linear coupling kernels, which generate distinct signatures in the bispectrum multipoles. We also include observational effects through Finger-of-God damping, modeled with a Gaussian velocity dispersion, and Poisson shot noise from galaxy discreteness.

We find that the kernels $F_2$ and $G_2$ in HS-$f(R)$ differ most strongly from their $\Lambda$CDM counterparts in stretched configurations and in linear triangle limits. These deviations grow toward low redshift, when screening becomes less efficient and the fifth force operates more strongly, while they remain suppressed at high redshift. This redshift evolution directly imprints on the bispectrum multipoles, with stronger signals expected at late times.

The monopole $B_0^0$ and quadrupole $B_2^0$ carry the clearest and strongest signatures of modified gravity. At $z=0.7$ with $f_{R0}=10^{-5}$ and $k_1=0.3 \, h \,{\rm Mpc}^{-1}$, deviations reach $\sim 2\%$ for $B_0^0$ and $\sim 8\%$ for $B_2^0$ in stretched configurations, even after FoG suppression. Higher multipoles such as $B_2^1$, $B_2^2$, $B_4^0$, and $B_4^1$ also show distinct patterns in $\mu-t$ space, but their amplitudes are typically reduced by factors of two to three once FoG effects are included. While individually weaker, these multipoles still add complementary information because their signatures do not overlap completely with those of lower-order multipoles or other cosmological parameters.

Another important outcome is the role of nuisance (or model) parameters. The bispectrum multipoles depend strongly on galaxy bias and velocity dispersion. The parameter $\gamma_2=b_2/b_1$ and the velocity dispersion $\sigma_p$ in particular generate degeneracies with $f_{R0}$. For instance, large values of $\sigma_p$ almost wash out the $f(R)$ signal in the reduced bispectrum monopole $Q_0^0$. This shows that realistic analyses must fit for $f_{R0}$ jointly with nuisance parameters, rather than treating them independently, to avoid biased constraints.

We also present forecasts of signal-to-noise ratios. For a \textit{Euclid}-like survey, the monopole $B_0^0$ achieves the highest SNR, reaching values of around $30$ for stretched and linear triangles when $f_{R0}=10^{-5}$ at $z=0.7$. The quadrupole $B_2^0$ follows with SNR around $15$. Higher multipoles rarely exceed SNR values of $3-5$. 
The quoted SNR values correspond to triangles with $k_1 = 0.3\,h\,\mathrm{Mpc}^{-1}$ at redshift $z = 0.7$ for all multipoles except the hexadecapoles, for which we use $k_1 = 0.4\,h\,\mathrm{Mpc}^{-1}$.
The overall detectability increases toward lower redshifts and is maximized for stretched and squeezed configurations, while equilateral configurations remain less sensitive. For $f_{R0}\lesssim10^{-7}$, the detectability falls below unity, and the model effectively converges to GR.

Taken together, these results show that the bispectrum multipoles, especially $B_0^0$ and $B_2^0$, are powerful probes of modified gravity. Their angular and configuration dependence provides signatures that $\Lambda$CDM or nuisance effects cannot easily reproduce. This makes the bispectrum a natural complement to the power spectrum, which often suffers from degeneracies with bias and FoG damping. By combining two- and three-point statistics, one can break degeneracies and achieve stronger constraints on $f(R)$ gravity.

At the same time, the analysis highlights the importance of systematic effects. FoG damping significantly reduces the amplitude of the MG signal, especially for the most sensitive linear configurations, making accurate modeling of velocity dispersion essential. Similarly, uncertainties in nonlinear galaxy bias affect the interpretation of bispectrum signals, so marginalization or simultaneous fitting is unavoidable in real analyses.

Finally, we note that our analysis is based on a tree-level model for the bispectrum multipoles. Importantly, the tree-level contribution remains present even at scales where loop corrections become significant ($k \gtrsim 0.1\,h\,\mathrm{Mpc}^{-1}$ at $z=0.7$), meaning our tree-level results represent a baseline estimate for the expected signal. By extending our forecasts to $k_1 = 0.3\,h\,\mathrm{Mpc}^{-1}$ (and $0.4\,h\,\mathrm{Mpc}^{-1}$ for the hexadecapole), we demonstrate that even this baseline tree-level signal exhibits strong discriminatory power for MG through the bispectrum multipoles. Exploring beyond tree-level corrections in future work will be important to fully quantify these findings, as baryonic effects and additional nuisance parameters on small scales can introduce complex degeneracies. A full treatment beyond tree-level would be required for rigorous parameter constraints from real survey data, which we identify as a natural and promising extension of this framework.
Simulations of MG cosmologies that calibrate higher-order kernels will also be important for reducing theoretical uncertainties. In the longer term, joint analyses that combine bispectrum multipoles with power spectrum multipoles, weak lensing, and other observables will maximize the sensitivity of upcoming surveys and ensure robust constraints.

In conclusion, the galaxy bispectrum multipoles open a promising avenue for testing gravity on cosmological scales. In HS-$f(R)$ gravity, the monopole and quadrupole carry detectable imprints of the fifth force in \textit{Euclid}-like surveys, with the strongest signals appearing in stretched and squeezed configurations at intermediate scales and redshifts. Higher multipoles add complementary signatures that, while weaker, enrich the parameter space coverage. Together, the bispectrum multipole hierarchy strengthens cosmological tests of GR and provides a complementary probe to two-point clustering, enhancing our ability to constrain deviations from standard gravity and to understand the physics behind cosmic acceleration.

%%%%%%%%%%%%%%%%%%%%%%%%%%%%%%%%%%%%%%%%%%%%%%%%%%%%%%%%%%%
\section*{Acknowledgements}
%%%%%%%%%%%%%%%%%%%%%%%%%%%%%%%%%%%%%%%%%%%%%%%%%%%%%%%%%%%
We thank Supratik Pal for useful discussions at the initial stages. The authors also thank the anonymous referee for valuable comments and insights that helped improve the manuscript.
SP acknowledges CSIR for financial support through Senior Research Fellowship (File no. 09/093(0195)/2020-EMR-I). 
DS acknowledges the support of the Canada 150 Chairs program, the Fonds de recherche du Québec Nature et Technologies (FRQNT) and the Natural Sciences and Engineering Research Council of Canada (NSERC) joint NOVA grant, and the Trottier Space Institute Postdoctoral Fellowship program.
AA ackonowledges SECIHTI grant CBF2023-2024-16 and UNAM-PAPIIT grants No. IA101825 and No.IG102123.

\bibliographystyle{apsrev4-2}
 \bibliography{refs}  

\appendix

\end{document}